\documentclass[referee]{aa}
\usepackage{graphicx}
\usepackage{aalongtable,lscape}
\bibliography{references,more_references}
\begin{document}
\title{Chemical abundances in LMC stellar populations I. The Inner disk sample
\thanks{Based on observations collected at the VLT UT2 telescope
(072.B-0608 and 066.B-0331 programs), Chile.}
}
\author{Luciana Pomp\'eia \inst{1}, \inst{2},
        Vanessa Hill \inst{3},
        Monique Spite \inst{3},
    Andrew Cole \inst{4,5},
    Francesca Primas \inst{6},
    Martino Romaniello \inst{6},
    Luca Pasquini \inst{6}
    Maria-Rosa Cioni \inst{7}
    \and
    Tammy Smecker Hane \inst{8}
        }
	\institute{IP\&D, Universidade do Vale do Para\'iba, Av. Shishima Hifumi, 2911, S\~ao,
         Jos\'e dos Campos, 12244-000 SP, Brazil\\
         \and
	 Instituto Astron\^omico e Geof\'isico (USP),
         Rua do Mat\~ao 1226, Cidade Universit\'aria, 05508-900 S\~ao Paulo, Brazil\\
     \email{pompeia@univap.br}
         \and
         Observatoire de Paris-Meudon, GEPI and CNRS UMR 8111, 92125 Meudon Cedex, France\\
         \email{Vanessa.Hill@obspm.fr}
         \email{Monique.Spite@obspm.fr}
     \and
     School of Mathematics and Physics, University of Tasmania, Private Bag 37, Hobart, TAS 7001, Australia
     \and
     Kapteyn Astronomical Institute, University of Groningen, Postbus 800, NL-9700 AV Groningen, Netherlands\\
         \email{cole@astro.rug.nl}
     \and
     European Southern Observatory, Karl Schwarschild Str. 2, 85748 Garching b. M\"unchen, Germany\\
         \email{fprimas@eso.org}
         \email{mromanie@eso.org}
         \email{lpasquin@eso.org}
     \and
     Edinburg
     SUPA, School of Physics, University of Edinburgh, IfA, Blackford Hill, Edinburgh EH9 3HJ, UK\\
         \email{mrc@roe.ac.uk}
     \and
     Department of Physics and Astronomy, 4129 Frederick Reines Hall, University of California, Irvine, CA 92697-4575\\
     \email{smecker@carina.ps.uci.edu}
     }
\offprints{L. Pompeia, \email{pompeia@univap.br}}
\date{Received/Accepted}

\abstract{
$Aims.$ We have used FLAMES (the Fibre Large Array Multi Element
Spectrograph) at the VLT-UT2 telescope to obtain spectra of a
large sample of red giant stars from the Inner Disk of the LMC, $\sim$2 kpc
from the center of the galaxy. We investigate the chemical abundances of
key elements for the understanding of the star formation and evolution of the
LMC disk: heavy and light [$s$-process/Fe] and [$\alpha$/Fe] give
constraints on the time-scales of formation of the stellar population.
Cu, Na, Sc and the iron-peak elements are also studied aiming to better understand
the build up of the elements of this population and the origin of these elements.
We aim to provide a more complete picture of the LMC's evolution by compiling a
large sample of field star abundances.

$Methods.$ LTE abundances are derived using line
spectrum synthesis or equivalent width analysis. We have used OSMARCS model atmospheres
and an updated line list.

$Results.$ We have found that the alpha-elements Ca, Si, and Ti show lower [X/Fe] ratios
than Galactic stars at the same [Fe/H], with most [Ca/Fe] being
subsolar. [O/Fe] and [Mg/Fe] ratios are slightly deficient,
with Mg showing some overlap with the Galactic distribution. Sc
and Na follow the underabundant behavior of Ca, with subsolar
distributions. For the light $s$-process elements Y and Zr, we have found
underabundant values compared to their Galactic counterparts. [La/Fe] ratios
are slightly overabundant relative to the galactic pattern showing
low scatter, while the [Ba/Fe] are enhanced, with a slight increasing trend
for metallicities [Fe/H] $>$ -1 dex. The [heavy-$s$/light-$s$]
ratios are high, showing a slow increasing trend with
metallicity. We were surprised to find an offset for three of the
iron-peak elements. We have found an offset for the
[iron-peak/Fe] ratios of Ni, Cr and Co, with an underabundant pattern
and subsolar values, while Vanadium ratios track
the solar value. Copper shows very low abundances in our sample
for all metallicities, compatible with those of the Galaxy only
for the most metal-poor stars. The overall chemical distributions
of this LMC sample indicates a slower star formation history relative to
that of the solar neighborhood, with a higher contribution from Type Ia
supernovae relative to Type II supernovae.

\keywords{Stars: abundances,
Galaxies: Magellanic Clouds, Galaxies: abundances, Galaxies:
evolution}
   }
\authorrunning{Pomp\'eia et al.}
\titlerunning{Abundances of Stars in the LMC Disk}
\maketitle
\section{Introduction}

During the last decade, due to the operation of the new class of
large telescopes, we have witnessed for the first time the
analysis of elemental abundances of large samples of individual stars in external
galaxies. Thanks to new optical technologies, objects fainter than
supergiant stars, planetary nebulae or HII regions are now
possible targets suitable for extragalactic research, allowing the study
of older objects and the exploration of earlier phases of galaxy evolution.
The abundance patterns of diverse elements in numerous
stars in a galaxy give information on different domains such as
the kinematic and chemical evolution, nucleosynthesis channels,
the star formation history (SFH) and the initial mass
function (IMF) of its stellar population(s).

One of the most interesting extragalactic objects in the study of
stellar populations is the Large Magellanic Cloud (LMC), our
nearest companion after the Sagittarius dwarf galaxy (that is
the process of merging with the Milky Way). The LMC is an irregular galaxy
located within 50 kpc from the Sun, with a kinematically-defined disk,
a bar and a thick disk or flattened halo (e.g. Westerlund 1997).
The almost face-on position of its disk, with a tilt relative to the plane
of the sky of $\sim$30$^{\rm o}$, gives us the precious opportunity to study
stars from its different components.

The star formation (SF) and cluster formation histories of this galaxy have been studied
for more than three decades (e.g. Butcher 1977, van den Bergh 1979, Olszewski
et al. 1996 and references therein, Cioni et al. 2006 and references therein)
although a final picture is far from complete (the current status
of the research deals with the detailed SF and cluster formation within
the different components and regions of this galaxy, e.g. Geha et al. 1998,
Smecker-Hane 2002, Subramaniam 2004, Javiel et al. 2005, Cole et al. 2005).
The clusters of the LMC show a ancient population with ages $>$ 11.5 Gyr,
followed by an hiatus when just one single cluster
seems to have formed (ESO 121-SC03) (e.g. van den Bergh 1998 and references
therein). Some 2-4 Gyr ago, a new formation event was triggered and some other
clusters have been built (e.g. Da Costa 1991). The
SF in the disk field shows a different evolution, with nearly
constant rate over most of the history of the LMC (Geha et al.
1998).  The SFR appears to have been enhanced some 1--4 Gyr ago, with
the timing and amplitude of the `burst' seeming to vary between
locations (Holtzman et al. 1999; Olsen et al. 1999).  The SFH of the
bar field appears to more closely track the cluster formation
history, with a strong burst $\approx$3-6 Gyr ago (Smecker-Hane et
al. 2002; Cole et al. 2005).   The lack of a field star age gap means
that field star properties can be used to trace the history of the
LMC during the 3-11 Gyr cluster age gap (Da Costa 1999; van den
Bergh 1999).

The elemental distributions of the LMC stars are still poorly known, due to
the paucity of data, but the present picture is in fast change due to new
observational programs (e.g. Evans et al. 2005, Dufton et al. 2006, Johnson et al. 2006).
We briefly sumarize here the results on elemental abundances in the LMC
(for a detailed discussion see Hill 2004). The abundance analysis of B stars and HII
regions (Garnett 1999, Korn et al. 2002, Rolleston et al. 2002)
show a deficient abundance of O, Mg and Si relative to their solar
neighborhood counterparts\footnote{taking as the solar value log (O/H) $\sim$ 8.83
(Grevesse \& Sauval 2000)}, with mean log (X/H) - log (X/H)$_{\odot}$ $\sim$ -0.2 dex for
oxygen, -0.2 dex for magnesium, and -0.4 dex for silicon abundances
(this last value is only for the B stars, HII regions show a much lower value of $\sim$
-0.8 dex), but compatible to galactic supergiant values. Russell \& Dopita (1992), Hill et al.
(1995) and Luck et al. (1998) studied samples of supergiants in the field of the LMC
and Hill \& Spite (1999) derived abundances for supergiants in clusters.
They found a similar behavior for the $\alpha$-elements when compared to the galactc disk
values, while for the heavy elements (those with Z $\geq$ 56), the abundance ratios are
enhanced by a factor of $\sim$ 2. Huter et al. (2007) derived C, Mg, O, Si and N abundances for three
globular clusters from the LMC, and found an average value 0.3 dex lower than that of the
Galactic Clusters for all the analysed elements, except for N.
Red giant branch stars from the field (Smith et al. 2000, hereafter
SM02) and from globular clusters (Hill et al. 2000, 2003, hereafter H00 and H03, and Johnson et al. 2006,
hereafter JIS06) have also been studied. A general behavior of low [$\alpha$/Fe] ratios compared to the
stars of the galactic disk with similar metallicities is detected (with the exception of Si and Mg in JIS06),
while for the heavy-elements, the same overabundant pattern found for the LMC supergiants has been derived. 
JIS06 inferred the [Y/Fe] ratios and found abundances compatible to the solar value.
Na abundances are different in field and clusters stars. While SM02 found low [Na/Fe]
ratios and [Sc/Fe] ratios close to zero, JIS06 found that [Sc/Fe] and [Na/Fe] ratios are
simillar to their galactic counterparts. JIS06 have derived the [iron-peak/Fe] abundances and found
that Ni, V and Cu abundances fall bellow their corresponding galactic values.

An observational project aiming at making the full analysis of the elemental abundances
of significant samples ($\sim$70-100) of stars from different locations in the LMC has been
developed, taking advantage of the FLAMES multiplex facility at the
VLT. We have obtained spectra from stars in three different regions of
the LMC: the Inner Disk (characterised by a galactocentric radius of
R$_{\rm C}$=2kpc); the Outer Disk (with R$_{\rm C}$=4kpc); and a field
near the optical center of the Bar. Stars have been selected based on
kinematics and metallicity data derived from the near-infrared calcium
triplet (CaT and CaT metallicities), trying to sample as evenly as possible
the whole metallicity range of this galaxy.
In the present paper we focus on a sample of Red Giant Branch (RGB) stars on the Inner Disk region, previosly studied by Smecker-Hane et al. (2002), who derived the ages, metallicities
(CaT) and kinematics of this sample. They have identified two kinematical groups in the
Inner Disk field, one with velocity dispersion of 13$\pm$4 km/s, characterizing
a thin disk, and one with velocity dispersion of 34$\pm$6 km/s, probably pertaining
to the flattened halo. The metallicities of these two groups are different: the low-dispersion
velocity group has mettalicities ranging -0.6 $\leq$ [Fe/H] $\leq$ -0.3 dex, while
the high-dispersion velocity component has -2 $\leq$ [Fe/H] $\leq$ -0.4. The ages derived for
this Inner Disk population has shown that stars have continuously formed during the last
$\sim$ 1 to 15 Gyr, with a possible enhancement in the star formation rate (SFR) some 3 Gyr
ago.

As the prototype galaxy of the Magellanic irregular class,
to learn the evolutionary history of the LMC is clearly
a vital step towards the global understanding of galaxies near
the dwarf-giant boundary.  Additionally, because the Magellanic
Clouds have evolved in such close proximity to the Milky Way,
their histories have been intimately tied to that of our own galaxy.
The ongoing impact of the LMC on the structure and kinematics
of the Milky Way is manifest in the warp of the Galactic disk and
possibly in the presence of the central bar (e.g., Weinberg 1999), while
Bekki \& Chiba (2005) have used N-body simulations to show that the
LMC could have made a significant contribution to the build up of the
Milky Way halo as a result of tidal stripping.

According to models of galaxy formation within a hierarchical
CDM scenario (D'Onghia \& Lake 2004; Moore et al.\ 1999), the
history of the Milky Way depends strongly on its interactions with
its environment.  It now seems that the abundance patterns in
dwarf spheroidal stars are dissimilar to those in Milky Way halo
stars (e.g., Shetrone et al.\ 2001; Tolstoy et al.\ 2003; Geisler et al.\ 2005),
ruling them out as analogues to the accreting fragments that built
up the halo.  Study of the LMC takes on added significance in
this light, because of the hypothesis by Robertson et al.\ (2005)
that the accretion of LMC-like fragments circumvents this difficulty
with the hierarchical accretion scenario.  Deeper knowledge of the
abundances in the oldest LMC stars therefore has direct bearing on
the evolution of our own Galaxy.

In the present paper, we focus on a sample of RGB stars in
the Inner Disk field, previously studied by Smecker-Hane et al.
(2002, hereafter SMH02), who derived the SFH of the region from  
Hubble Space Telescope color-magnitude diagrams; they find stars in
this field to have formed continuously over the whole life of the
LMC, with a slight enhancement in the star formation rate (SFR) $
\approx$3 Gyr ago.  Smecker-Hane et al. (2007)
obtained CaT spectra for a large number of red giants to measure
their kinematics and overall heavy element abundances, finding the
most metal-rich stars to belong to a kinematically cold population
and the metal-poor stars to be more kinematically hot, possibly
belonging to a flattened halo or very thick disk population.
We focus our work on the Inner Disk region, presenting abundance results for
iron-peak, heavy and light $s$-process elements, and $\alpha$
elements for a total of 59 stars. With
this detailed information in hand, we aim to shed light
on the following questions: (i) what are the chemical
abundance patterns of the Inner Disk of the LMC?; (ii) what do these
elemental distributions tell us about the formation and evolution
of the LMC?; (iii) are they similar to any component of the
Milky Way?; (iv) or to the populations of other Local Group
galaxies?; (v) based on the elemental distributions, is a merging
scenario with LMC debris a likely solution for the Galactic Halo
formation?

The paper is organized as follows: in Sect. 2 the observations and
the reduction procedure are described; in Sect. 3 the calculation
of stellar parameters is presented; Sect. 4 describes the
abundance determination procedures; Sect. 5 reports the results
for the abundance ratios, comparing to Milky Way samples; in
Sect. 6 we compare our results to those for the dSph galaxies; we
discuss the results in Sect. 7; and finally in Sect. 8 a summary
of the work is given.

\section{Sample selection, observations and reductions}

\subsection{Sample selection}

To best measure the elemental abundances of the LMC disk and their
evolution along time, we selected a field located $1.7^{\circ}$
southwest of the LMC Bar, in the Bar's minor axis direction to
ensure a negligible contribution of its stellar populations.
An HST color-magnitude diagram study of this field (SMH02)
found it to have experienced a rather
smooth and continuous history of star formation over the past 13 Gyr,
with a possibly increased star-formation rate over the last 2 Gyr.
This stands in contrast to the history of the Bar itself, in which
significant star-formation episodes are seen to have commenced 4--6
Gyr ago (SMH02; see also Holtzman et al.\ 1999 and references therein).
This field has also more recently been the target of a low-resolution
spectroscopy campaign (Cole et al.\ 2000; SMH), using the CaT
to derive its metallicity distribution and break the
age-metallicity degeneracy inherent to color magnitude diagram CMD analyses.

\begin{figure}[H]
\centerline{%
\begin{tabular}{c@{\hspace{1pc}}c}
\includegraphics[width=2 in, angle=-90]{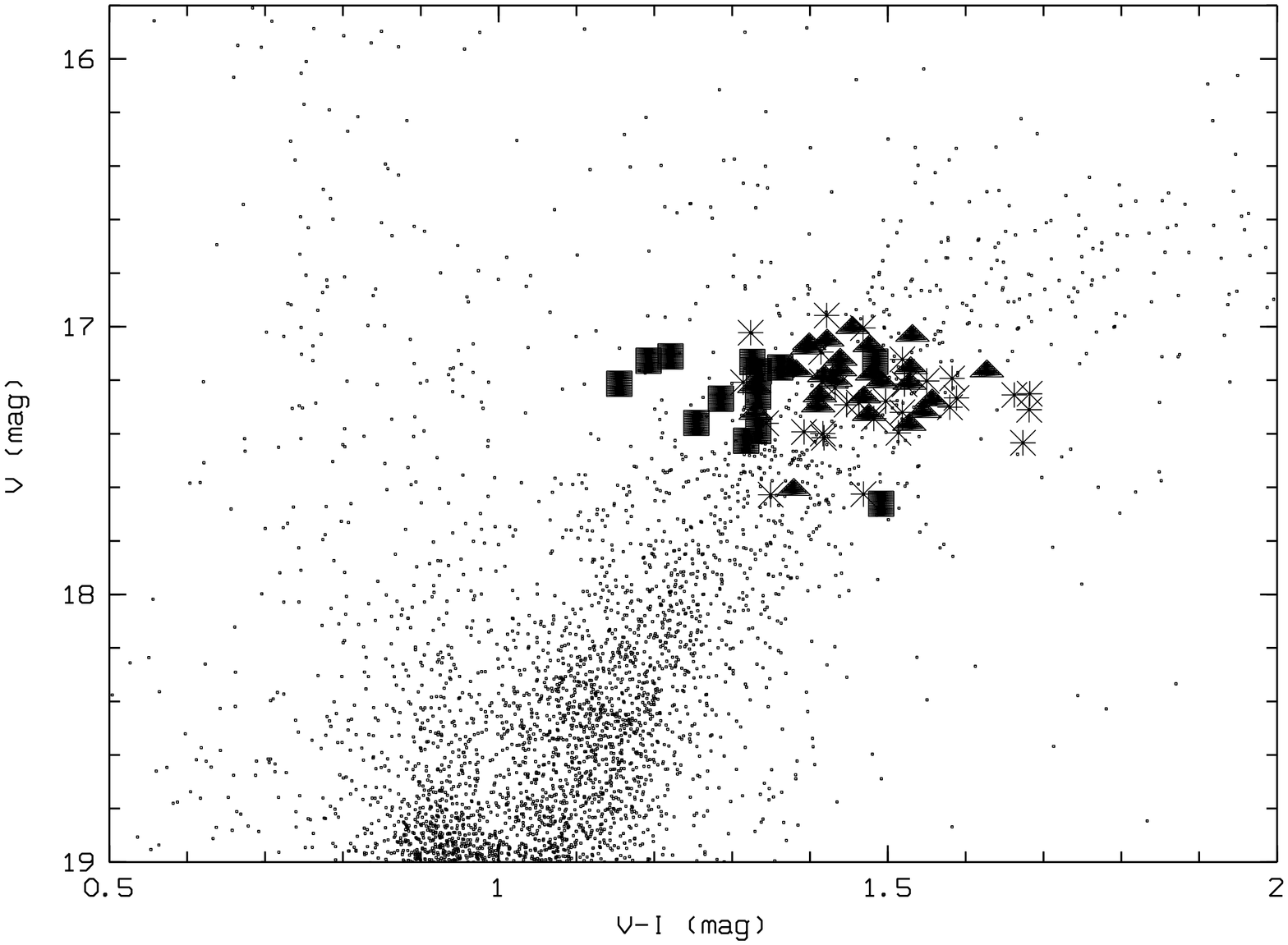} \\
\end{tabular}
}
\caption{V (V$-$I) color-magnitude diagram of the Disk region
(following SMH02), with our sample stars overplotted:
{\it asterisks} are stars with [Fe/H]$_{CaT} \geq - 0.5$~dex,
{\it triangles}$-1.0 \leq$[Fe/H]$_{CaT} < - 0.5$~dex,
{\it squares} [Fe/H]$_{CaT} < - 1.0$~dex.}
 \label{CMD}
\end{figure}

We have used these infrared CaT metallicities from SMH to
select a sample of red giant branch members of the LMC
(based on their radial velocities) distributed
uniformly (i.e., with the same number of stars
in each metallicity bin) over the whole metallicity
range of the LMC disk.
In this way, we have been able to sample
the lower metallicity bins of the LMC very efficiently. The most metal-poor stars
convey essential information on the evolution of the elements of this galaxy,
but they are rare, hence their number would have been significantly
lower if we had selected our sample by picking stars randomly
across the RGB.
The final sample consists of 67 stars with CaT metallicities
ranging from $-$1.76 to $-$0.02~dex (including 13 stars
with metallicities below $-$1.0~dex), drawn from the 115-star
sample of SMH. In Fig.~\ref{CMD} we show the sample stars overplotted on
the color-magnitude diagram of the LMC inner disk region
(CTIO photometry from SMH).
The sample mean magnitude is V=17.25~mag, bright enough to
allow reasonable S/N high-reslution spectra to be acquired.

\subsection{Observations and reductions}

The observations were made at the VLT Kueyen (UT2) telescope at
Paranal during the Science Verification of FLAMES/GIRAFFE (Pasquini
et al. 2000) in
January, February and March, 2003, complemented by one night of the
Paris Observatory Guaranteed Time Observations in January, 2004.
In its MEDUSA mode, GIRAFFE is a multiobject spectrograph
with 131 fibers of which 67 were used for the present
project. The remaining fibers were allocated to targets
of other Science Verification projects in the LMC.
The detector is a 2048 $\times$ 4096 EEV CCD with 15$\mu$m pixels.
We used the high resolution grating of GIRAFFE in three different
setups: (i) H14 $\lambda$638.3 - $\lambda$662.6 nm with R=28800;
(ii) H13 $\lambda$612.0 - $\lambda$640.6 nm with R=22500, and
(iii) H11 $\lambda$559.7 - $\lambda$584.0 nm with R=18529.
Exposure times are 6 hours for H14 and H13 setups and 7h30
for H11. The
setups were chosen in order to cover the maximum number of key
elements such as Fe I and Fe II for spectroscopic calculations of
stellar parameters, and $\alpha$, iron-peak and $s$-process
elements, for the abundance analysis. The average signal to
noise ratio of the spectra is S/N$\sim$80 per resolution element.

The data reduction was carried out using the BLDRS (GIRAFFE Base-Line
Data Reduction Software http://girbldrs.sourceforge.net/) and
consists of bias subtraction, localization and
extraction of the spectra, wavelength calibration and rebinning.
We have also used the MIDAS packages for sky subtraction and
co-addition of individual exposures.

\section{Determination of stellar parameters}

\subsection{Photometric stellar parameters}

A first guess of the stellar parameters was made using photometric
data of CTIO (V,I from SHM) and 2MASS (J,H,K). Bolometric magnitudes and
effective temperatures were derived from calibrations of Bessell,
Castelli \& Plez (1998, hereafter BCP). The observed CTIO and 2MASS colors
were transformed into the corresponding photometric systems using
Fernie (1983, V$-$I Cousins to Johnson) and Carpenter (2001, K, V$-$K
\& J$-$K 2MASS). Photometric data are given in Table \ref{photometry},
while Table \ref{stellpar} gives the derived effective
temperatures (T$_{\rm phot}$) and surface gravities (log g$_{\rm phot}$):
T$_{\rm phot}$ is derived using the BCP calibration of the deredenned V$-$I
and V$-$K colors, and the surface gravity is computed using the following relation:
$$\log {\rm g_{phot}=4.44 +log(M) +4 \times log(T_{phot}/5790.) +0.4 \times (M_{bol}-4.75)},$$
 where $\rm M_{bol}$ is computed from the dereddened K magnitude of the star,
the bolometric correction BC$_{\rm K}$ taken from BCP, and the mass of the stars (M)
are assumed to be 2M$\sun$. A distance modulus based on Hipparcos data and the
period-luminosity relations from LMC Cepheids of 18.44 $\pm$ 0.05 mag is assumed
(Westerlund 1997, Madore \& Freedman 1998). Uncertainties of this value stems from the
specific subsets of the Cepheids chosen for the comparison (Madore \& Freedman 1998).
For the reddening, two values were checked:
E(B$-$V)~=~0.03, which was derived by SMH02 for the sample of the Inner Disk, using
Str\"omgren photometry, and E(B$-$V)~=~0.06, a mean value for the whole disk (Bessell 1991).
We adopted CaT metallicities from SMH as our initial guesses
and reported them in the [Fe/H]$_{\rm CaT}$ column of Table \ref{stellpar}.

We have derived temperatures from V$-$I, V$-$K and J$-$K colors. We have
found some trends when comparing temperatures from different
colors: T$_{\rm eff}$(V$-$I) is 65K hotter than T$_{\rm eff}$(V$-$K)
in the mean, with $\sigma$=59K; T$_{\rm eff}$(J$-$K) is 21K hotter than
T$_{\rm eff}$(V$-$K) and shows a highly dispersed relation, with
$\sigma$=118K (these numbers vary only slightly when choosing
a reddening of E(B$-$V)=0.06 or 0.03). As initial values of our stellar
temperatures we have chosen to use a weighted mean of the estimates
from V$-$I and V$-$K, omitting the less sensitive J$-$K color.
We assign higher weight to the more temperature-sensitive (V$-$K),
according to the following expression:
$${\rm T_{eff} = (T_{eff}(V-I)+ 2 \times T_{\rm eff}(V-K))/3.}$$
In Table \ref{stellpar} the inferred temperatures for the two values of
reddening, T$_{\rm photLow}$ and T$_{\rm phot}$ for E(B$-$V) = 0.03 and
0.06 respectively, are given.

\subsection{Spectroscopic parameters}

\begin{figure}[H]
\centerline{%
\begin{tabular}{c@{\hspace{1pc}}c}
\includegraphics[width=2 in, angle=270]{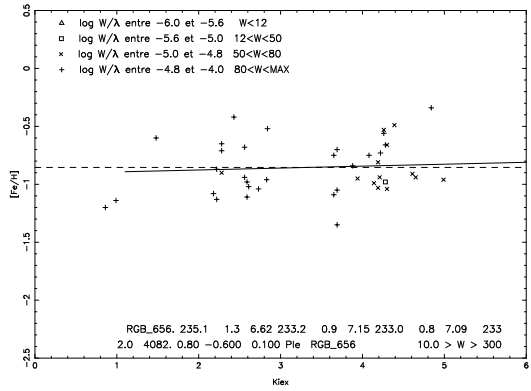} \\
\end{tabular}
}
\caption{Example of the temperature calculation for RGB\_625:
[Fe I/H] vs. $\chi_{exc}$.}
\label{temp625}
\end{figure}

\begin{figure}[H]
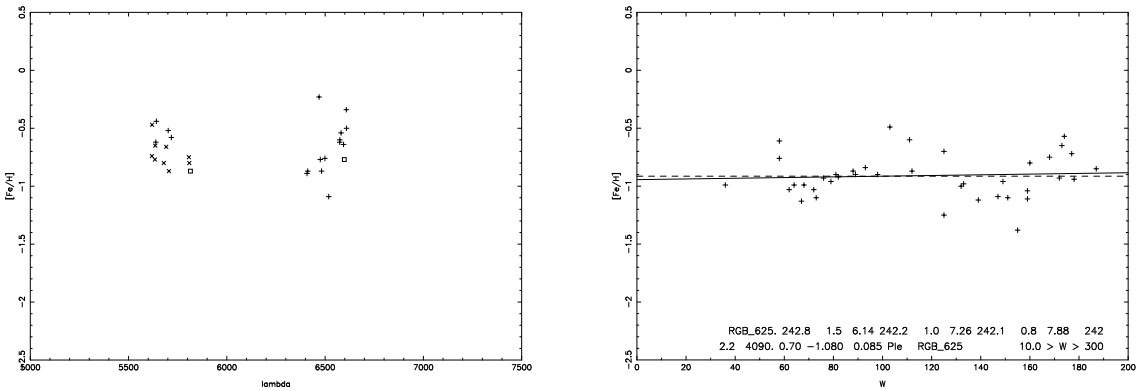

\centerline{%
\begin{tabular}{c@{\hspace{2.5pc}}c}
\includegraphics[width=2in,angle=270]{pompeiaLMC_fig2.ps} &
\includegraphics[width=2in,angle=270]{pompeiaLMC_fig3.ps} \\
\end{tabular}}
\caption{Examples of microturbulence velocity calculation for RGB\_710:
[Fe I/H] vs. $\lambda$ (left); and [Fe I/H] vs. EW (right). The different values
for the reduced EW in the left panel are given with different symbols:
1) squares: -5.6 $\leq$ log W/$\lambda$ $\leq$ -5.0 and 12$<$W$<$50;
2) crosses: - -4.8 $\leq$ log W/$\lambda$ $\leq$ -4.0  and 80$<$W$<$300;
and 3) times: -5.0 $\leq$ log W/$\lambda$ $\leq$ -4.8  50$<$W$<$80.
 }
\label{veloc710}
\end{figure}

The final stellar parameters used for the abundance determination of
the sample stars were derived spectroscopically using abundances derived
from the equivalent widths (EW) of iron lines. Although 67 stars were observed,
8 of them have one or two setups with low S/N, compromising the determination of stellar
parameters. These stars have not been included in the abundance analysis.
Due to low S/N ratios, the H13 setup has not been
used for the following stars: RGB\_601, RGB\_646, RGB\_672, RGB\_699,
RGB\_705, RGB\_710, RGB\_720, RGB\_731, RGB\_748, RGB\_756, RGB\_773 and
RGB\_775; and for RGB\_666 the H11 setup has been discarded. We have estimated
the stellar parameters as follows: effective temperatures are calculated by
requiring no slope in the A(Fe I) vs. $\chi_{exc}$ (excitation potential) plot ($\chi_{exc}$ is the excitation
potention of the line); microturbulent
velocities, $\rm V_{\rm t}$, are derived demanding that lines of different
EW give the same iron abundance, also checking for no slope in the
[Fe/H] vs. log(W/$\lambda$) plot (iron abundance vs. the reduced equivalent
width); and surface gravities are determined by forcing the agreement between
Fe~I and Fe~II iron abundances (within the accuracy of the abundance determination
of Fe~II). For the temperature and surface gravity ranges covered by our current
sample of stars, T$_{\rm eff}$ and log g
determinations are well correlated and the calculation of stellar
parameters is made iteratively. In Fig.~\ref{temp625} we show an example of the excitation
equilibrium calculation for RGB\_625, and in Fig.~\ref{veloc710}, the [Fe/H] vs. $\lambda$
with the log(W/$\lambda$) check, and the [Fe/H] vs. EW are given for RGB\_710.
The spectroscopic and photometric parameters of all our stars are
reported in Table \ref{stellpar}, together with the barycentric radial velocities
calculated from the spectra.

\subsubsection{Equivalent widths, line list and model atmospheres}

The EW of the lines and the radial velocities (RV, in km/s, reported in Table
\ref{stellpar}) of the stars are computed using the program
DAOSPEC\footnote{The documentation and details about this program can be found in
http://cadcwww.dao.nrc.ca/stetson/daospec/.}
written by Stetson (Stetson and Pancino, in preparation). The line list and
the atomic data were assembled from the literature and the oscillator strengths references
are given in Table \ref{linelist}. DAOSPEC has already been used to measure the
EW of spectra for different types of stars yielding reliable results (e.g. Pasquini
et al. 2004, Barbuy et al. 2006, Sousa et al. 2006). We have made a study of the DAOSPEC EW
estimations using GIRAFFE spectra. In the Appendix A we show a comparison of
DAOSPEC EW with those made by hand using the Splot - Iraf task for six of our
sample stars. We have found a very good agreement between the two methods for
the analysis of the GIRAFFE spectra within the expected uncertainties.

MARCS 1D plane-parallel atmospheres models Gustafsson et al. 1975, Plez et al.  1992,
Gustafsson et al. 2003) were kindly provided by B. Plez (private communication).

\subsubsection{Comparison with UVES analysis}

In a previous observing run (066.B-0331), we obtained UVES spectra (in slit mode)
for one of our sample stars, RGB\_666. UVES is an echelle spectrograph also mounted on
the VLT Kueyen telescope with a higher resolving power: R\,=\,45000 (with a slit of
1$\arcsec$) and a much wider wavelength coverage (in the case of our chosen set-up,
4800-6800\AA), and therefore with a better performance to derive equivalent widths.
We have used this spectrum to evaluate DAOSPEC performance to derive EW from low
resolution spectra. In Fig. 4, equivalent widths derived with DAOSPEC from UVES spectra from
5800 to 6800{\AA} are compared to those measured also by this program on the GIRAFFE spectra
of the same star using the same line list. In the
top of the plot, the mean differences between analyses are given together with the dispersion
and the number of lines used (lines of all elements are plotted in this comparison). We can see
from this plot that GIRAFFE EW are only slightly higher than UVES EW. Such a
difference is probably due to a better definition of the continuum for the UVES spectra,
as well as the increased blending at the lower resolution of GIRAFFE.
Using the EW of this figure, we have inferred the stellar parameters for UVES to compare the
analysis from both spectrographs. We have found that the stellar parameters are almost identical
to those given for RGB\_666 in Table \ref{stellpar}, except for V$_{\rm t}$ for which we have found
V$_{\rm t}$ = 1.8~kms$^{-1}$ (a difference of $\Delta$ V$_{\rm t}$ = +0.1~kms$^{-1}$) .
Comparing the results from the two spectrographs, we have:
[FeI/H]$_{\rm UVES}$-[FeI/H]$_{\rm GIRAFFE}$ = -0.11 dex and
[FeII/H]$_{\rm UVES}$-[FeII/H]$_{\rm GIRAFFE}$ = 0.00 dex. Therefore it is possible that a systematic
 uncertainty of [Fe/H] $\sim$ 0.1~dex may be present in the following abundance analysis, although robust
statements on this uncertainty would require better statistics. Let us further note that this 0.1~dex difference is within the errorbar that we quote for our GIRAFFE metallicities.

\begin{figure}
\centerline{%
\begin{tabular}{c@{\hspace{1pc}}l}
\includegraphics[width=2in]{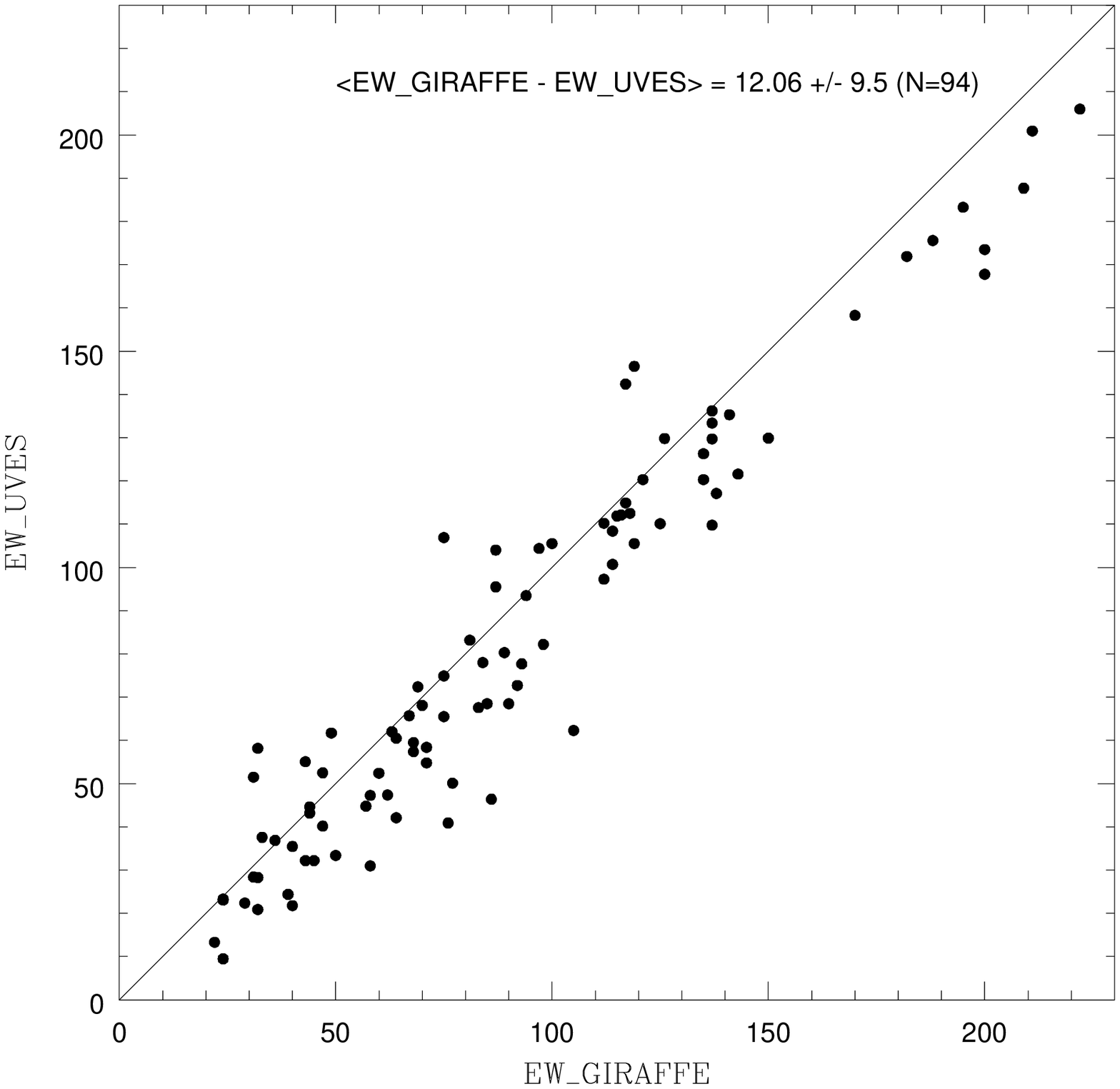}
\end{tabular}}
\caption{Comparison between UVES and GIRAFFE spectra analyses for RGB\_666.}
\end{figure}
\hspace{\fill}

\subsection{Behavior of stellar parameters}

\begin{figure}[H]
\centerline{%
\begin{tabular}{c@{\hspace{1pc}}c}
\includegraphics[width=4 in]{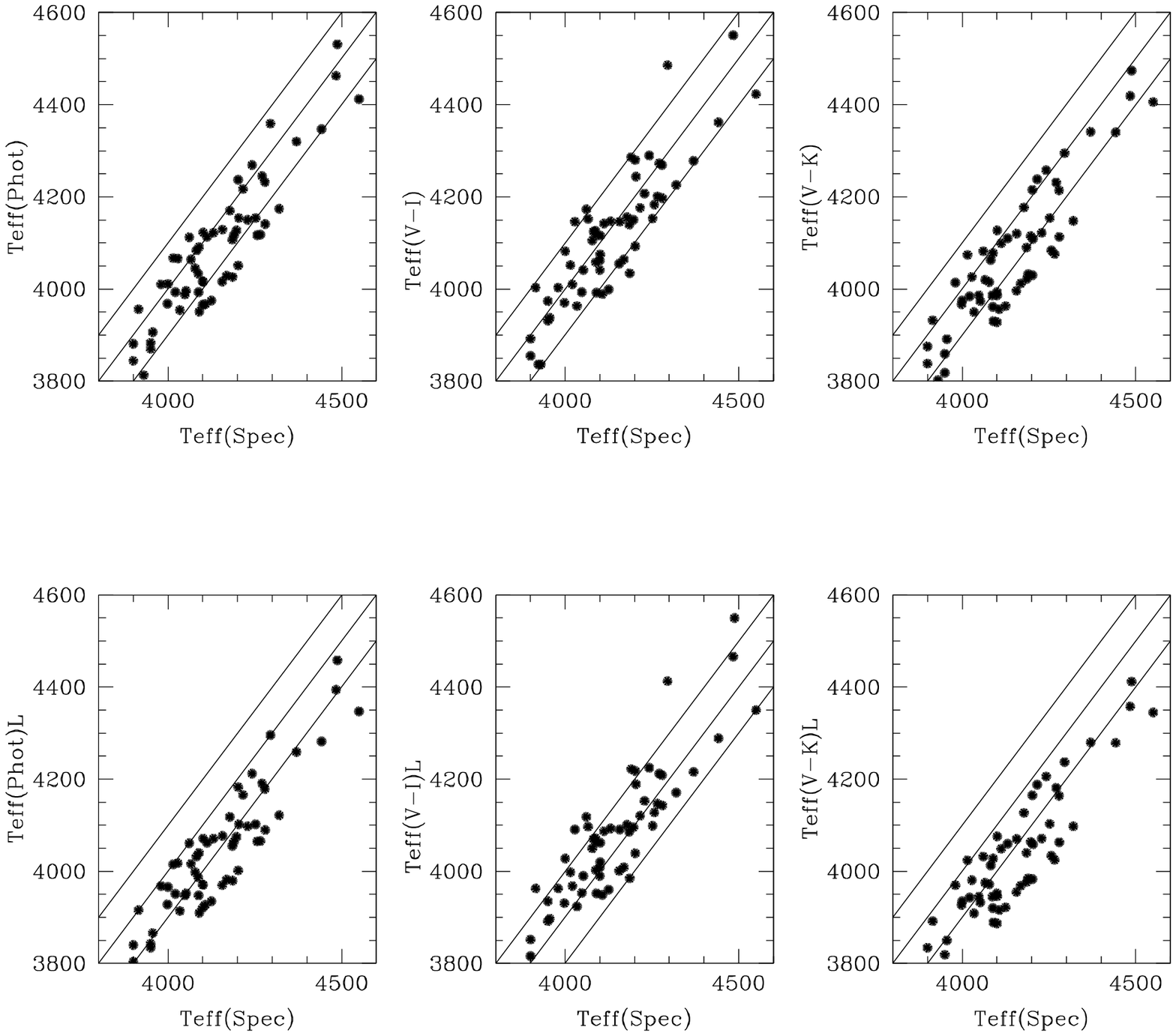} \\
\end{tabular}}
\caption{Comparison between photometric and spectroscopic
temperatures (see text). On the bottom plots, photometric temperatures
are derived with E(B$-$V)=0.03 (SMH02), while on the upper plots,
photometric temperatures are derived with a higher reddening value, E(B$-$V)=0.06
(Bessel 1991). Solid lines represent T$_{\rm eff(spec)}$ =
T$_{\rm eff(phot)}$ and T$_{\rm eff(spec)}$ = T$_{\rm eff(phot)}$ $\pm$ 100K.}
\label{photSpec_temp}
\end{figure}

We found good agreement between spectroscopic and photometric
temperatures. Our spectroscopic temperatures are hotter than
photometric temperatures derived using the low reddening value,
T$_{\rm photLow}$, by 113 K, with $\sigma$=91K, and by 54K than
T$_{\rm phot}$ (higher reddening value) with {$\sigma$=64K}. An
interesting result is depicted in Fig.~\ref{photSpec_temp} where 
we compare the
spectroscopic temperatures T$_{\rm eff}$(spec) with those derived
from colors, T$_{\rm eff}$(V$-$I) and T$_{\rm eff}$(V$-$K), and
from the equation given in Sect. 3.1.1, T$_{\rm eff}$(phot), for
both values of reddening (E(B-V)=0.06 in the upper panels,
and E(B-V)=0.03 in the lower panels). This figure shows that
photometric temperatures inferred using E(B$-$V) = 0.06 are in
much better agreement with spectroscopic temperatures than those
derived with the lower E(B$-$V). Provided that the photometric
temperatures and the excitation temperature
scale show a good agreement, this could indicate that E(B$-$V)=0.06
is a better reddening value for this region.

On average, spectroscopic surface gravities are lower than
the photometric estimates
by $\rm \Delta (\log g_{spec} - \log g_{phot}) = -0.38$ dex,
as might be expected if NLTE overionization effects are at work (Korn
et al. 2003).
This systematic effect in log g corresponds to a 0.2 dex
difference between FeI and FeII.

The metallicities that we derive differ on average from those derived
from the CaT by $\rm \Delta ([Fe/H]_{CaT} - [Fe/H]_{spec}) =
+0.13$ dex with $\sigma$= 0.27 dex. In fact, most of this effect comes
from the high-metallicity end of the sample: for
[Fe/H]$_{\rm CaT}>-0.6$~dex, CaT seems to overestimate the
metallicity systematically by 0.27~dex ($\sigma$= 0.19 dex), whereas for the
metal-poor end of the sample, there is almost no systematic effect
($\rm \Delta [Fe/H]_{CaT} - [Fe/H]_{spec}) = -0.04$~dex with
$\sigma$= 0.24 dex.

\begin{figure}[H]
\centerline{\includegraphics[width=3in]{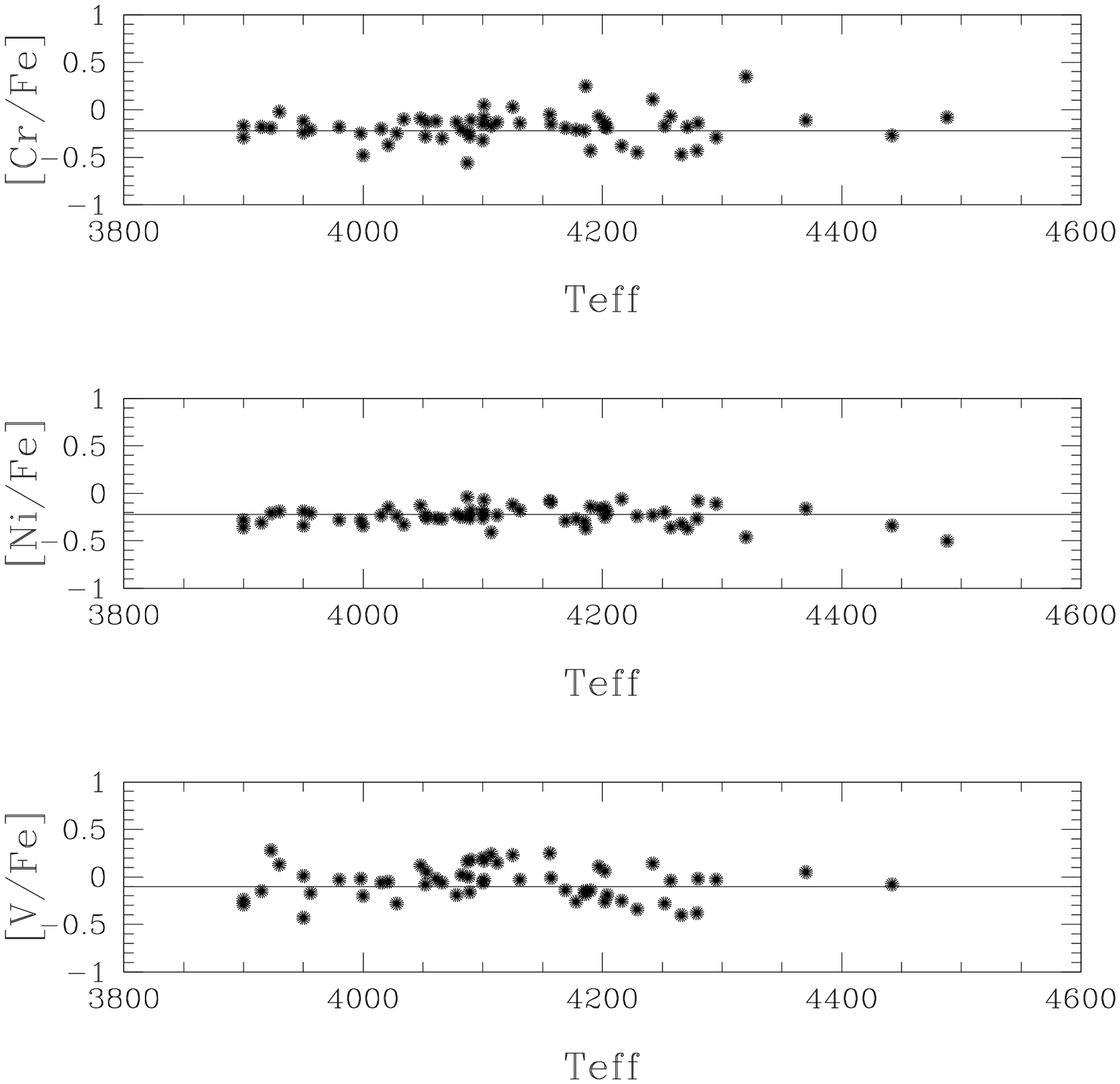} }
\caption{Abundance ratios against temperatures. From top to
bottom: [Cr/Fe] vs. T$_{\rm eff}$, [Ni/Fe] vs. T$_{\rm eff}$ and
[V/Fe] vs. T$_{\rm eff}$.}
\label{ElTef}
\end{figure}

Finally, in Fig.~\ref{ElTef}, abundance ratios of different species, [Cr/Fe],
[Ni/Fe] and [V/Fe], against temperatures are plotted in order to
check the quality of the spectroscopic temperatures. As can be
seen from this picture, there is no trend of the abundances of the
elements with temperature, which means that our temperatures are
well defined. Our final sample comprises 59 red giant stars within
$-1.7 < {\rm[Fe/H]} < -0.30$ dex and temperatures ranging from 3900 K
to 4500 K.

\section{Abundance determination}

We have selected a list of lines covering the chosen setups in
order to sample as much as possible the most important elements:
iron-peak, neutron-capture and $\alpha$ elements. Abundances are
derived from EW mesurements for eight elements (in parenthesis the
average number of lines used in the analysis): Fe (45), Ni (7), Cr
(4), V (11), Si (3), Ca (10), Ti (7) and Na (3). We have also
derived abundances by using line synthesis for nine elements (in
parenthesis the lines used in the synthesis): O ([O I] 6300$\rm
\AA$); Mg~I (5711 $\rm \AA$), Co~I (5647 $\rm \AA$), Cu~I (5782 $\rm
\AA$), Sc~II (5657 $\rm \AA$), La~II (6320 $\rm \AA$), Y~II (6435 $\rm
\AA$), Ba~II (6496 $\rm \AA$), and Zr~I (6134 $\rm \AA$).
The code used for the abundance analysis was developed by
Monique Spite (1967) and has been improved over the years.
We note that both model atmospheres and the line synthesis program are
in spherical geometry, so errors due to geometry inconsistencies are
minimized (Heiter \& Eriksson 2006).
For the synthesis of the [O I] line in 6300.311 $\rm \AA$, we have taken
into account the blend with Ni I 6300.336 $\rm \AA$ (line data
from Allende Prieto 2001), but no difference have been detected between
results with or without such blend.
Hyperfine structures (HFS) are taken into account for the
following elements (the line sources are given in parenthesis): Ba~II
(Rutten 1978, and the isotopic solar mix following McWilliam 1998);
La~II (Lawler et al. 2001 with log gf from
Bord et al. 1996); Cu (Biehl 1976), and Co I and Sc II (Prochaska et
al. 2000). In Fig. 7 the fitting procedure is shown for the Y~I
6435$\rm \AA$~line in RGB\_752 and the La II line 6320$\rm \AA$~in
RGB\_690. Abundances are given relative to solar abundances of
Grevesse \& Sauval (2000). Atomic lines for the
synthesis have been chosen according to the quality of the
synthetic fit in the Solar Flux Atlas of Kurucz et al. (1984).
In Tables 4 to 7 the derived abundances are given.

\begin{figure}
\centerline{%
\begin{tabular}{c@{\hspace{1pc}}l}
\includegraphics[width=1.5in]{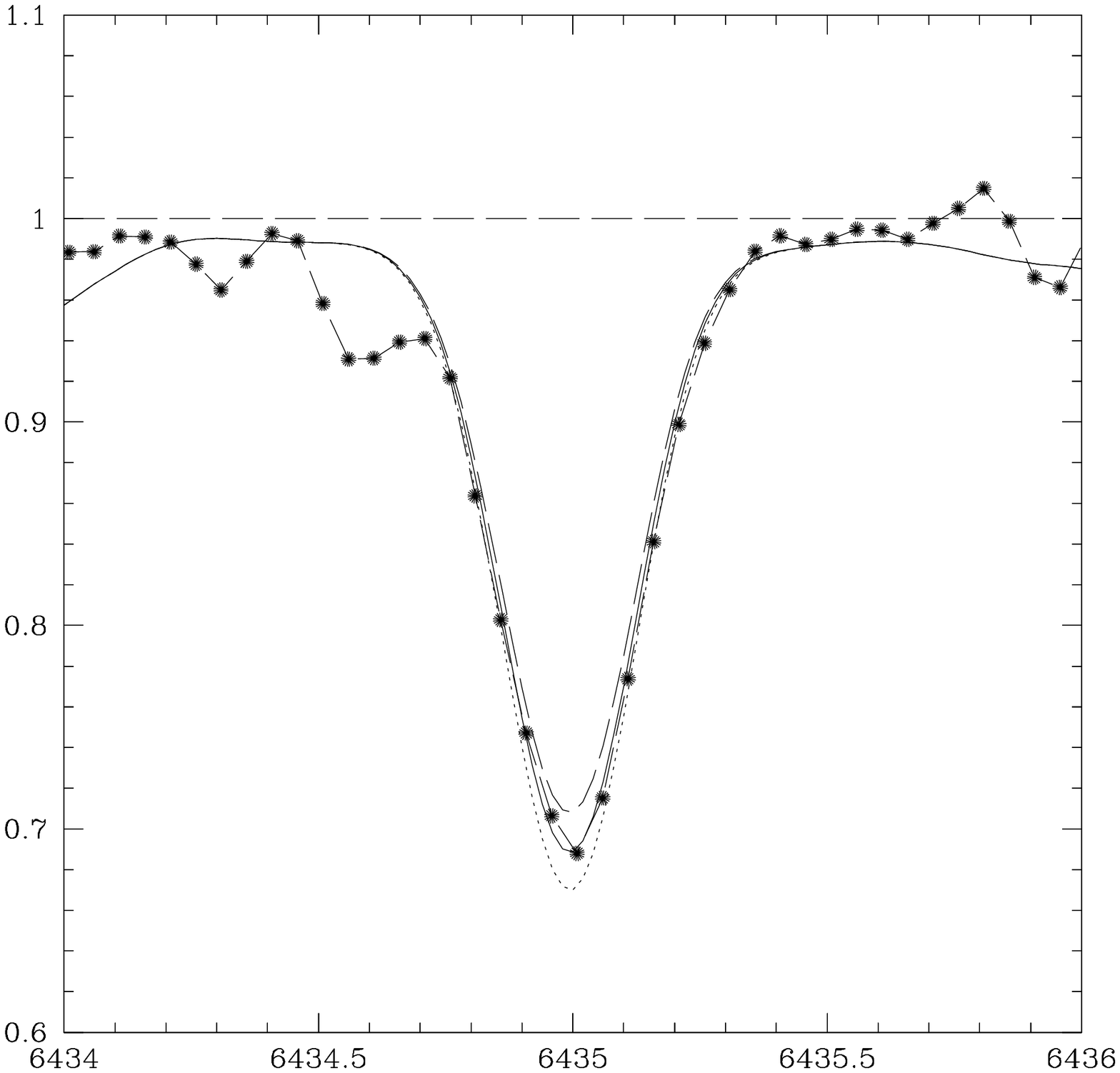} &
\includegraphics[width=1.5in]{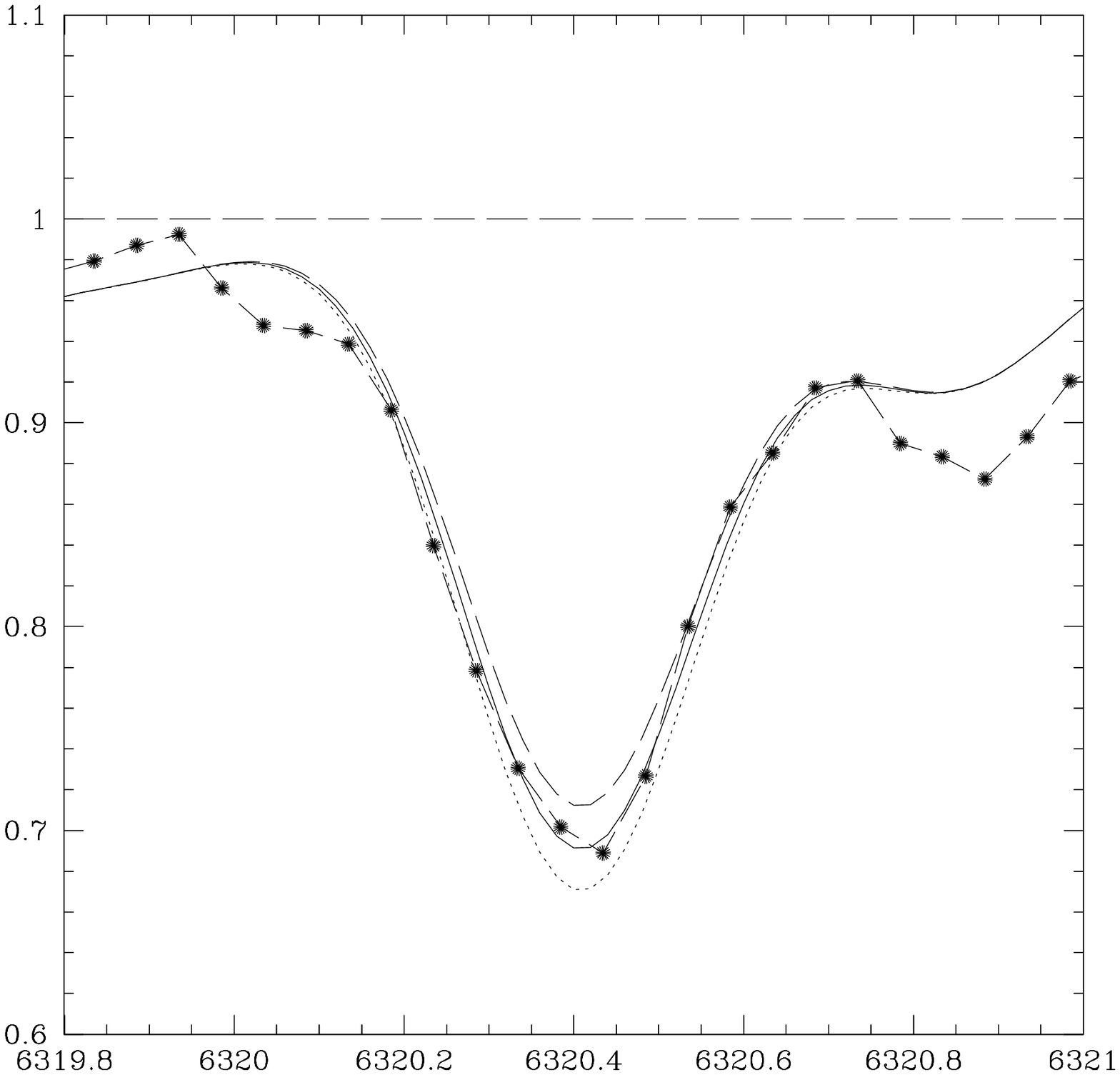} \\
\end{tabular}}
\caption{Example of the line synthesis procedure for the Y I and La II
lines: left panel: Y I $\lambda$6435$\rm \AA$~line fitting
for RGB\_752; right panel: La II$\rm \AA$6320~line fitting for
RGB\_690. The black circles depict the observed spectra and the
lines are the synthetic spectra. Abundances of the synthetic
spectra are: [Y/Fe] = $-$0.55 (dashed line), $-$0.45 (continuous
line - best fit), $-$0.25 (dotted line); and [La/Fe] = 0.56 (dashed line),
0.66 (continuous line - best fit), and 0.76 (dotted line).}
\end{figure}

Errors in the derived abundances have three main sources: the
uncertainties in the stellar parameters, the uncertainties in the
measurements of the EW (or spectrum synthesis fitting) and the
uncertainties on the physical data of the lines (mainly $\log$
gf). The errors due to stellar parameters uncertainties have been chosen as
the maximum range each parameter could change not to give unrealistic models 
atmospheres. The errors $\rm
\delta([X/Fe])_{model}$, are given in Table 8, assuming the
following uncertainties in each of the stellar parameters:
$\Delta$(T$_{\rm eff}$) = $\pm$100K, $\Delta$(log g) = $\pm$0.4 dex,
$\Delta$(V$_{\rm t}$) = $\pm$0.2 km/s and $\Delta$([Fe/H]) = $\pm$0.15
dex.

Errors in the EW measurement are computed by DAOSPEC during the fitting
procedure, then propagated into an abundance uncertainty for each line, and
then combined into an abundance error on the mean abundance for each element
($\delta_{\rm DAOSPEC}$).
Errors due to the combined uncertainties on the line data
and line measurement are reflected in the abundance dispersion
observed for each element, provided that the number of lines is
large enough to measure this dispersion in a robust way (N$\geq$3).
We therefore combined these error estimates in a conservative
way as given bellow:

\begin{equation}
{
 \begin{array}{ll}
 \rm  N_X < 3: & \delta([X/H]) = \delta_{DAOSPEC}, \\
 \rm N_X \geq 3: & \delta([X/H]) = Max(\delta_{DAOSPEC}, \frac{\sigma(X)}{\sqrt{N_X}}) \\
 \end{array}
}
\end{equation}
where $\rm N_X$ is the number of lines of the element X
and $\sigma(X)$ the dispersion among lines.

These errors are calculated for each element and given in
Tables 4 to 6 together with the abundances derived from
the EW.
For elements measured by synthesis
spectrum fitting, an error estimate has been carried out of the typical
abundance change for which two different synthetic spectra (i.e. computed with
two slightly different abundances) still fit satisfactorily the same
line. On average, these values are the following for each element:
${\delta}$[Zr/H] = 0.15 dex;
${\delta}$[Y/H] = 0.15 dex; ${\delta}$[La/H] = 0.20 dex;
${\delta}$[Ba/H] = 0.25 dex; ${\delta}$[Co/H] = 0.10 dex;
${\delta}$[Cu/H] = 0.20 dex; ${\delta}$[Sc/H] = 0.10 dex;
${\delta}$[Mg/H] = 0.15 dex and ${\delta}$[O/H] = 0.20 dex.
 For the error bars reported in our
 abundance plots (always shown in the lower left corner of Figs. 8-12) we
 have adopted two error sources. The first, due to stellar parameter
 uncertainties (leftmost side of the plots), comes directly from Table 8,
 whereas the second (more to the right side) represents the error associated
 with the abundance analysis. For those abundances derived from the EW, this is the mean
 error of Tables 4,5, and 6, and for those elements with abundances derived
 from spectrum synthesis, it is the value described earlier on in this
 Section.

\section{Abundance Distributions and comparison to Galactic samples}

In Figs. 8 to 12 we depict the elemental distributions for the
 $\alpha$-elements, the iron-peak group, Na, Sc, Cu and $s$-elements
for our stars compared to different samples of the Galaxy and the LMC.
Our data are represented as dots. The references of the disk are:
Fulbright 2000 (crosses); Reddy et al. 2003 (open squares);
Allende Prieto et al. 2004 (open stars);
Prochaska et al. 2000 (open triangles); Burris et al.
2000 (stars - only for the heavy-elements plots); Johnson \&
Bolte 2002 (open trianlges - only for the heavy-elements plots);
Simmerer et al. 2004 (open hexagons); Nissen \& Shuster 1997
(asterisks, only stars with low [$\alpha$/Fe] ratios);
Nissen et al. 2000 (asterisks - Sc abundances for the
low-$\alpha$ stars ); and Bensby et al. 2004 (open squares - only
for the oxygen plot). LMC globular clusters (GC) stars from Hill
et al. (2000, hereafter HI00) for O, and Hill (2004 hereafter
HI04) for Na, Mg, Ca and Si are plotted as downward-pointing, open
triangles; LMC GC stars from JIS06 are represented as open diamonds;
field LMC red giants of SM02 are depicted as open pentagons. Error bars
as described in Sect.4.0.1 are shown in the lower left side of the plots.

\subsection{Ca, Si and Ti}

In Fig. 8, the elemental distributions for Ca~I, Si~I, and Ti~I
are depicted. We have found that [Si/Fe] follows roughly the solar
ratio with some scatter. [Ca/Fe] shows a slight decrease with
metallicity. Compared to the distribution of the galactic halo,
both silicon and calcium mean abundances are deficient by a
factor of 3. Ti~I ratios are also underabundant relative to
galactic disk and galactic halo samples, and agree very well with the
results of SM02, who derived titanium abundances from neutral
lines for a sample of red giants from LMC disk. There is a hint
of a decreasing trend of Ti abundances for higher metallicity
stars, especially when SM02 datapoints are taken into account
together with our sample. Compared to the LMC GC of H04, we have
found that the star of our sample with  metallicity similar to
those of those of Hill et al. (2004) also has similar [Ca/Fe] ratio. The JIS06
sample of LMC GC stars seems to overlap our [Ca/Fe] and [Ti/Fe] distributions,
while their [Si/Fe] ratios are enhanced.

A very interesting result emerges when comparing our data with those
of Nissen \& Shuster 1997 (hereafter NS97, asterisks). NS97 discovered a sample
of stars from the galactic halo with abnormal abundances: low [$\alpha$/Fe], [Na/Fe]
and [Ni/Fe] ratios compared to ``standard" halo stars. Such chemically peculiar or
``low-$\alpha$" halo stars have an important role in
elucidating the possible merging history of the galactic halo. Because of their
chemical properties, they indicate that this group have formed in
another stellar system that evoveld separately, and which has been
captured or ejected to the halo.
Comparing our LMC distribution to the low-$\alpha$ stars, we have found
that NS97 stars show a slightly enhanced mean $\alpha$ abundance relative
to our LMC stars.

 Si, Ca and Ti are predicted to be produced in
intermediate mass Type II SNe (SNe~II) with a smaller contribution
from Type Ia SNe (SNe~Ia) (e.g. Tsujiomoto et al. 1995, Thielemann
et al. 2002), while Fe is mostly produced by SNe~Ia (e.g.
Thielemann et al. 2001, Iwamoto et al. 1999). The low [$\alpha$/Fe]
ratios observed indicate that SNe~Ia have contributed more to the ISM
content in the past than the SNe~II.

\begin{figure}[H]
\centerline{\includegraphics[width=5in]{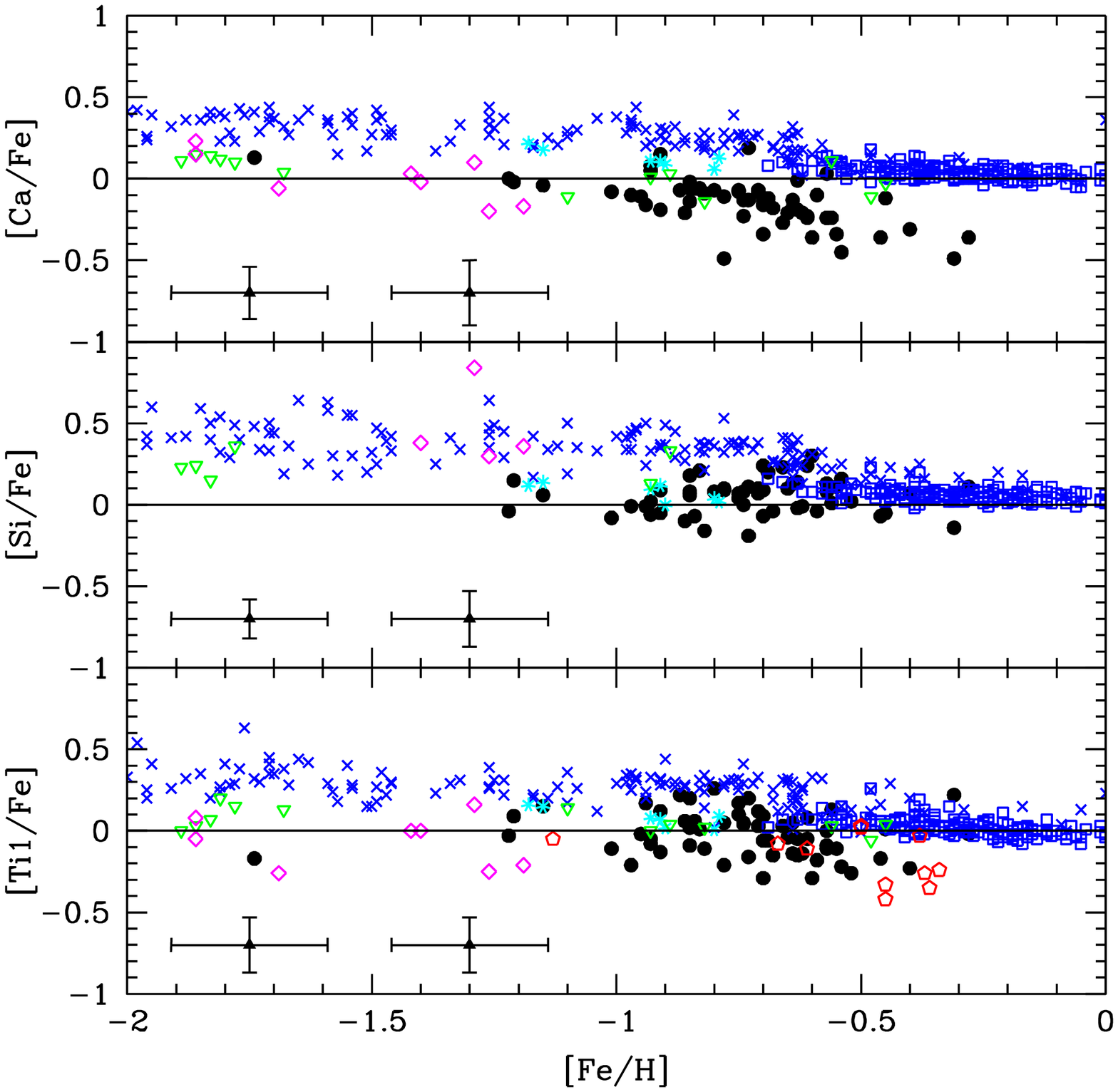} }
\caption{Abundance distributions for the Inner Disk LMC stars:
[$\alpha$/Fe] vs. [Fe/H] (blue dots). LMC samples are depicted
 with polygons (downward-pointing green triangles  - Hill et al.
 2000; red pentagons - Smith et al. 2002; magenta diamonds - Johnson et al. 2006);
 and the remaining symbols (all in blue) are data for the galactic stars (crosses -
 Fulbright 2000; open squares - Reddy et al. 2003; cyan asterisks -  Nissen \& Schuster
 1997). Error bars depict:
 a. leftmost side of the plots - errors due to stellar parameter
 uncertainties (Table 8); and b. more to the right side - errors associated
 with the abundance analysis - for those derived from the EW, is the mean
 error of Tables 4,5, and 6; for those elements with abundances derived
 from spectrum synthesis, is the value described in Sect. 4.}
\end{figure}

\subsection{Mg, O, Na and Sc}

In Fig. 9, abundance ratios are given for O, Mg, Sc and Na.
Nucleosynthetic predictions attribute the main source of O, Mg and
Na to high-mass stars, with M $>$ 25 M${_\odot}$, which explode as
SNe~II (Woosley \& Weaver 1995, hereafter WW95), with Na
production controlled by the neutron excess. Although WW95 have
attributed the origin of Sc to SNe~II, the main source of Sc
production is still unclear (e.g. McWilliam 1997, Nissen et al.
2000).

As can be seen in the upper panel of Fig. 9, oxygen ratios fall
in the lower envelope of the galactic halo and disk distributions.
For higher metallicities, it shows a faster decline with metallicity
compared to stars from the galactic disk. In
the second plot we see that the [Mg/Fe] ratios for the LMC Inner
Disk overlap those of the Galaxy, but with smaller mean values.
In contrast, Na and Sc behaviors are similar to those of the $\alpha$-elements
Ti, and Si. Both elements are deficient and show smaller values
for higher metallicities, while for the metal-poor tail, a match
to the Galactic samples is observed. From this figure we see that the
different LMC samples agree very well for all elements, Mg, O, Na
and Sc, even the LMC globular clusters of H00, H04
and JIS06. A few stars in the NS00 sample of
low-$\alpha$ stars show small [Sc/Fe] ratios and overlap our sample, but
most of them show solar [Sc/Fe] values, higher than in our LMC sample.
Sodium abundances in NS97 sample are similar to our values,
although with a higher mean abundance.
It is important to notice that sodium
abundances in giants are still uncertain. Pasquini et al. (2004)
found that [Na/Fe] ratios in giant stars are slightly higher than
those from dwarf stars in the same cluster. High [Na/Fe] ratios
were also inferred from giants in M67 (Tautvai\v{s}iene et al. 2000).
But such results have not been confirmed in the reanalysis of [Na/Fe] in
giants and dwarfs of M67 (Randich et al. 2006).

Nissen et al. (2000) also found that Sc behaves similarly to Na,
showing lower [Sc/Fe] ratios in their low-$\alpha$ stars,
suggesting a correlation among those elements. In order to test the
hypothesis of a correlation among Na and $\alpha$-elements, and Sc and
$\alpha$-elements we have applied a statistical test to check for the
existance and significance of such correlation, calculating the linear
correlation coeficient, which varies from 1 or -1 (maximum correlation
or anti-correlation) to 0 (no correlation).  We have found that the
correlations are weak: for Na-Ca, a correlation coefficient $\phi$ = -0.06
is found, and for Sc-Ca, $\phi$ = 0.39.

\begin{figure}[H]
\centerline{\includegraphics[width=5in]{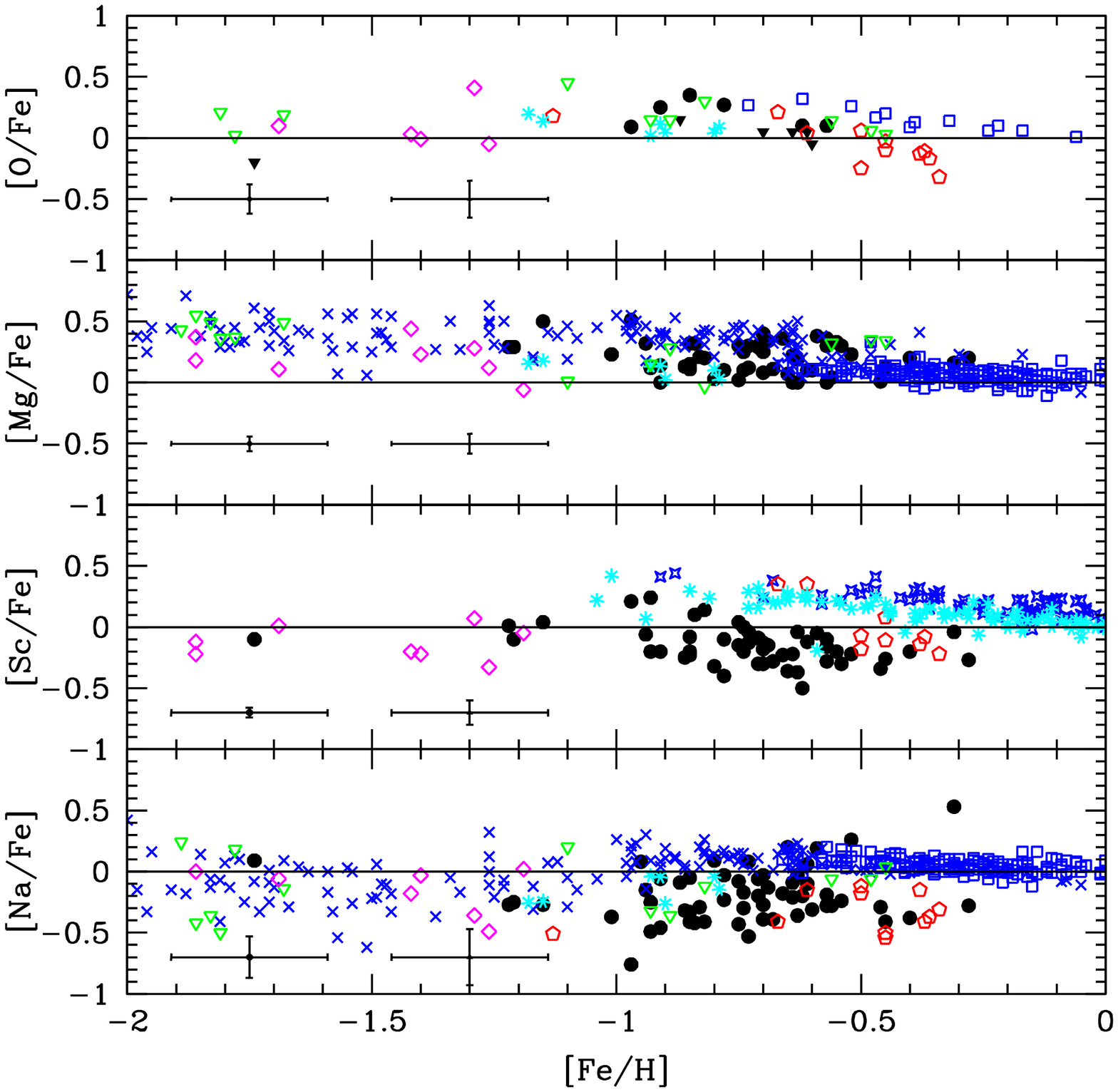} }
\caption{Abundance distributions for the Inner Disk LMC stars: [
O, Mg, Na, Sc/Fe] vs. [Fe/H] (symbols are the same as in Fig. 8,
and we added: blue open stars - Allende Prieto et al. 2004, blue crosses -
Bensby et al. 2004 only for oxygen; cyan asterisks - Nissen et al. 2000 for
Sc and Nissen \& Schuster 1997 for the other elements).}
\end{figure}

\subsection{Iron-peak elements}

Abundance distributions for the iron-peak elements are shown in
Fig. 10.  The iron-peak elements Co, Ni and Cr display a
very distinct pattern in the LMC Inner Disk stars, with underabundant values compared to the
Galactic distributions and many subsolar ratios. [Co/Fe], [Cr/Fe] and [Ni/Fe]
show a flat trend for most of the metallicity range, with mean abundances of
$\sim$ -0.18 dex for Cr, $\sim$ -0.24 for Ni, and $\sim$ -0.14 dex for Co.
The [V/Fe] ratios are similar to the galactic halo and disk patterns and track the
solar value, with a group of stars showing smaller values.
Results from the LMC GC of JIS06 seem to agree with our samples for
Co, Ni and Cr. Vanadium in their sample shows an offset, with abundance
ratios corresponding to the stars with smaller values in our sample.
NS97 low-$\alpha$ stars overlap our sample for Ni and Cr,
but lie in the high abundance envelope of the distributions.

According to nucleosynthetic predictions, iron-peak elements are
mainly produced in SNe~Ia (Iwamoto 1999, Travaglio et al. 2005):
while each SN~Ia produces $\approx$ 0.8 M$_{\odot}$ of the solar
iron-peak elements, SN~II produce $\approx$ 0.1 M$_{\odot}$ each
(Timmes et al. 2003). The difference in the distributions
from one environment to the other are an evidence that the
production factors for each iron-peak element are not the same in
the different types of SNe and depend on the SFH of the parent
population. This will be further discussed in Sect. 7.

\begin{figure}[H]
\centerline{\includegraphics[width=5in]{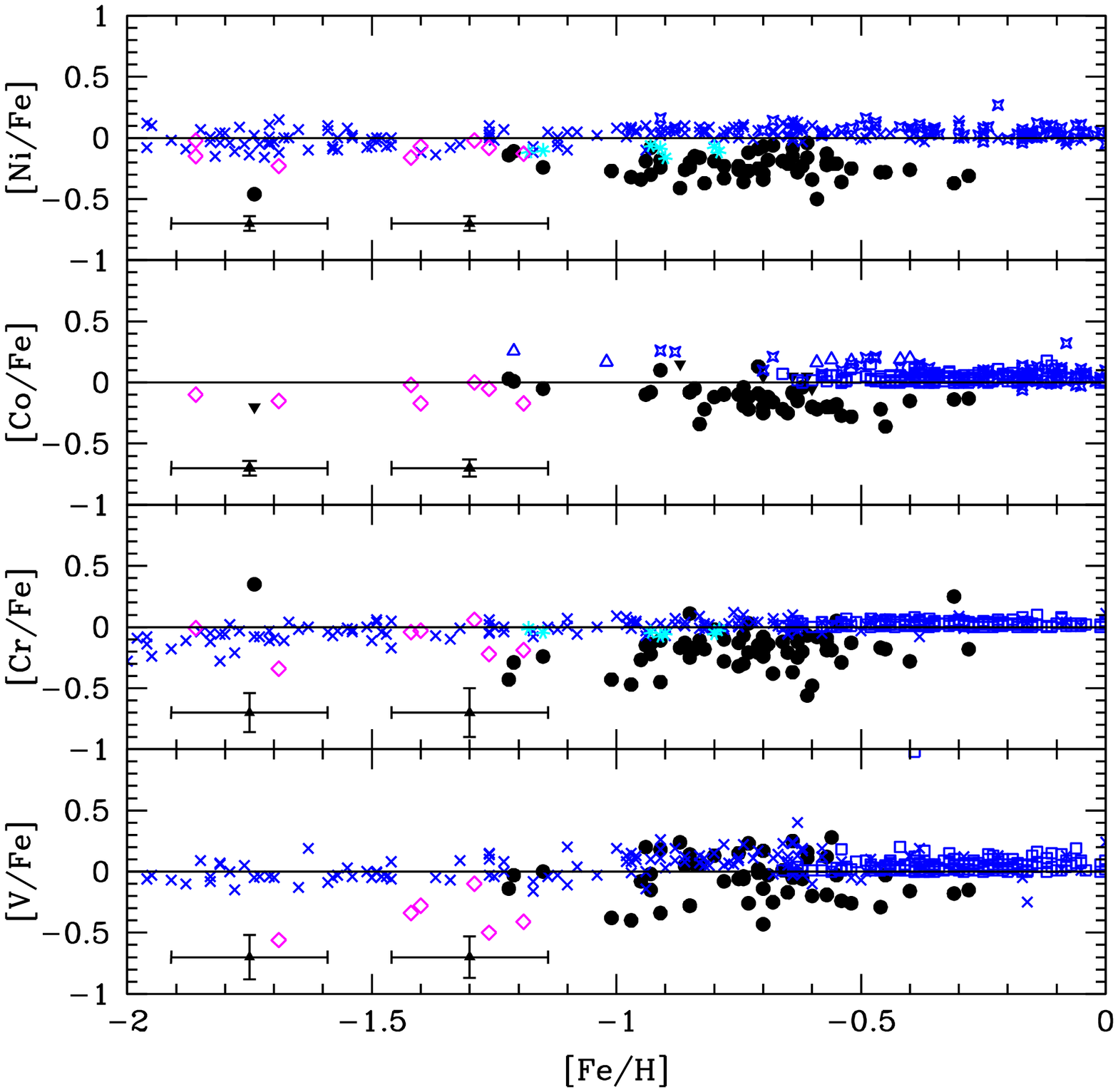} }
\caption{Abundance distributions for the Inner Disk LMC stars:
[Iron-peak/Fe] vs. [Fe/H] (symbols are the same as in Figs. 8
and 9, and the solid downtriangles depict upper limits for our sample stars).}
\end{figure}

\subsection{Copper}

In Fig. 11 we show the plot for Cu. We have found that in the
Inner Disk LMC stars, the copper distribution is flat, with a
mean value of [Cu/Fe] = $-$0.68 dex. Comparing to the Galaxy, there
is an overlap between the LMC and Halo stars at the metal-poor
end ([Fe/H] {$< -1.3$ dex}); for the higher
metallicity range, the distributions diverge, with LMC stars
showing a clear underabundance with respect to the Galactic Disk.
JIS06 also found an offset in their [Cu/Fe], compatible to our
abundance ratios.

Although originally associated with the iron-peak elements, the origin of
copper is still much-debated (e.g. Bisterzo et al. 2004, Cunha
et al. 2004, Mishenina et al. 2002). Sometimes its main source is
attributed to SNe~Ia (Matteucci et al. 1993, Cunha et al. 2002,
Mishenina et al. 2002) and sometimes to SNe~II, particularly to a
metallicity dependent mechanism (Bisterzo et al. 2004; McWilliam \&
Smecker-Hane 2005). If the
elemental behavior of the present sample, with low [$\alpha$/Fe],
low [iron-peak/Fe] ratios, is due to a higher contribution from
SNe~Ia, the overall low [Cu/Fe] pattern indicates that thermonuclear
supernovae cannot be the main source of Cu production.

\begin{figure}[H]
\centerline{\includegraphics[width=3.5in]{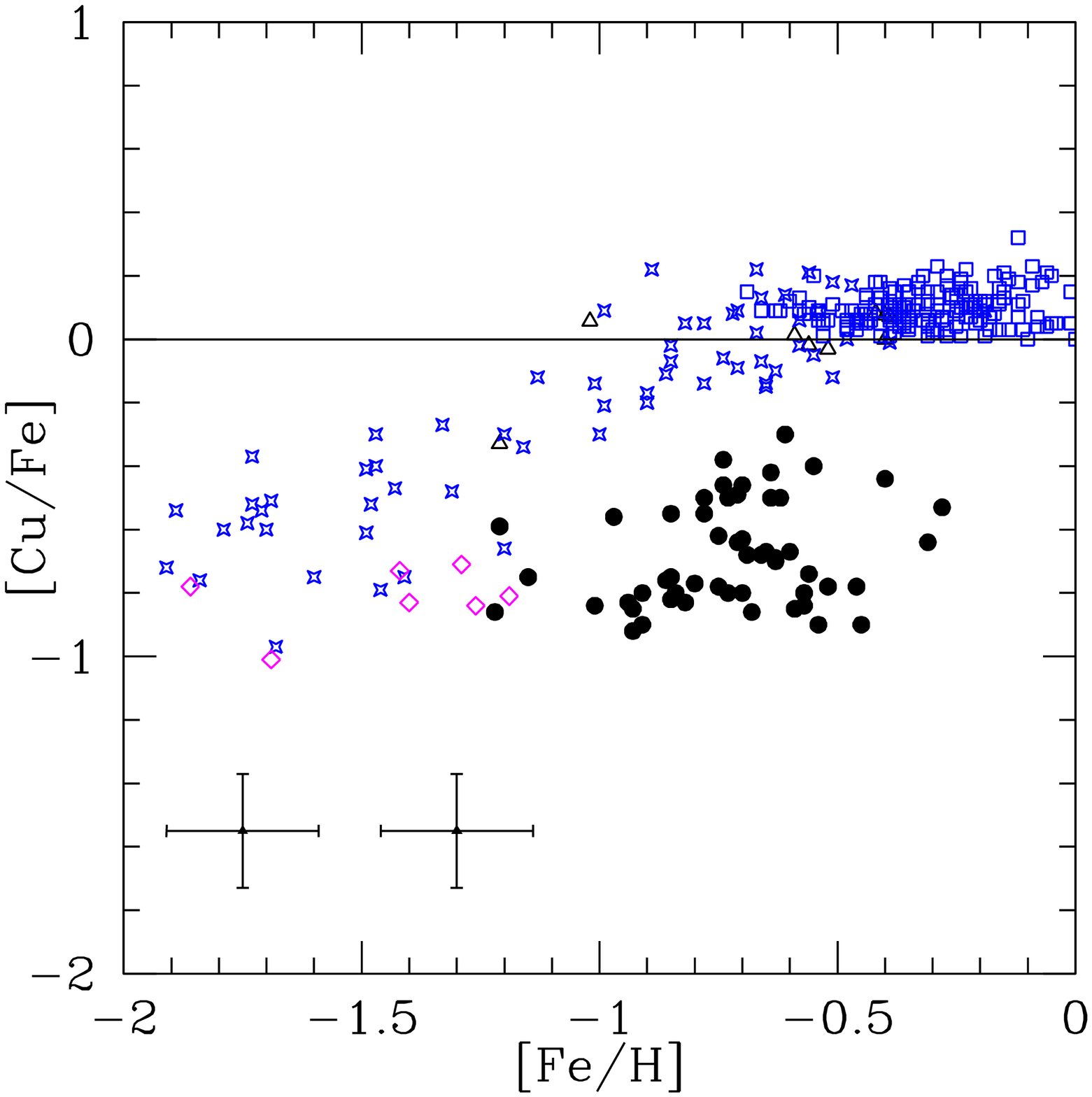}}
\caption{Abundance distributions for Inner Disk LMC stars:
Copper. The symbols are the data from: our sample stars (dots);
Mishenina et al. 2002 (blue stars); Prochaska et al. 2002 (open blue triangles);
Reddy et al. 2003 (open blue squares); Johnson et al. 2006 (magenta diamonds).}
\end{figure}

\subsection{$s$-process elements}

We have found interesting elemental distributions for the $s$-process elements
for our sample stars (Fig. 12). While the light $s$-process elements (hereafter $ls$: elements made by the $s$-process
with atomic number lower than $\sim$45) Zr and Y, show subsolar ratios with mean abundances
 the heavy $s$-process elements (hereafter $hs$: elements made by the $s$-process with
atomic number higher than $\sim$50) La and Ba show supersolar values with enhanced
pattern compared to those of the Galaxy.
The underabundance of $ls$ elements is quite strong, [Y/Fe] = $-$0.33 dex and [Zr/Fe] = $-$0.48 dex,
and Zr shows a hint of decreasing with increasing metallicities.
Of the $hs$ elements, Ba has a peculiar
behavior with a high value for one metal-poor star ([Fe/H] $<
-1.4$ dex), mild enhancements until [Fe/H] $\sim -$1.15 dex, increasing again
towards higher metallicities. La shows no
trend with metalicity, with mild enhancements everywhere.
One star, RGB\_1118, has particularly high
La and Ba abundances ([Ba/Fe] and [La/Fe]$\geq$+1.0 dex) and
could be a star enriched in $s$-process elements (via mass-transfer from a former
AGB companion), although it is not possible from our present
data to discriminate between enhancements of $s$-process or $r$-process
elements. The $s$-process elements in JIS06 sample are different when compared
to our results. While they have found no offset for the $ls$
elements compared to the galactic distribution, showing therefore
a higher abundance compared to our stars, their $hs$ elements
(Ba and La) are less enhanced than ours.
Comparing NS97 low-$\alpha$ stars with our
sample, we find that these stars show abundances nearer
those of normal disk stars for Ba and Y than the LMC stars.

The $hs/ls$ ratios are high, showing large scatter, with a mean value
of [$hs/ls$] = +0.77 dex, as can be seen in Fig. 13.
This is very different from what is observed for
the galactic halo and disk stars, which fall around -0.2 to
+0.2 dex (e.g. Pagel \& Tautvai\v{s}iene 1997; Travaglio et al.
2004). A slow increasing trend with metallicity is observed.

High abundances of elements heavier than Zr were also
derived for LMC and SMC supergiants (Russell \& Bessell 1989;
Spite et al. 1993; Hill et al. 1995). Hill et al. (1995) for example, found
that the light $s$-elements Zr and Y show solar composition in
LMC supergiants while heavier $s$-elements (Ba, La, Nd) as well as the
$r$-process element Eu are enhanced by +0.30~dex. As discussed by these authors,
the overabundance of the heavier $s$-process and $r$-process elements seems to be a
characteristic of the Magellanic Clouds, and indicate a particular
evolution of that galactic system, although no satisfactory
explanation was proposed for it.

In order to evaluate the $r$-process and $s$-process contributions within our
sample we analysed the $r$-process content of one of our sample stars for
which we have UVES spectra that cover the Eu $\lambda$ 6645 $\rm \AA$ line.
Eu and Ba abundances were derived from these spectra in the same
way as was done for GIRAFFE spectra. For RGB\_666 we find respectively
[Ba/Fe] = $+$0.52,
and [Eu/Fe] = $+$0.40~dex. The corresponding [Ba/Eu] ratio of 0.12
(to be compared with the solar $r$-process [Ba/Eu]=$-$0.55 and the
solar $s-$process [Ba/Fe]=+1.55, following Arlandini et al. 1999), indicate that
this star contains a significant $r$-process contribution at a value close to the
solar $s$/$r$ mix at intermediate metallicities (RGB\_666: [Fe/H]=-1.10).

A high content of $r$-process elements seems to be in contradiction with the
observed low [$\alpha$/Fe] ratios (both being produced in massive stars). More
data on Eu abundances are needed to confirm this high content of $r$-process
elements, and in particular, the trend of the $s$/$r$ fraction
(traced by [Ba/Eu]) as a function of metallicity will help to constrain the source
of the high content of heavy $s$-process elements in the LMC disk. We intend to
tackle this issue in the two other fields (Bar and Outer Disk) of our LMC
program, since one of the MEDUSA wavelength ranges covers the Eu line for these fields.

\begin{figure}[H]
\centerline{\includegraphics[width=5in]{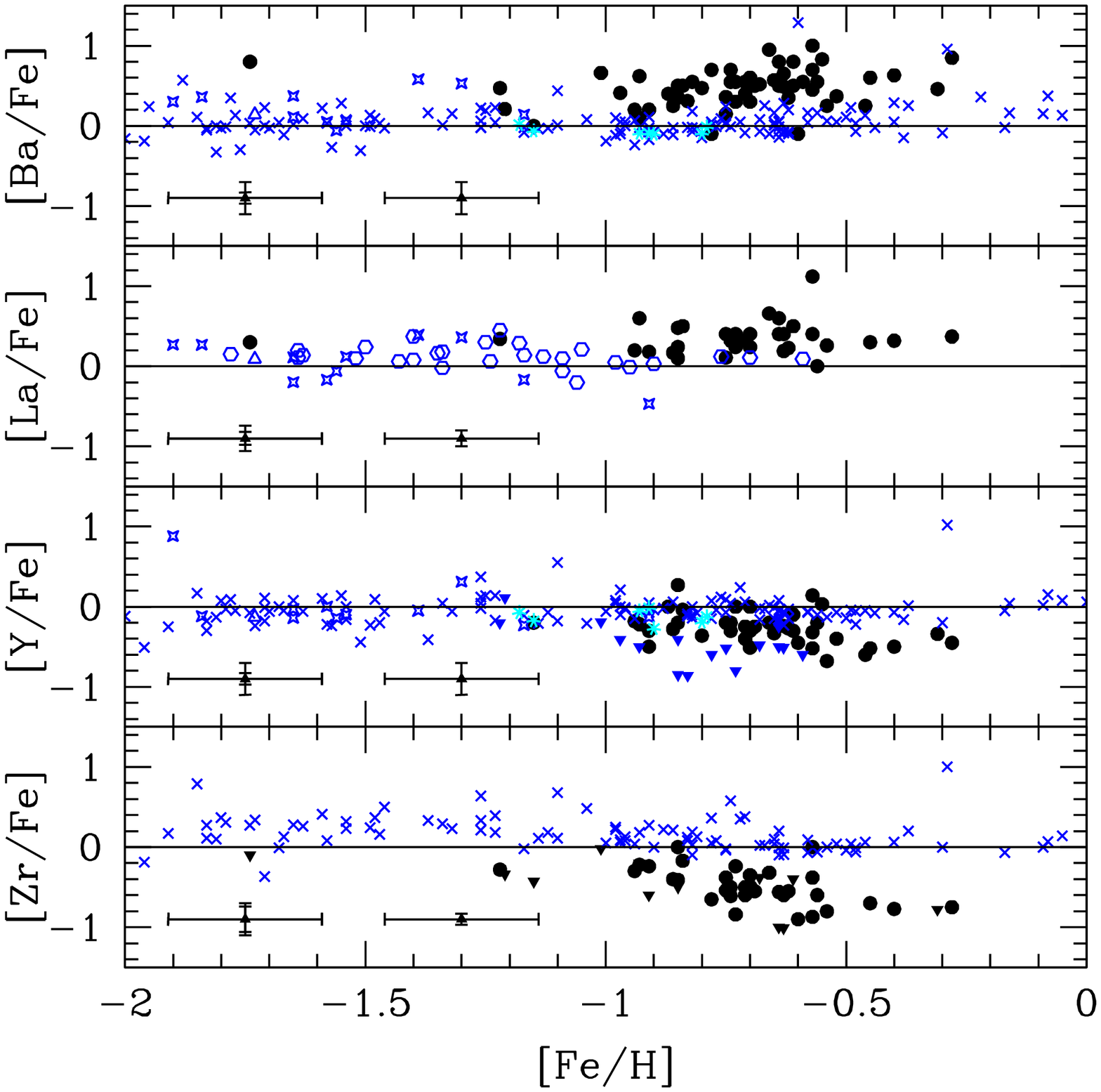} }
\caption{Abundance distributions for Inner Disk LMC stars:
[$s$-elements/Fe] vs. [Fe/H]. The large dots (or downtriangles
for upper limits) depict our sample stars, while open blue symbols represent galactic samples:
symbols as in Figure 8, plus Burris et al. 2000 (blue stars); Johnson \& Bolte 2002 (blue trianlges);
Simmerer et al. 2004 (blue hexagons); Nissen \& Schuster 1997 (cyan asterisks).}
\end{figure}

\begin{figure}[H]
\centerline{\includegraphics[width=3.5in]{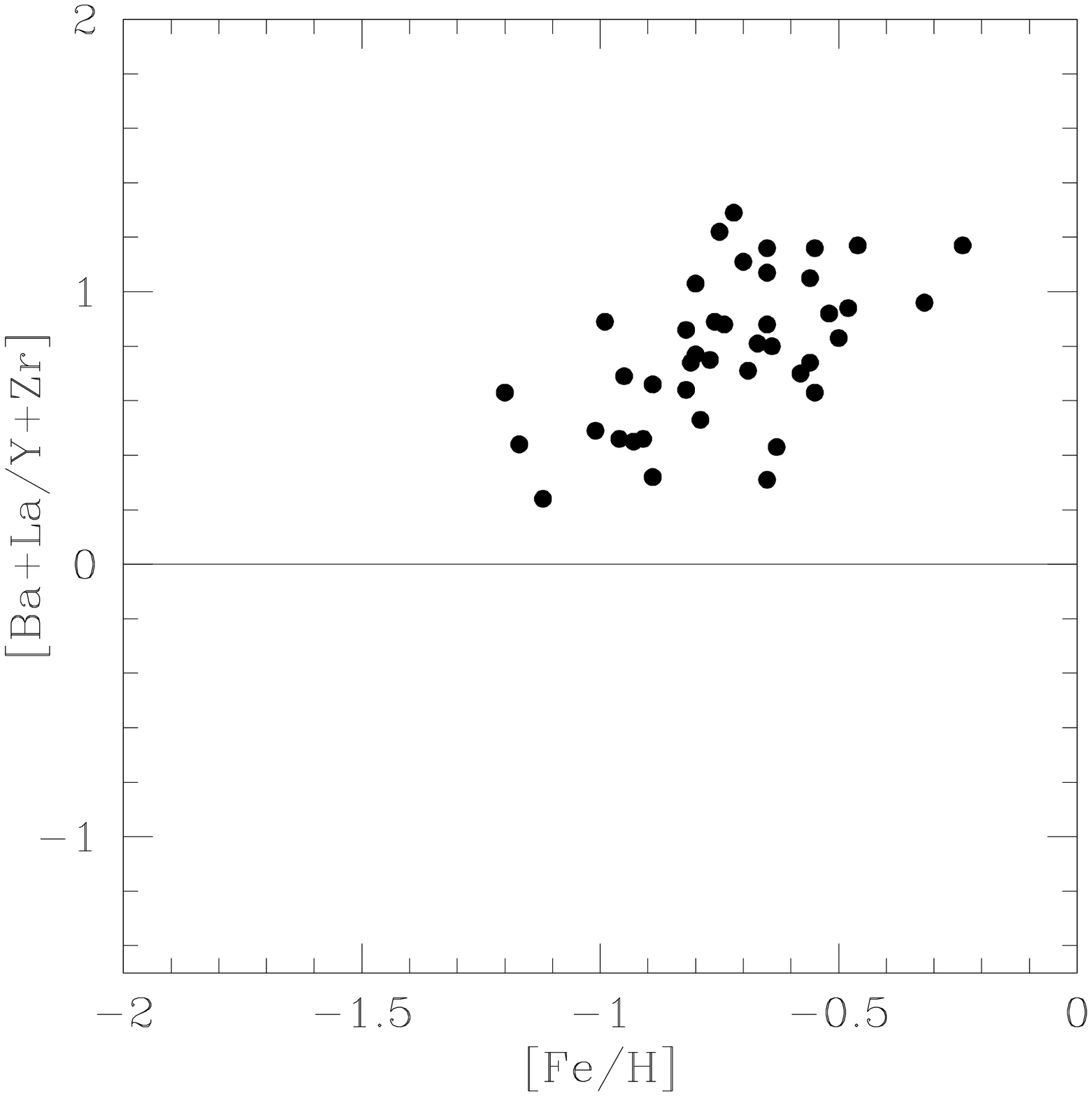} }
\caption{Observed abundance ratios [$hs$/$ls$] = [Ba+La/Y+Zr].}
\end{figure}

\subsubsection{The NaMg, NaNi relations}

In the paper by NS97 the authors found a correlation between
Na and Ni for their halo stars (both ``normal" and ``low-$\alpha$"
stars). Such correlation has been confirmed for a group of stars
in the Dwarf Spheroidal Galaxies (Shetrone et al. 2003, SH03,
Tolstoy et al. 2003, TO03, Venn et al. 2004). To evaluate this
trend, we plot in Fig. 14 the [Ni/Fe] vs. [Na/Fe] relation for our
sample stars (dots) together with NS97 low-$\alpha$ stars. We see
that the LMC stars also show a correlation between Na and Ni,
although with a flatter pattern than the increasing
trend observed for the NS97 sample. According to Tsujimoto et al.
(1995), Ni can be produced in SNe~Ia without Na production;
therefore, a higher contribution from SNe~Ia would flatten the
NaNi relation\footnote{however Travaglio et al. 2005 found that
some Na and Mg are also produced in SNe~Ia} (Venn et al. 2004) and
could explain the behavior of the LMC stars. In Fig. 14 we also
analyze the correlation between Na and Mg and we find
decreasing [Na/Mg] ratios for increasing
[Mg/H] ratios. The NS97 low-$\alpha$ stars seem a continuation of
the observed trend.

\begin{figure}[H]
\centerline{\includegraphics[width=5in]{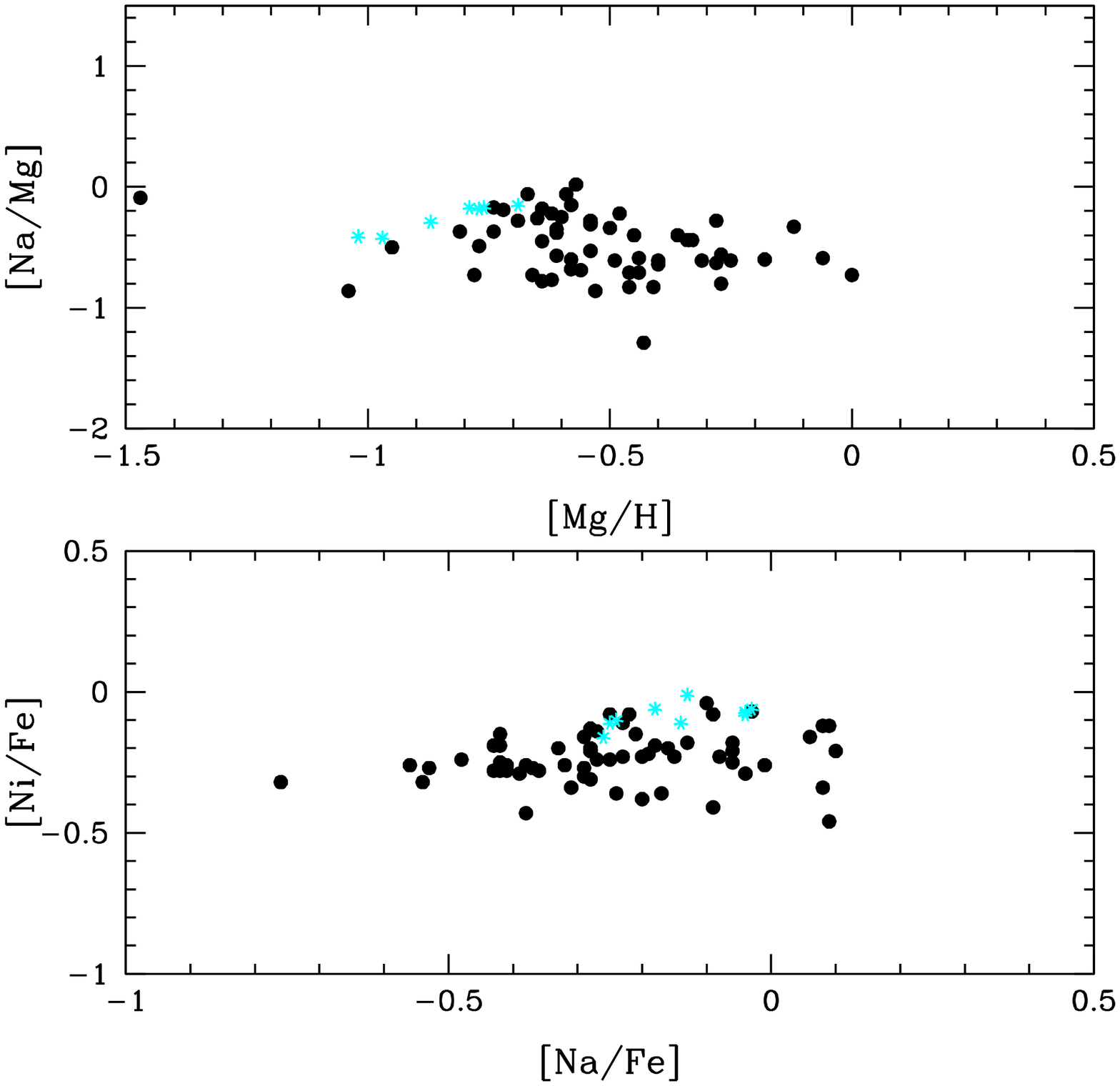} }
\caption{The NaNi and NaMg abundance relations. Our sample stars are
depicted as dots and NS97 low-$\alpha$ stars as starred symbols.}
\end{figure}

\section{Comparison to the Dwarf Spheroidal Galaxies}

In Figs. 15 and 16 we show a comparison of the chemical distributions
of our LMC sample to those of the dSph galaxies of Shetrone et al. (2003) 
and Tolstoy et al. (2003), and the Sagittarius dwarf galaxy (Sgr)
of Bonifacio et al. (2004) and Sbordone et al. (2007). The elemental distributions of
most dSph galaxies are more concentrated in the metallicity range
for which we have the lowest number of stars, [Fe/H] $< -1.2$, so
the present analysis is not ideal. In Fig. 15 the distributions for
the $\alpha$-elements Ca, Ti and Si and for Cu are depicted (the description
of the different symbols are given in the figure captions). As can be seen
from these figures, and observed for also for O, Mg, Na and Sc,
there is an overlap among the LMC abundance ratios and those of the dSph
galaxies. The same occurs for the iron-peak elements Cr, V and Ni and for
Cu (with the exception of Fornax, which shows higher values for Cu).
Particularlly, the agreement among our data and those of the Sagittarius
dwarf galaxy is very good, except that this galaxy shows [Ti/Fe] and [Mg/Fe]
ratios slightly underabundant relative to our values.

For the $s$-process elements, depicted in Fig. 16, the dSph galaxies
show enhanced $hs$ and deficient $ls$ compared to the Galactic
behavior, although the general pattern is less discrepant
than that showed by the LMC inner disk stars, except for Sgr,
which shows striking similar ratios when compared to our data.
Fornax has a more metal-rich star (Fnx21) with high $s$ content, which may be
an $s$-enriched star. The [Ba/Y] ratios show a large offset relative to galactic
samples, of the same order magnitude we have found. Venn et al. (2004)
attribute such offset to primary $s$-process production by low-metallicity
AGB stars.

The very similar elemental distributions of the Sgr galaxy indicate that this
galaxy must have been very similar to LMC, i.e., with a higher mass content,
which may be nowadays hidden in streams and/or dynamically mixed to the Galaxy.

\begin{figure}[H]
\centerline{\includegraphics[width=5in]{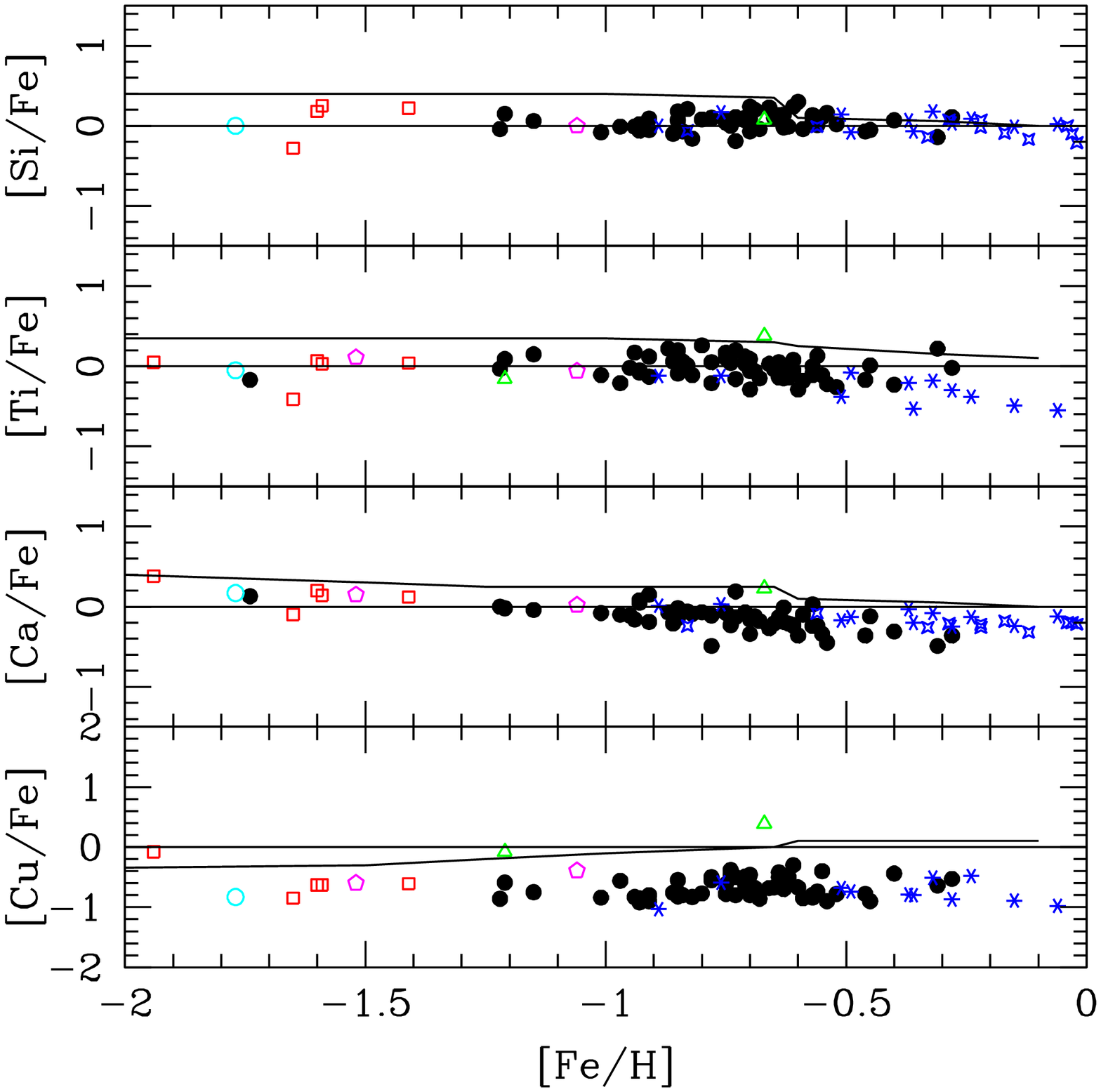} }
\caption{Comparison of the Inner Disk LMC stars with stars from
the dwarf spheroidal galaxies and the Sgr galaxy.
1. Alpha elements. The symbols are: our sample (dots), Leo I
(magenta pentagons), Sculptor (open cyan dots), Fornax (green triangles),
Carina (red squares), and Sgr (Bonifacio et al. 2005 - blue stars; Sbordone et al.
2007 - blue asteriks). Solid lines depict mean values of the Galactic distributions
for each element.
}
\end{figure}

\begin{figure}[H]
\centerline{\includegraphics[width=5in]{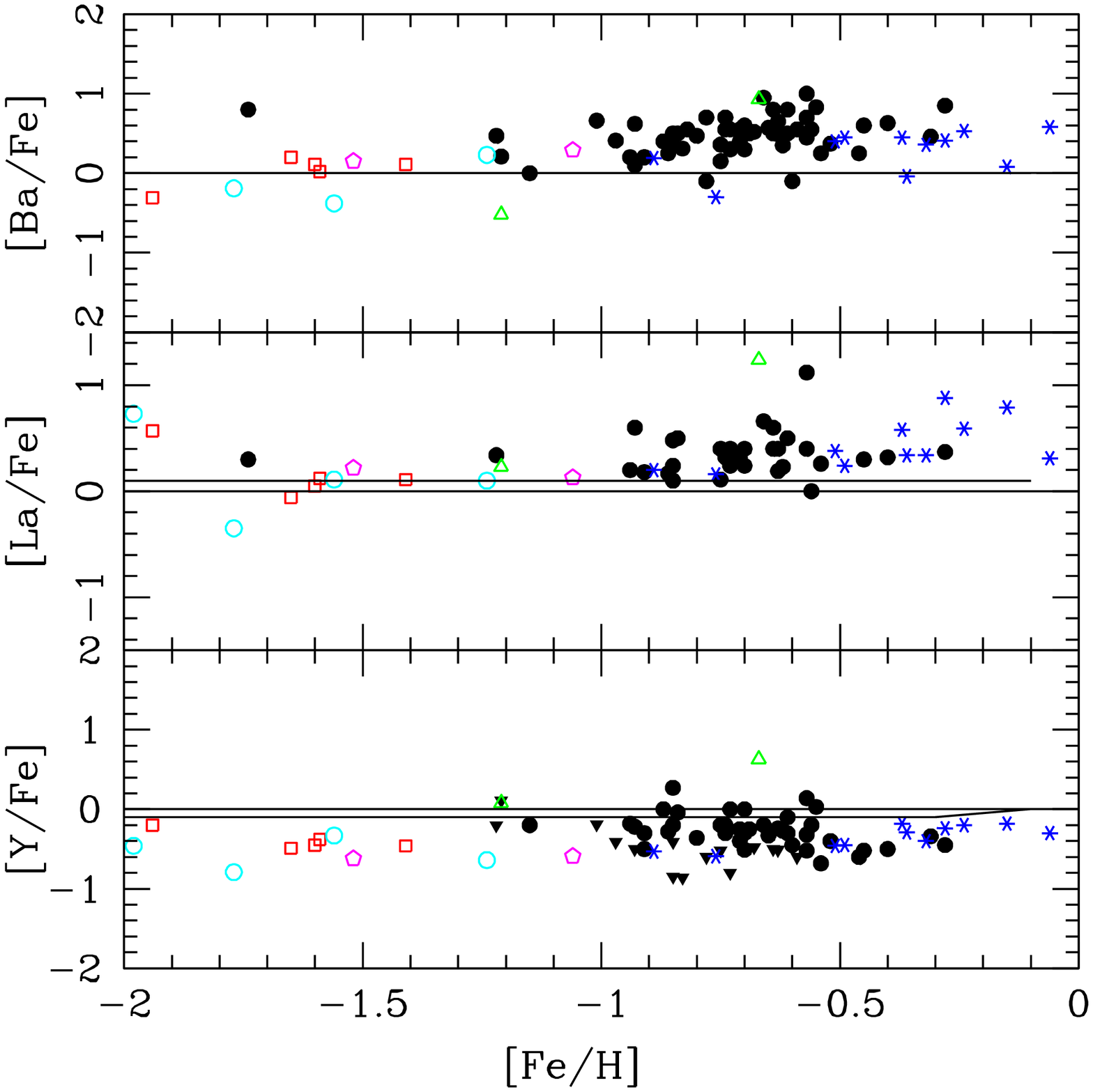} }
\caption{Comparison of the Inner Disk LMC stars with stars from
the dwarf spheroidal galaxies. 2. $s$-process elements (symbols
are the same as in Fig. 15).}
\end{figure}

\section{Discussion}

It is an amazing opportunity to have so much data on the
amount of various elements of stars in an external galaxy. With this unique
dataset, we can now explore in more detail the SFH and better understand
the evolution of the LMC disk. The overall low [X/Fe] ratios
indicate that such stars have undergone a global process which is different from
that experienced by the average halo and disk stars in the Galaxy.
In this section we discuss the possible explanations for such
behavior.

We have found an overall low abundance pattern for the $\alpha$-elements,
in agreement with many previous works with stars in this galaxy (Sect. 1).
The heavy $s$-elements show an enhancement relative to the Galactic
disk distributions, as inferred before for supergiants and red giants
in the LMC. New results from the present work include low light-$s$
abundance ratios ([Y/Fe] and [Zr/Fe]), with most of the stars showing
subsolar values, and an unexpected offset for the iron-peak elements Ni,
Cr and Co, and in some stars, also for V. Na and Sc are deficient
with many subsolar ratios relative to iron, and copper shows a very
low abundance in all stars from the present sample, with mean [Cu/Fe] $\sim$
-0.7 dex, and no trend with metallicity.

As seen in previous sections, small [$\alpha$/Fe] ratios have already been observed
in other stellar systems such as the chemically peculiar halo stars (NS97, NS00), the
dSph galaxies of the Local Group (Shetrone 2003, Tolstoy 2003), the Sagittarius
galaxy (Smecker-Hane \& McWilliam 2002, Bonifacio et al. 2004, Monaco et al. 2005,
Sobordone et al. 2007), as in samples in the LMC (e.g. Hill et al. 2000, 2003; SM02,
Garnett 2000, Korn et al. 2002).
It is interesting to notice that the $s$-process trends in the dSph galaxies
(enhanced $hs$ and deficient $ls$ ratios) are the
same as for our stars. Correlations between abundances of iron-peak elements and
$\alpha$-elements were observed also in other stellar systems. A pattern of slightly
deficient Ni and Cr has been observed for the low-$\alpha$ stars of NS97.
Bensby et al. (2003) found a correlation among the [iron-peak/Fe] and
[Na/Fe] vs. [$\alpha$-elements/Fe] abundance ratios, i.e., sligtly higher
[Cr/Fe], [Ni/Fe] and [Na/Fe] ratios in thick disk stars with enhanced
[$\alpha$-element/Fe] ratios (see their Fig. 13). Sbodorne et al. (2007)
found subsolar ratios for Na, Sc, Co, Ni and V in their analysis of the Sagittarius dwarf
galaxy stars, which has also low [$\alpha$/Fe] ratios. Such behavior may tell us interesting
details about the formation of these elements and give clues about low-mass galaxy formation.

Many interpretations have been given for the small [$\alpha$/Fe]
ratios observed. One hypothesis is that the star formation (SF)
developed slowly, in short bursts, followed by long quiescent periods without SF,
during which the SNe~Ia contaminated the ISM and increased the Fe
content (e.g. Gilmore \& Wyse 1991). Smaller SNe~II/SNe~Ia ratios,
therefore a higher frequency of SNe~Ia relative to SNe~II, have
also been invoked, within a bursty or continuous regime, and with
or without galactic winds (e.g. Pagel \& Tautvai\v{s}iene 1997,
Smith et al. 2002); a steepened IMF relative to that of the solar
neighborhood has been proposed by Tsujimoto et al. (1995) and de
Freitas Pacheco (1998), whereas alpha-enriched galactic winds,
which would lower the [alpha/Fe] content, have been suggested by Pilyugin (1996);
and finally, a small (low-mass) star-formation event that would effectively
truncate the IMF, yieding fewer high-mass SNe~II than produced by normal
SF events has been suggested (Tolstoy et al. 2003).
To find explanations for the behavior of the iron-peak
elements is more puzzling, since they are predicted to be
basically produced in SNe~Ia (e.g. Travaglio et al. 2005). A
possible explanation is that the yields of the SNe~Ia are
metallicity dependent (Timmes et al. 2003).

The abundance distributions observed for the $hs$ and the
$ls$ elements, with $hs/ls$=[Ba+La/Y+Zr], are in agreement with
the hypothesis that the $s$-process in AGB stars
is metallicity dependent (Busso et al. 1999 and references
therein; Busso et al. 2001; Abia et al. 2003, Travaglio et al.
2004). It has been noticed that, due to details of the
nucleosynthesis of the $s$-process, $hs$-elements (e.g. Ba, La and
Nd) are preferentially produced by metal-poor AGB stars compared to
 $ls$ elements (e.g. Y, Zr and Sr), which are most efficiently produced
 at [Fe/H] $\approx$ -0.1 (e.g. Fig. 1 of Travaglio et al. 2004). If
 the SF is slow, low-metallicity AGB stars have enough time to
contaminate the ISM, leaving noticeable chemical signatures for
the next generations.

Nevertheless, Venn et al. (2004) discuss the possibility that the
abundances of these elements (including Y) in dSph cannot be accounted for solely
by the $s$-process, requiring a strong contribution from the $r$-process.
Also, according to Richtler et al. (1989) and
Russell \& Dopita (1992), the most probable explanation for the
high Ba and La abundances observed in the Magellanic Clouds is an
additional $r$-process component. This would mean that $hs$ and
$ls$ elements are produced in different rates by the $r$-process
nucleosynthesis, probably in different sites. Therefore, the analysis of the
behavior of the $s$-elements in the given metallicity range is
complex and must take into account both the $r$ and the $s$
contributions.

\subsection{Galaxy Formation and Evolution}

One of the most debated themes about galaxy formation in the
Universe under a $\Lambda$CDM hierarchical scenario concerns the
problem of overprediction of galaxy counts at low-$z$ and
underprediction at high-$z$ (Cimatti et al. 2002). One of the
consequences for the Local Group is a larger number of small galaxies
than is actually observed although the number of dwarf
galaxies observed around the Milky Way and M31 has lately grown
significantly (eg. Belokurov et al. 2007). According to these models, numerous
merging and accretion events play an important role in the formation process
of massive galaxies (e.g. Moore et al. 1999), although not all
dark matter clumps are predicted to host star formation and
thereby become visible galaxies (e.g. Bullock \& Johnston 2005). The
quest for signatures of possible accreted stars from nearby galaxies in the
Galactic halo and disk have been carried out, without definite
conclusions (NS97, NS00, Ivans et al. 2003, Venn et al. 2004). A
careful inspection of the elemental distributions of the different
Galactic components reveals a low dispersion in the abundance
ratios at each metallicity bin and smooth transitions between
them (see e.g. plots from Venn et al. 2004). This seems to indicate a
different process: that the Galaxy, including the halo, has grown in a
holistic way, rather than by many independent accreting events, even for the
galactic halo (see Gilmore \& Wyse 2004). Another possibility is that
the merging events occurred very early in the building process of our Galaxy,
involving mostly dark matter and primordial gas. Such observational features
hint for a common history within the same environment rather than a mix of SFHs.
The results from the present work strongly support this idea, showing that an
LMC-like SFH results in a quite distinctive elemental pattern not seen in any
galactic stellar population.

We have found that the elemental compositions of the
LMC Inner Disk stars show a different pattern when compared to
their galactic counterparts (if we exclude the low-alpha stars
of NS97). This indicates that possible acreting events of LMC and
LMC-like fragments (Bekki \& Chiba 2005, Robertson et al. 2005),
from which our Galactic halo could have been buit, are unlikely,
but strong conclusions are still not
possible because more representative samples are needed, from both
halo and LMC stars. However, we stress here that the stellar
populations probed in the LMC are mostly intermediate age, and would
not have been merged into a Milky Way halo or disk if the accretion of an
LMC-like galaxy occurred early on (z$>$1). Strong conclusions
concerning the possible early accretion of LMC-type systems therefore
still await detailed analysis of the elemental abundances of representative samples of
the oldest populations in the LMC. The elemental distributions of the LMC Inner Disk
also hint for a different process of galaxy formation, showing that the
galactic local environment is fundamental for the the amount of various elements of its
components.

\section{Summary}

In the present paper we report abundance ratios for a series of
elements, including $\alpha$, $s$- and iron-peak elements, Na, Sc
and Cu for a sample of 59 RGB stars of the inner LMC disk. We have
found a very different behavior for most of the elements relative
to stars from the Galaxy with similar metallicity, hinting at a
very different evolutionary history. On the other hand, there is a
good overall agreement between the the elemental
distributions of our sample stars and previous results of the LMC GC and
field stars of Hill et al. (2000, 2003), Smith et al. (2002) and Johnson et al. (2006)
The main results are summarized as follows:

\begin{itemize}

\item {} [$\alpha$/Fe] ratios show an overall deficient pattern
relative to Galactic distributions, in agreement with a slower
star-formation history in the LMC, leading to a stronger Type Ia
supernovae influence. However, all $\alpha$-elements do not show the same degree
of deficiency: while O/Fe and Mg/Fe are hardly different in the LMC
and Milky-Way disks, Si, Ca and Ti are strongly underabundant. This
illustrates that all $\alpha$-elements are not alike from the
nucleosynthesis point of view

\item Cu is strongly depleted with respect to iron, [Cu/Fe] $\simeq$ -0.70 dex,
 with no apparent trend with metallicity. This also hints at a strong
 contribution of Type Ia supernovae to the creation of copper

\item the [X/Fe] deficiency of the $\alpha$-elements is also displayed
by Na, Sc, and, in an unexpected behavior, by the iron-peak
elements Ni, Cr and Co. The iron peak elements underabundances are not
expected in any standard chemical evolution model (i.e. currently not
predicted by SNe yields)

\item we have found relationships between Na-Ni and Na-Mg, in agreement
to those derived by Nissen \& Schuster (1997) for a sample of low-$\alpha$
halo stars. As Na is predicted to be mainly produced by SNe~II, together
with O and Mg, a relationship Na-Mg is expected, althoug Na production is
also controlled by the neutron excess during carbon burning in massive stars
(Umeda et al. 2000). The Na-Ni relationship is also expected if Ni is also
produced in SN~II, with yields dependent on the neutron excess (Thielemann et
al. 1990)

\item heavy neutron capture elements fall into two well-defined groups: while high-mass $s$-process
elements (Ba and La) present an enhanced pattern, low-mass $s$-process elements
(Y and Zr) are deficient relative to the galactic samples. Such behavior
has been observed before in LMC and SMC F supergiants and in
dSph galaxy RGB stars. It could reflect a strong contribution of
metal-poor AGB stars to the metal-enrichment of these systems, as
low-metallicity AGB stars preferentially produce the heavier $s$-process
elements over the lighter ones (see Travaglio et al. 2004 for the theoretical
side and de Laverny et al. 2006 for the observation of low metallicity AGBs)

\item we have derived Eu abundances for one of our intermediate-metallicity stars (RGB\_666: [Fe/H]=-1.10), and combined with the measured Ba abundance for this star, this enabled us to disantangle the respective $r$- and $s$-process contributions to heavy neutron-capture elements: this star
contains a solar mix of $r$- and $s$-process elements. Although a single measurement is obviously not enough to conclude, we thereby confirm that the high abundances of $ls$ elements observed at
intermediate metallicity should be attributed to the $s$-process.

For the next two fields of our program (see Introduction) the wavelength range of the spectra
covers a Eu line and a better evaluation of such contributions will be possible

\item compared to the dSph galaxies, similar abundance ratios for almost all the
elements have been derived, with slight enhancements of La, Ba, Na and Y, although
the match in metallicity among our sample and the dSph samples is not ideal. LMC
Inner Disk abundances of Ca, Si, Ti and Cu are also similar to those of the Sagittarius
dwarf galaxy. The commonalities between the LMC inner disk population and
the samples in dSph galaxies indicate that all these galaxies
may have undergone similar SFH

\end{itemize}

The overall pattern of the elemental distributions for the LMC Inner Disk population can
be explained by a higher contribution of Type Ia SNe, indicating that the build up of this
population has been slower than that of the solar neighborhood stars. A higher contribution
from metal-poor ABG stars is also proposed. The present results support the hypothesis that the elemental distributions of the stars are directly related to galaxy they pertain.

\begin{acknowledgements}
L. P. acknowledges CAPES and FAPESP fellowships \#0606-03-0 and \#01/14594-2.
We greatfully thank Peter Stetson for the availability of the DAOSPEC
program.

\end{acknowledgements}


\begin{longtable}{cccccc}
\caption{Photometric Data}\label{photometry} \\
\hline \hline
Star Reference & 2MASS number & V   &    I &  J   &  K    \\
               &              & mag & mag            & mag&  mag          \\
\hline \hline
\endhead
RGB\_1055&  05113508-7112309 & 17.599 & 16.219 & 15.070 & 14.172  \\
RGB\_1105&  05125047-7107463 & 17.661 & 16.170 & 14.972 & 13.952  \\
RGB\_1118&  05104862-7109301 & 17.628 & 16.278 & 15.298 & 14.258  \\
RGB\_499 &  05130497-7115406 & 17.023 & 15.699 & 14.624 & 13.905  \\
RGB\_512 &  05105703-7111340 & 16.994 & 15.539 & 14.452 & 13.489  \\
RGB\_522 &  05112258-7107277 & 17.005 & 15.537 & 14.360 & 13.420  \\
RGB\_533 &  05131266-7118005 & 16.958 & 15.537 & 14.562 & 13.571  \\
RGB\_534 &  05123774-7118119 & 17.111 & 15.890 & 14.807 & 14.045  \\
RGB\_546 &  05112068-7108113 & 17.041 & 15.619 & 14.546 & 13.639  \\
RGB\_548 &  05130454-7113055 & 17.095 & 15.680 & 14.502 & 13.573  \\
RGB\_565 &  05111922-7112564 & 17.061 & 15.585 & 14.468 & 13.489  \\
RGB\_576 &  05120852-7116597 & 17.132 & 15.806 & 14.668 & 13.631  \\
RGB\_593 &  05132454-7109519 & 17.168 & 15.688 & 14.484 & 13.529  \\
RGB\_599 &  05124460-7109195 & 17.174 & 15.756 & 14.574 & 13.663  \\
RGB\_601 &  05111325-7120037 & 17.112 & 15.673 & 14.633 & 13.772  \\
RGB\_606 &  05133509-7109322 & 17.152 & 15.791 & 14.694 & 13.850  \\
RGB\_611 &  05114888-7111492 & 17.122 & 15.603 & 14.478 & 13.589  \\
RGB\_614 &  05145465-7113031 & 17.023 & 15.492 & 14.459 & 13.375  \\
RGB\_620 &  05142327-7107446 & 17.205 & 15.790 & 14.608 & 13.845  \\
RGB\_625 &  05103395-7112074 & 17.144 & 15.614 & 14.473 & 13.440  \\
RGB\_629 &  05104928-7110057 & 17.140 & 15.766 & 14.723 & 13.792  \\
RGB\_631 &  05134131-7118477 & 17.054 & 15.655 & 14.638 & 13.638  \\
RGB\_633 &  05120481-7113402 & 17.131 & 15.647 & 14.527 & 13.702  \\
RGB\_640 &  05100529-7112259 & 17.154 & 15.772 & 14.747 & 13.791  \\
RGB\_646 &  05140805-7117297 & 17.071 & 15.674 & 14.653 & 13.922  \\
RGB\_651 &  05114466-7107176 & 17.152 & 15.713 & 14.672 & 13.729  \\
RGB\_655 &  05143617-7109412 & 17.202 & 15.674 & 14.521 & 13.616  \\
RGB\_656 &  05122551-7112106 & 17.191 & 15.758 & 14.637 & 13.743  \\
RGB\_658 &  05100845-7109582 & 17.229 & 15.797 & 14.728 & 13.643  \\
RGB\_664 &  05100659-7115514 & 17.156 & 15.529 & 14.438 & 13.336  \\
RGB\_666 &  05104728-7119320 & 17.167 & 15.833 & 14.763 & 13.977  \\
RGB\_671 &  05114880-7113428 & 17.197 & 15.705 & 14.555 & 13.585  \\
RGB\_672 &  05130066-7116289 & 17.193 & 15.611 & 14.460 & 13.407  \\
RGB\_679 &  05123409-7113324 & 17.203 & 15.653 & 14.481 & 13.578  \\
RGB\_690 &  05144229-7110108 & 17.266 & 15.678 & 14.488 & 13.413  \\
RGB\_699 &  05095252-7115084 & 17.214 & 16.058 & 15.243 & 14.399  \\
RGB\_700 &  05113581-7113336 & 17.284 & 15.821 & 14.717 & 13.676  \\
RGB\_701 &  05124208-7110018 & 17.214 & 15.693 & 14.590 & 13.579  \\
RGB\_705 &  05141536-7107463 & 17.215 & 15.886 & 14.866 & 14.026  \\
RGB\_710 &  05110701-7108413 & 17.308 & 15.762 & 14.424 & 13.368  \\
RGB\_720 &  05103055-7116158 & 17.314 & 15.984 & 14.901 & 14.313  \\
RGB\_728 &  05142677-7119303 & 17.249 & 15.836 & 14.777 & 13.955  \\
RGB\_731 &  05120180-7117002 & 17.255 & 15.593 & 14.382 & 13.357  \\
RGB\_748 &  05122530-7119025 & 17.279 & 15.781 & 14.804 & 13.760  \\
RGB\_752 &  05144969-7110095 & 17.320 & 15.801 & 14.566 & 13.621  \\
RGB\_756 &  05143449-7112462 & 17.251 & 15.568 & 14.349 & 13.278  \\
RGB\_758 &  05111461-7118573 & 17.269 & 15.983 & 14.996 & 14.266  \\
RGB\_766 &  05111734-7115235 & 17.343 & 15.861 & 14.726 & 13.779  \\
RGB\_773 &  05115657-7108489 & 17.264 & 15.707 & 14.632 & 13.602  \\
RGB\_775 &  05095756-7116288 & 17.261 & 15.927 & 14.879 & 14.100  \\
RGB\_776 &  05111615-7116401 & 17.287 & 15.877 & 14.826 & 14.033  \\
RGB\_782 &  05104950-7107338 & 17.291 & 15.844 & 14.758 & 13.761  \\
RGB\_789 &  05121657-7108570 & 17.310 & 15.629 & 14.382 & 13.267  \\
RGB\_793 &  05110667-7111205 & 17.319 & 15.843 & 14.751 & 13.783  \\
RGB\_834 &  05112287-7116589 & 17.355 & 15.828 &   -    &  -      \\
RGB\_854 &  05102155-7118506 & 17.415 & 15.997 & 14.922 & 14.064  \\
RGB\_855 &  05124558-7116301 & 17.393 & 16.001 & 14.901 & 13.981  \\
RGB\_859 &  05112287-7116589 & 17.397 & 15.883 & 14.730 & 13.806  \\
RGB\_900 &  05130400-7113289 & 17.400 & 15.983 & 14.915 & 14.032  \\
\end{longtable}

\begin{longtable}{ccccccccccc}
\caption{Stellar Parameters}\label{stellpar} \\
\hline\hline
Star  &  T$_{\rm photLow}$ &  T$_{\rm phot}$ & T$_{\rm spec}$ & log g$_{\rm phot}$
& log g$_{\rm spec}$ & [Fe/H]$_{\rm spec}$ & [Fe/H]$_{\rm CaT}$ & [FeII/H] & Vt & Rv \\
\hline
\endfirsthead
\caption{continued}\\
\hline\hline
Star  & T$_{\rm photLow}$ &  T$_{\rm phot}$ & T$_{\rm spec}$  & log g$_{\rm phot}$
& log g$_{\rm spec}$ & [FeI/H]$_{\rm spec}$ & [FeI/H]$_{\rm CaT}$ &  [FeII/H] & Vt & Rv\\
\hline
\endhead
\hline
\endfoot
RGB\_1055  & 4066  & 4118 & 4266 & 1.5  & 0.90 & -0.96 & -0.87 & -0.87  & 1.2 & 177\\
RGB\_1105  & 3921  & 3965 & 4100 & 1.4  & 0.90 & -0.71 & -1.15 & -0.69  & 1.6 & 243\\
RGB\_1118  & 4102  & 4154 & 4204 & 1.5  & 1.30 & -0.57 & -0.25 & -0.65  & 1.8 & 208\\
RGB\_499   & 4212  & 4269 & 4242 & 1.4  & 1.00 & -0.85 & -0.44 & -0.89  & 2.2 & 220\\
RGB\_512   & 4002  & 4051 & 4202 & 1.2  & 0.80 & -0.84 & -0.78 & -0.89  & 1.7 & 247\\
RGB\_522   & 3971  & 4016 & 4101 & 1.2  & 1.01 & -0.70 & -0.37 & -0.77  & 2.0 & 270\\
RGB\_533   & 4062  & 4113 & 4112 & 1.2  & 0.80 & -0.75 & -0.43 & -0.82  & 2.0 & 243\\
RGB\_534   & 4295  & 4359 & 4295 & 1.5  & 1.20 & -1.22 & -1.11 & -1.12  & 1.6 & 246\\
RGB\_546   & 4055  & 4107 & 4185 & 1.3  & 0.80 & -0.96 & -0.91 & -0.99  & 1.7 & 260\\
RGB\_548   & 4016  & 4064 & 4066 & 1.2  & 0.90 & -0.74 & -0.31 & -0.80  & 2.0 & 247\\
RGB\_565   & 3970  & 4016 & 4100 & 1.2  & 0.70 & -0.94 & -0.60 & -0.96  & 1.9 & 242\\
RGB\_576   & 4064  & 4117 & 4190 & 1.3  & 0.80 & -1.24 & -1.03 & -1.20  & 1.6 & 305\\
RGB\_593   & 3948  & 3993 & 4088 & 1.2  & 0.70 & -1.15 & -0.58 & -1.17  & 1.9 & 234\\
RGB\_599   & 4018  & 4066 & 4028 & 1.3  & 0.80 & -0.84 & -0.71 & -0.81  & 1.8 & 241\\
RGB\_601   & 4071  & 4123 & 4101 & 1.3  & 1.01 & -0.55 & -0.77 & -0.44  & 2.0 & 242\\
RGB\_606   & 4122  & 4174 & 4320 & 1.4  & 0.80 & -1.74 & -1.63 & -1.72  & 1.0 & 183\\
RGB\_611   & 3968  & 4010 & 3980 & 1.2  & 0.70 & -0.45 & -0.42 & -0.55  & 1.6 & 244\\
RGB\_614   & 3756  & 3967 & 4107 & 1.1  & 0.70 & -0.87 & -0.71 & -0.84  & 2.2 & 241\\
RGB\_620   & 4075  & 4127 & 4197 & 1.4  & 1.30 & -0.61 & -0.28 & -0.74  & 2.0 & 197\\
RGB\_625   & 3910  & 3951 & 4090 & 1.2  & 0.70 & -0.91 & -0.86 & -0.91  & 2.2 & 242\\
RGB\_629   & 4099  & 4150 & 4229 & 1.3  & 0.80 & -0.91 & -0.97 & -0.95  & 1.7 & 188\\
RGB\_631   & 4061  & 4112 & 4061 & 1.3  & 0.80 & -0.64 & -0.90 & -0.75  & 1.7 & 256\\
RGB\_633   & 4015  & 4067 & 4015 & 1.3  & 0.90 & -0.62 & -1.21 & -0.55  & 1.9 & 194\\
RGB\_640   & 4089  & 4141 & 4280 & 1.3  & 0.80 & -0.93 & -0.82 & -0.93  & 1.9 & 219\\
RGB\_646   & 4166  & 4218 & 4216 & 1.4  & 1.20 & -0.72 & -0.69 & -0.63  & 1.9 & 236\\
RGB\_651   & 4039  & 4091 & 4089 & 1.3  & 1.10 & -0.40 & -0.51 & -0.46  & 1.8 & 247\\
RGB\_655   & 3948  & 3988 & 4048 & 1.2  & 0.80 & -0.57 & -0.66 & -0.50  & 1.8 & 226\\
RGB\_656   & 4032  & 4084 & 4082 & 1.3  & 0.80 & -0.71 & -0.56 & -0.65  & 2.0 & 233\\
RGB\_658   & 3987  & 4033 & 4087 & 1.3  & 1.10 & -0.61 & -0.40 & -0.57  & 2.0 & 231\\
RGB\_664   & 3840  & 3881 & 3900 & 1.1  & 0.70 & -0.54 & -0.58 & -0.48  & 1.9 & 251\\
RGB\_666   & 4179  & 4233 & 4279 & 1.4  & 1.00 & -1.02 & -1.02 & -1.01  & 1.7 & 225\\
RGB\_671   & 3952  & 3996 & 4052 & 1.2  & 0.90 & -0.78 & -0.55 & -0.70  & 1.9 & 249\\
RGB\_672   & 3866  & 3906 & 3956 & 1.2  & 0.70 & -0.68 & -0.38 & -0.66  & 1.9 & 251\\
RGB\_679   & 3928  & 3968 & 3998 & 1.2  & 0.80 & -0.63 & -0.34 & -0.67  & 2.0 & 253\\
RGB\_690   & 3843  & 3883 & 3950 & 1.2  & 0.90 & -0.66 & -0.23 & -0.70  & 2.0 & 296\\
RGB\_699   & 4458  & 4531 & 4488 & 1.6  & 1.20 & -0.64 & -1.15 & -0.70  & 1.4 & 230\\
RGB\_700   & 3966  & 4011 & 4000 & 1.3  & 1.01 & -0.60 & -0.37 & -0.56  & 2.0 & 282\\
RGB\_701   & 3934  & 3975 & 4125 & 1.2  & 0.70 & -0.73 & -0.33 & -0.65  & 2.1 & 257\\
RGB\_705   & 4182  & 4237 & 4202 & 1.4  & 1.20 & -0.55 & -0.72 & -0.50  & 1.6 & 250\\
RGB\_710   & 3834  & 3870 & 3950 & 1.1  & 0.80 & -0.75 & -0.65 & -0.53  & 1.9 & 265\\
RGB\_720   & 3814  & 4320 & 4370 & 1.6  & 1.40 & -0.82 & -0.90 & -0.85  & 1.7 & 200\\
RGB\_728   & 4102  & 4153 & 4252 & 1.4  & 0.90 & -0.85 & -0.80 & -0.76  & 2.1 & 270\\
RGB\_731   & 3804  & 3844 & 3900 & 1.1  & 0.80 & -0.48 & -0.23 & -0.38  & 1.8 & 278\\
RGB\_748   & 3980  & 4026 & 4186 & 1.3  & 0.90 & -0.35 & -0.17 & -0.32  & 1.5 & 223\\
RGB\_752   & 3915  & 3956 & 3915 & 1.2  & 1.01 & -0.28 & -0.08 & -0.24  & 1.8 & 225\\
RGB\_756   & 3780  & 3813 & 3930 & 1.1  & 0.70 & -0.82 & -0.46 & -0.75  & 2.0 & 254\\
RGB\_758   & 4282  & 4347 & 4442 & 1.6  & 1.20 & -0.95 & -1.22 & -0.92  & 1.7 & 257\\
RGB\_766   & 3971  & 4016 & 4156 & 1.3  & 0.90 & -0.64 & -0.46 & -0.65  & 1.9 & 282\\
RGB\_773   & 3914  & 3954 & 4034 & 1.2  & 0.80 & -0.87 & -0.51 & -0.76  & 2.4 & 232\\
RGB\_775   & 4191  & 4245 & 4271 & 1.5  & 1.01 & -0.82 & -1.28 & -0.82  & 1.2 & 241\\
RGB\_776   & 4118  & 4170 & 4178 & 1.4  & 1.01 & -0.73 & -0.75 & -0.72  & 1.7 & 241\\
RGB\_782   & 3998  & 4045 & 4078 & 1.3  & 0.90 & -0.57 & -0.34 & -0.52  & 1.8 & 249\\
RGB\_789   & 3763  & 3796 & 3923 & 1.1  & 0.60 & -0.56 & -0.36 & -0.58  & 1.8 & 245\\
RGB\_793   & 3982  & 4029 & 4169 & 1.3  & 0.80 & -0.70 & -0.53 & -0.80  & 1.9 & 241\\
RGB\_834   & 3953  & 3993 & 4053 & 1.3  & 0.80 & -0.86 & -0.64 & -0.79  & 2.0 & 197\\
RGB\_854   & 4077  & 4129 & 4157 & 1.4  & 1.20 & -0.70 & -0.10 & -0.82  & 2.0 & 313\\
RGB\_855   & 4065  & 4117 & 4257 & 1.4  & 1.20 & -0.74 & -0.02 & -0.73  & 1.9 & 217\\
RGB\_859   & 3951  & 3992 & 4021 & 1.3  & 1.01 & -0.64 & -0.22 & -0.56  & 1.9 & 244\\
RGB\_900   & 4071  & 4123 & 4131 & 1.4  & 1.01 & -0.69 & -0.27 & -0.64  & 2.1 & 276\\
\end{longtable}

\begin{longtable}{cccccccc}
\caption{Line List}\\
\label{linelist} \\
\hline\hline Wavelength  &
Element & $\chi_{exc}$  & log gf & Wavelength  &  Element
& $\chi_{exc}$  & log gf \\
\hline
\endfirsthead
\caption{continued.}\\
\hline\hline Wavelength  &  Element  & $\chi_{exc}$  & log gf &
Wavelength  &  Element
& $\chi_{exc}$  & log gf \\
\hline
\endhead
\hline
\endfoot
6496.900 & BA2 & 0.604   &  -0.380   & 6393.610 & FE1 & 2.430   &  -1.580  \\
6572.800 & CA1 & 0.000   &  -4.300   & 6344.160 & FE1 & 2.430   &  -2.920  \\  
6162.190 & CA1 & 1.900   &  -0.090   & 6593.870 & FE1 & 2.437   &  -2.420  \\   
6169.560 & CA1 & 2.520   &  -0.270   & 5701.560 & FE1 & 2.560   &  -2.220  \\   
6169.040 & CA1 & 2.520   &  -0.540   & 6609.120 & FE1 & 2.560   &  -2.690  \\   
5601.290 & CA1 & 2.520   &  -0.690   & 6475.630 & FE1 & 2.560   &  -2.940  \\ 
6493.790 & CA1 & 2.521   &   0.140   & 6137.700 & FE1 & 2.590   &  -1.400  \\
6166.440 & CA1 & 2.521   &  -0.900   & 6322.690 & FE1 & 2.590   &  -2.430  \\ 
6499.650 & CA1 & 2.523   &  -0.590   & 6575.040 & FE1 & 2.590   &  -2.710  \\ 
6161.300 & CA1 & 2.523   &  -1.030   & 6200.320 & FE1 & 2.610   &  -2.440  \\ 
6455.610 & CA1 & 2.523   &  -1.360   & 6180.210 & FE1 & 2.730   &  -2.650  \\ 
6439.080 & CA1 & 2.526   &   0.470   & 6518.370 & FE1 & 2.830   &  -2.300  \\ 
6471.670 & CA1 & 2.526   &  -0.590   & 6355.040 & FE1 & 2.840   &  -2.290  \\ 
6508.840 & CA1 & 2.526   &  -2.110   & 6411.660 & FE1 & 3.650   &  -0.720  \\ 
5647.240 & CO1 & 2.280   &  -1.560   & 6301.510 & FE1 & 3.650   &  -0.600  \\ 
6330.100 & CR1 & 0.940   &  -2.910   & 6302.500 & FE1 & 3.690   &  -0.910  \\ 
6362.880 & CR1 & 0.940   &  -2.700   & 6336.830 & FE1 & 3.690   &  -1.050  \\
5787.930 & CR1 & 3.320   &  -0.080   & 6408.030 & FE1 & 3.690   &  -1.000  \\
5783.070 & CR1 & 3.320   &  -0.500   & 5809.220 & FE1 & 3.883   &  -1.690  \\
5782.130 & CU1 & 1.642   &  -1.720   & 6188.020 & FE1 & 3.940   &  -1.720  \\
6358.690 & FE1 & 0.860   &  -4.470   & 6157.730 & FE1 & 4.076   &  -1.110  \\
6498.950 & FE1 & 0.960   &  -4.700   & 6165.360 & FE1 & 4.142   &  -1.470  \\
6574.250 & FE1 & 0.990   &  -5.020   & 6380.750 & FE1 & 4.190   &  -1.380  \\
6581.220 & FE1 & 1.480   &  -4.860   & 5618.630 & FE1 & 4.209   &  -1.260  \\
6430.860 & FE1 & 2.180   &  -2.010   & 5638.270 & FE1 & 4.220   &  -0.870  \\
6151.620 & FE1 & 2.180   &  -3.300   & 5635.820 & FE1 & 4.256   &  -1.740  \\
6335.340 & FE1 & 2.200   &  -2.180   & 5641.450 & FE1 & 4.260   &  -1.180  \\
6297.800 & FE1 & 2.220   &  -2.740   & 5814.810 & FE1 & 4.283   &  -1.820  \\
6173.340 & FE1 & 2.220   &  -2.880   & 5717.830 & FE1 & 4.284   &  -0.980  \\
6421.350 & FE1 & 2.279   &  -2.010   & 5705.470 & FE1 & 4.301   &  -1.360  \\
6481.880 & FE1 & 2.280   &  -2.980   & 5691.500 & FE1 & 4.301   &  -1.370  \\
6392.540 & FE1 & 2.280   &  -4.030   & 5619.610 & FE1 & 4.390   &  -1.700  \\
6608.040 & FE1 & 2.280   &  -4.030   & 5806.730 & FE1 & 4.607   &  -0.900  \\
6494.990 & FE1 & 2.400   &  -1.270   & 5679.020 & FE1 & 4.651   &  -0.770  \\
6597.560 & FE1 & 4.795   &  -0.920   & 5793.070 & SI1 & 4.930   &  -2.060  \\
6469.190 & FE1 & 4.835   &  -0.620   & 6599.110 & TI1 & 0.900   &  -2.085  \\
5633.950 & FE1 & 4.990   &  -0.270   & 6126.220 & TI1 & 1.070   &  -1.420  \\
6516.080 & FE2 & 2.890   &  -3.450   & 6261.110 & TI1 & 1.430   &  -0.480  \\
61432.68 & FE2 & 2.890   &  -3.708   & 6554.240 & TI1 & 1.440   &  -1.220  \\
6149.250 & FE2 & 3.889   &  -2.724   & 6303.770 & TI1 & 1.440   &  -1.570  \\
6247.560 & FE2 & 3.890   &  -2.329   & 6258.100 & TI1 & 1.443   &  -0.350  \\
6416.920 & FE2 & 3.891   &  -2.740   & 6556.080 & TI1 & 1.460   &  -1.080  \\
6456.390 & FE2 & 3.900   &  -2.075   & 5648.580 & TI1 & 2.490   &  -0.250  \\
6320.430 & LA2 & 0.170   &  -1.520   & 6559.590 & TI2 & 2.048   &  -2.190  \\
5711.090 & MG1 & 4.346   &  -1.833   & 6491.560 & TI2 & 2.061   &  -1.793  \\
5688.220 & NA1 & 2.100   &  -0.460   & 6606.950 & TI2 & 2.061   &  -2.790  \\
5682.650 & NA1 & 2.100   &  -0.700   & 6274.660 &  V1 & 0.270   &  -1.670  \\
6160.750 & NA1 & 2.100   &  -1.230   & 6285.170 &  V1 & 0.280   &  -1.510  \\
6154.230 & NA1 & 2.100   &  -1.530   & 6199.190 &  V1 & 0.290   &  -1.290  \\
6327.600 & NI1 & 1.680   &  -3.150   & 6292.820 &  V1 & 0.290   &  -1.470  \\
6128.980 & NI1 & 1.680   &  -3.330   & 6224.510 &  V1 & 0.290   &  -2.010  \\
6314.670 & NI1 & 1.930   &  -1.770   & 6251.820 &  V1 & 0.290   &  -1.300  \\
6482.810 & NI1 & 1.930   &  -2.630   & 6150.150 &  V1 & 0.300   &  -1.790  \\
6532.890 & NI1 & 1.935   &  -3.390   & 6135.370 &  V1 & 1.050   &  -0.750  \\
6586.320 & NI1 & 1.950   &  -2.810   & 6119.530 &  V1 & 1.060   &  -0.320  \\
6175.370 & NI1 & 4.090   &  -0.530   & 6452.320 &  V1 & 1.190   &  -1.210  \\
6300.310 &  O1 & 0.000   &  -9.770   & 6531.410 &  V1 & 1.218   &  -0.840  \\
5657.150 & SC2 & 1.500   &  -0.603   & 6357.290 &  V1 & 1.849   &  -0.910  \\
5665.560 & SI1 & 4.920   &  -1.720   & 6435.010 &  Y1 & 0.070   &  -0.820  \\
5690.430 & SI1 & 4.930   &  -1.870   & 6134.570 & ZR1 & 0.000   &  -1.280  \\
\end{longtable}

\begin{longtable}{cccccc}\label{AbAlpha1} \\
\caption{Abundance ratios of the elements. Iron and Si, Ca, Ti1 and Ti2.}\\
\hline\hline
Star & [Fe/H]   & [Si/Fe]  & [Ca/Fe]  & [Ti1/Fe]  & [Ti2/Fe]   \\
\hline
\endfirsthead
\caption{continued.}\\
\hline\hline
Star & [Fe/H]  &  [Si/Fe]  & [Ca/Fe]  & [Ti1/Fe] &  [Ti2/Fe]   \\
\hline
\endhead
\hline
\endfoot
RGB\_1055& -0.96 $\pm$ 0.16 & -0.01 $\pm$ 0.09 &  -0.10 $\pm$ 0.05 & -0.21 $\pm$ 0.04 & -0.06 $\pm$ 0.12 \\
RGB\_1105& -0.73 $\pm$ 0.16 &  0.04 $\pm$ 0.21 &  -0.07 $\pm$ 0.11 &  0.10 $\pm$ 0.12 &  0.01 $\pm$ 0.09 \\
RGB\_1118& -0.57 $\pm$ 0.16 &  0.13 $\pm$ 0.07 &   0.03 $\pm$ 0.12 &  0.00 $\pm$ 0.06 &  0.13 $\pm$ 0.03 \\
RGB\_499 & -0.85 $\pm$ 0.16 &  0.18 $\pm$ 0.05 &  -0.06 $\pm$ 0.09 &  0.20 $\pm$ 0.05 & -0.06 $\pm$ 0.13 \\
RGB\_512 & -0.84 $\pm$ 0.16 & -0.07 $\pm$ 0.08 &  -0.06 $\pm$ 0.08 &  0.06 $\pm$ 0.04 & -0.04 $\pm$ 0.03 \\
RGB\_522 & -0.70 $\pm$ 0.16 &  0.24 $\pm$ 0.04 &  -0.13 $\pm$ 0.07 &  0.09 $\pm$ 0.02 &  0.22 $\pm$ 0.03 \\
RGB\_533 & -0.75 $\pm$ 0.16 &  0.07 $\pm$ 0.04 &  -0.08 $\pm$ 0.10 &  0.17 $\pm$ 0.04 & -0.06 $\pm$ 0.12 \\
RGB\_534 & -1.21 $\pm$ 0.16 &  0.15 $\pm$ 0.13 &  -0.02 $\pm$ 0.05 &  0.09 $\pm$ 0.05 &  0.17 $\pm$ 0.14 \\
RGB\_546 & -0.93 $\pm$ 0.16 & -0.06 $\pm$ 0.16 &   0.05 $\pm$ 0.09 & -0.08 $\pm$ 0.04 & -0.07 $\pm$ 0.04 \\
RGB\_548 & -0.74 $\pm$ 0.16 &  0.08 $\pm$ 0.08 &  -0.23 $\pm$ 0.09 &  0.04 $\pm$ 0.04 & -0.09 $\pm$ 0.09 \\
RGB\_565 & -0.95 $\pm$ 0.16 & -0.01 $\pm$ 0.03 &  -0.16 $\pm$ 0.03 &  0.17 $\pm$ 0.03 &  0.00 $\pm$ 0.06 \\
RGB\_576 & -1.24 $\pm$ 0.16 & -0.04 $\pm$ 0.12 &   0.00 $\pm$ 0.04 & -0.03 $\pm$ 0.03 &  0.05 $\pm$ 0.04 \\
RGB\_593 & -1.15 $\pm$ 0.16 &  0.06 $\pm$ 0.13 &  -0.04 $\pm$ 0.10 &  0.15 $\pm$ 0.05 &  0.00 $\pm$ 0.27 \\
RGB\_599 & -0.85 $\pm$ 0.16 &  0.06 $\pm$ 0.05 &  -0.14 $\pm$ 0.08 & -0.09 $\pm$ 0.02 &  0.05 $\pm$ 0.14 \\
RGB\_601 & -0.52 $\pm$ 0.17 &  0.11 $\pm$ 0.09 &  -0.34 $\pm$ 0.11 & -0.11 $\pm$ 0.05 & -0.09 $\pm$ 0.05 \\
RGB\_606 & -1.74 $\pm$ 0.16 &   -   $\pm$  -   &   0.13 $\pm$ 0.08 & -0.17 $\pm$ 0.08 &  0.21 $\pm$ 0.25 \\
RGB\_611 & -0.45 $\pm$ 0.16 & -0.05 $\pm$ 0.04 &  -0.12 $\pm$ 0.14 &  0.01 $\pm$ 0.08 & -0.14 $\pm$ 0.05 \\
RGB\_614 & -0.87 $\pm$ 0.17 &   -   $\pm$  -   &  -0.07 $\pm$ 0.08 &  0.22 $\pm$ 0.04 & -0.01 $\pm$ 0.11 \\
RGB\_620 & -0.61 $\pm$ 0.17 &  0.24 $\pm$ 0.10 &  -0.23 $\pm$ 0.08 &  0.08 $\pm$ 0.09 &  0.28 $\pm$ 0.08 \\
RGB\_625 & -0.91 $\pm$ 0.16 &  0.09 $\pm$ 0.05 &   0.15 $\pm$ 0.08 &  0.12 $\pm$ 0.04 &  0.09 $\pm$ 0.10 \\
RGB\_629 & -0.91 $\pm$ 0.16 & -0.05 $\pm$ 0.09 &  -0.19 $\pm$ 0.06 & -0.13 $\pm$ 0.03 &  0.05 $\pm$ 0.06 \\
RGB\_631 & -0.63 $\pm$ 0.16 & -0.02 $\pm$ 0.12 &  -0.01 $\pm$ 0.12 & -0.05 $\pm$ 0.06 &  0.04 $\pm$ 0.03 \\
RGB\_633 & -0.62 $\pm$ 0.16 & -0.01 $\pm$ 0.14 &  -0.21 $\pm$ 0.07 & -0.14 $\pm$ 0.06 &  0.06 $\pm$ 0.16 \\
RGB\_640 & -0.93 $\pm$ 0.16 &  0.02 $\pm$ 0.11 &   0.08 $\pm$ 0.09 & -0.05 $\pm$ 0.05 &  0.07 $\pm$ 0.08 \\
RGB\_646 & -0.69 $\pm$ 0.16 & -0.04 $\pm$ 0.05 &  -0.18 $\pm$ 0.15 & -0.15 $\pm$ 0.07 & -0.01 $\pm$ 0.08 \\
RGB\_651 & -0.40 $\pm$ 0.16 &  0.07 $\pm$ 0.08 &  -0.31 $\pm$ 0.10 & -0.23 $\pm$ 0.04 & -0.11 $\pm$ 0.05 \\
RGB\_655 & -0.57 $\pm$ 0.16 &  0.13 $\pm$ 0.03 &  -0.24 $\pm$ 0.08 & -0.09 $\pm$ 0.05 & -0.02 $\pm$ 0.09 \\
RGB\_656 & -0.71 $\pm$ 0.16 &  0.07 $\pm$ 0.06 &  -0.07 $\pm$ 0.09 &  0.03 $\pm$ 0.11 &  0.02 $\pm$ 0.23 \\
RGB\_658 & -0.61 $\pm$ 0.16 &   -   $\pm$  -   &  -0.24 $\pm$ 0.07 & -0.05 $\pm$ 0.05 & -0.06 $\pm$ 0.26 \\
RGB\_664 & -0.54 $\pm$ 0.16 &  0.16 $\pm$ 0.08 &  -0.45 $\pm$ 0.07 & -0.22 $\pm$ 0.07 &  0.00 $\pm$ 0.18 \\
RGB\_666 & -1.02 $\pm$ 0.16 & -0.08 $\pm$ 0.13 &  -0.08 $\pm$ 0.08 & -0.11 $\pm$ 0.09 & -0.06 $\pm$ 0.06 \\
RGB\_671 & -0.78 $\pm$ 0.16 &  0.10 $\pm$ 0.08 &  -0.11 $\pm$ 0.11 & -0.21 $\pm$ 0.02 & -0.06 $\pm$ 0.03 \\
RGB\_672 & -0.65 $\pm$ 0.17 &  0.10 $\pm$ 0.05 &  -0.21 $\pm$ 0.22 & -0.04 $\pm$ 0.07 &  0.09 $\pm$ 0.13 \\
RGB\_679 & -0.63 $\pm$ 0.16 &  0.14 $\pm$ 0.04 &  -0.19 $\pm$ 0.09 & -0.15 $\pm$ 0.07 & -0.08 $\pm$ 0.05 \\
RGB\_690 & -0.66 $\pm$ 0.16 &  0.23 $\pm$ 0.04 &  -0.27 $\pm$ 0.09 &  0.03 $\pm$ 0.03 & -0.10 $\pm$ 0.05 \\
RGB\_699 & -0.59 $\pm$ 0.16 & -0.04 $\pm$ 0.17 &  -0.10 $\pm$ 0.09 & -0.18 $\pm$ 0.04 & -0.09 $\pm$ 0.24 \\
RGB\_700 & -0.60 $\pm$ 0.16 &  0.30 $\pm$ 0.14 &  -0.36 $\pm$ 0.10 & -0.29 $\pm$ 0.03 & -0.10 $\pm$ 0.12 \\
RGB\_701 & -0.73 $\pm$ 0.17 &  0.11 $\pm$ 0.09 &   0.19 $\pm$ 0.13 &  0.20 $\pm$ 0.07 &  0.00 $\pm$ 0.09 \\
RGB\_705 & -0.52 $\pm$ 0.16 &  0.02 $\pm$ 0.06 &     -  $\pm$ 0.16 & -0.26 $\pm$ 0.04 & -0.06 $\pm$ 0.05 \\
RGB\_710 & -0.70 $\pm$ 0.16 &  0.09 $\pm$ 0.11 &  -0.34 $\pm$ 0.13 & -0.29 $\pm$ 0.10 &  0.02 $\pm$ 0.06 \\
RGB\_720 & -0.83 $\pm$ 0.16 &  0.21 $\pm$ 0.10 &  -0.06 $\pm$ 0.08 &  0.01 $\pm$ 0.04 &  0.15 $\pm$ 0.12 \\
RGB\_728 & -0.85 $\pm$ 0.16 &  0.08 $\pm$ 0.08 &  -0.02 $\pm$ 0.07 &  0.02 $\pm$ 0.06 &  0.35 $\pm$ 0.13 \\
RGB\_731 & -0.46 $\pm$ 0.17 & -0.07 $\pm$ 0.09 &  -0.36 $\pm$ 0.13 & -0.17 $\pm$ 0.12 & -0.11 $\pm$ 0.16 \\
RGB\_748 & -0.31 $\pm$ 0.17 & -0.14 $\pm$ 0.10 &  -0.49 $\pm$ 0.08 &  0.22 $\pm$ 0.07 &    -  $\pm$ 0.09 \\
RGB\_752 & -0.28 $\pm$ 0.16 &  0.11 $\pm$ 0.04 &  -0.36 $\pm$ 0.09 & -0.02 $\pm$ 0.07 & -0.01 $\pm$ 0.08 \\
RGB\_756 & -0.80 $\pm$ 0.17 &  0.08 $\pm$ 0.05 &  -0.07 $\pm$ 0.15 &  0.26 $\pm$ 0.12 & -0.04 $\pm$ 0.08 \\
RGB\_758 & -0.95 $\pm$ 0.16 &   -   $\pm$  -   &  -0.11 $\pm$ 0.08 & -0.02 $\pm$ 0.10 & -0.05 $\pm$ 0.14 \\
RGB\_766 & -0.64 $\pm$ 0.17 &  0.13 $\pm$ 0.15 &  -0.18 $\pm$ 0.07 &  0.05 $\pm$ 0.04 & -0.02 $\pm$ 0.04 \\
RGB\_773 & -0.78 $\pm$ 0.17 &  0.09 $\pm$ 0.09 &  -0.49 $\pm$ 0.13 &  0.05 $\pm$ 0.07 &  0.15 $\pm$ 0.16 \\
RGB\_775 & -0.82 $\pm$ 0.16 & -0.16 $\pm$ 0.11 &  -0.09 $\pm$ 0.10 & -0.11 $\pm$ 0.07 & -0.10 $\pm$ 0.18 \\
RGB\_776 & -0.73 $\pm$ 0.16 & -0.19 $\pm$ 0.08 &  -0.13 $\pm$ 0.08 & -0.16 $\pm$ 0.05 &  0.03 $\pm$ 0.03 \\
RGB\_782 & -0.57 $\pm$ 0.16 &  0.08 $\pm$ 0.09 &  -0.24 $\pm$ 0.09 & -0.11 $\pm$ 0.03 & -0.07 $\pm$ 0.07 \\
RGB\_789 & -0.56 $\pm$ 0.16 &  0.01 $\pm$ 0.03 &  -0.24 $\pm$ 0.08 &  0.13 $\pm$ 0.07 &  0.04 $\pm$ 0.09 \\
RGB\_793 & -0.70 $\pm$ 0.16 & -0.07 $\pm$ 0.06 &  -0.16 $\pm$ 0.08 & -0.06 $\pm$ 0.05 &  0.01 $\pm$ 0.06 \\
RGB\_834 & -0.86 $\pm$ 0.16 & -0.10 $\pm$ 0.06 &  -0.21 $\pm$ 0.05 &  0.06 $\pm$ 0.07 &  0.11 $\pm$ 0.03 \\
RGB\_854 & -0.70 $\pm$ 0.16 &  0.10 $\pm$ 0.08 &  -0.08 $\pm$ 0.08 &  0.12 $\pm$ 0.14 & -0.06 $\pm$ 0.33 \\
RGB\_855 & -0.74 $\pm$ 0.16 &  0.00 $\pm$ 0.09 &  -0.13 $\pm$ 0.09 &  0.05 $\pm$ 0.06 &  0.11 $\pm$ 0.18 \\
RGB\_859 & -0.64 $\pm$ 0.16 &  0.13 $\pm$ 0.09 &  -0.13 $\pm$ 0.12 & -0.14 $\pm$ 0.04 &  0.17 $\pm$ 0.05 \\
RGB\_900 & -0.69 $\pm$ 0.16 &  0.20 $\pm$ 0.05 &  -0.12 $\pm$ 0.09 & -0.06 $\pm$ 0.03 &  0.09 $\pm$ 0.10 \\
\end{longtable}

\begin{longtable}{cccccc}\label{AbAlpha2} \\
\caption{Abundance ratios of the elements. Na, Sc, Cu and the $\alpha$-elements Mg and O.}\\
\hline\hline
Star & [O/Fe]   & [Mg/Fe]  & [Na/Fe]  & [Sc/Fe] & [Cu/Fe]  \\
\hline
\endfirsthead
\caption{continued.}\\
\hline\hline
Star & [O/Fe]   & [Mg/Fe]  & [Na/Fe]  & [Sc/Fe] & [Cu/Fe]  \\
\hline
\endhead
\hline
\endfoot
RGB\_1055&  0.10    & 0.50 & -0.76 $\pm$ 0.05 &   0.20  & -0.57 \\
RGB\_1105&   -      & 0.02 & -0.43 $\pm$ 0.07 &   0.04  & -0.78 \\
RGB\_1118&  0.10    & 0.00 & -0.28 $\pm$ 0.10 &  -0.16  & -0.84 \\
RGB\_499 &  0.40    & 0.15 & -0.05 $\pm$ 0.07 &  -0.23  & -0.55 \\
RGB\_512 &    -     & 0.32 & -0.42 $\pm$ 0.07 &   0.10  & -0.80 \\
RGB\_522 &    -     & 0.40 & -0.03 $\pm$ 0.15 &  -0.18  & -0.46 \\
RGB\_533 &    -     & 0.30 & -0.08 $\pm$ 0.07 &  -0.15  & -0.62 \\
RGB\_534 &    -     & 0.28 & -0.25 $\pm$ 0.05 &  -0.11  & -0.60 \\
RGB\_546 &    -     & 0.10 & -0.49 $\pm$ 0.13 &   0.22  & -0.94 \\
RGB\_548 &    -     & 0.25 & -0.30 $\pm$ 0.10 &   0.00  & -0.46 \\
RGB\_565 &    -     & 0.32 & -0.15 $\pm$ 0.07 &  -0.06  & -0.83 \\
RGB\_576 &    -     & 0.31 & -0.27 $\pm$ 0.05 &   0.03  & -0.84 \\
RGB\_593 &    -     & 0.50 & -0.27 $\pm$ 0.06 &   0.04  & -0.75 \\
RGB\_599 &    -     & 0.30 & -0.41 $\pm$ 0.06 &  -0.10  & -0.84 \\
RGB\_601 &    -     & 0.33 &    -  $\pm$ 0.07 &  -0.20  & -0.40 \\
RGB\_606 & $<$-0.20 &    - &  0.09 $\pm$ 0.08 &  -0.10  &    -  \\
RGB\_611 &    -     & 0.12 & -0.41 $\pm$ 0.13 &  -0.26  & -0.90 \\
RGB\_614 & $<$0.15  &    - & -0.09 $\pm$ 0.05 &   -     &    -  \\
RGB\_620 & $<$0.05  & 0.10 &  0.06 $\pm$ 0.17 &  -0.12  & -0.30 \\
RGB\_625 &    -     & 0.00 & -0.06 $\pm$ 0.07 &  -0.20  & -0.80 \\
RGB\_629 &  0.30    & 0.14 & -0.46 $\pm$ 0.10 &  <0.00  & -0.90 \\
RGB\_631 &    -     & 0.20 & -0.03 $\pm$ 0.10 &  -0.03  & -0.68 \\
RGB\_633 &  0.10    &   -  & -0.20 $\pm$ 0.07 &  -0.50  & -0.50 \\
RGB\_640 &    -     & 0.12 & -0.25 $\pm$ 0.05 &  -0.20  & -0.85 \\
RGB\_646 &    -     & 0.17 & -0.39 $\pm$ 0.08 &  -0.22  & -0.80 \\
RGB\_651 &    -     & 0.15 & -0.38 $\pm$ 0.14 &  -0.20  & -0.44 \\
RGB\_655 &    -     & 0.36 & -0.28 $\pm$ 0.14 &  -0.10  & -0.80 \\
RGB\_656 &    -     & 0.30 & -0.06 $\pm$ 0.12 &  -0.30  & -0.64 \\
RGB\_658 &    -     &  -   & -0.10 $\pm$ 0.09 &   -     &   -   \\
RGB\_664 &    -     & 0.30 & -0.24 $\pm$ 0.12 &  -0.30  & -0.90 \\
RGB\_666 &    -     & 0.22 & -0.37 $\pm$ 0.07 &   -     & -0.85 \\
RGB\_671 &    -     &   -  & -0.23 $\pm$ 0.08 &  -0.10  & -0.50 \\
RGB\_672 &    -     & 0.10 &  0.20 $\pm$ 0.14 &  -0.33  & -0.64 \\
RGB\_679 &    -     & 0.00 & -0.36 $\pm$ 0.09 &  -0.37  & -0.70 \\
RGB\_690 &    -     & 0.36 & -0.18 $\pm$ 0.12 &  -0.23  & -0.68 \\
RGB\_699 &    -     & 0.42 &  0.19 $\pm$ 0.08 &   0.00  & -0.80 \\
RGB\_700 & $<$-0.05 & 0.10 & -0.31 $\pm$ 0.11 &$<$-0.20 & -0.67 \\
RGB\_701 &    -     & 0.30 &  0.08 $\pm$ 0.10 &  -0.04  & -0.50 \\
RGB\_705 &    -     & 0.26 &  0.26 $\pm$ 0.09 &  -0.19  & -0.75 \\
RGB\_710 &    -     & 0.30 & -0.27 $\pm$ 0.08 &  -0.08  & -0.58 \\
RGB\_720 & $<$0.35  & 0.20 & -0.29 $\pm$ 0.11 &$<$-0.30 &   -   \\
RGB\_728 &    -     & 0.11 & -0.33 $\pm$ 0.07 &  -0.20  & -0.75 \\
RGB\_731 &   0.00   & 0.03 & -0.29 $\pm$ 0.08 &  -0.32  & -0.76 \\
RGB\_748 & $<$0.00  & 0.20 &  0.53 $\pm$ 0.20 &   0.00  & -0.60 \\
RGB\_752 &    -     & 0.20 & -0.28 $\pm$ 0.12 &  -0.27  & -0.53 \\
RGB\_756 &    -     & 0.05 &  0.09 $\pm$ 0.08 &$<$-0.30 & -0.75 \\
RGB\_758 &    -     &  -   &  0.08 $\pm$ 0.12 &   -     &   -   \\
RGB\_766 & $<$0.10  & 0.00 & -0.09 $\pm$ 0.13 &   -     & -0.50 \\
RGB\_773 &   0.25   & 0.02 & -0.03 $\pm$ 0.07 &  -0.32  & -0.57 \\
RGB\_775 &    -     & 0.20 & -0.41 $\pm$ 0.04 &   0.14  & -0.83 \\
RGB\_776 &    -     & 0.12 & -0.53 $\pm$ 0.10 &  -0.13  & -0.80 \\
RGB\_782 &    -     & 0.30 & -0.19 $\pm$ 0.13 &  -0.28  & -0.80 \\
RGB\_789 &    -     & 0.04 & -0.28 $\pm$ 0.14 &  -0.30  & -0.74 \\
RGB\_793 & $<$0.10  & 0.08 & -0.39 $\pm$ 0.10 &  -0.30  & -0.80 \\
RGB\_834 &    -     & 0.13 & -0.32 $\pm$ 0.04 &  -0.25  & -0.76 \\
RGB\_854 &    -     & 0.30 & -0.20 $\pm$ 0.12 &  -0.10  & -0.50 \\
RGB\_855 &    -     & 0.10 & -0.17 $\pm$ 0.09 &  -0.20  & -0.38 \\
RGB\_859 &    -     & 0.22 & -0.21 $\pm$ 0.10 &  -0.22  & -0.42 \\
RGB\_900 &    -     & 0.34 & -0.13 $\pm$ 0.11 &  -0.15  & -0.68 \\
\end{longtable}

\begin{longtable}{ccccccc}\label{AbIron} \\
\caption{Abundance ratios of the elements (continuation). FeII and Iron-peak elemens.}\\
\hline\hline
Star  &   [Cr/Fe]  & [V/Fe] &  [Ni/Fe] &  [Co/Fe]  \\
\hline
\endfirsthead
\caption{continued.}\\
\hline\hline
Star  &   [Cr/Fe]  & [V/Fe] &  [Ni/Fe] &  [Co/Fe]  \\
\hline
\endhead
\hline
\endfoot
RGB\_1055&  -0.47 $\pm$ 0.12 & -0.40 $\pm$ 0.04 & -0.32 $\pm$ 0.08 &  $<$0.00   \\
RGB\_1105&  -0.32 $\pm$ 0.11 & -0.06 $\pm$ 0.05 & -0.26 $\pm$ 0.08 &  -0.10   \\
RGB\_1118&  -0.19 $\pm$ 0.09 & -0.19 $\pm$ 0.08 & -0.20 $\pm$ 0.06 &  -0.20   \\
RGB\_499 &   0.11 $\pm$ 0.12 &  0.14 $\pm$ 0.06 & -0.23 $\pm$ 0.09 &  -0.08   \\
RGB\_512 &  -0.18 $\pm$ 0.11 &  0.06 $\pm$ 0.08 & -0.15 $\pm$ 0.06 &  -0.05   \\
RGB\_522 &  -0.08 $\pm$ 0.11 &  0.17 $\pm$ 0.06 & -0.07 $\pm$ 0.05 &  -0.12   \\
RGB\_533 &  -0.13 $\pm$ 0.10 &  0.15 $\pm$ 0.06 & -0.23 $\pm$ 0.07 &  -0.10   \\
RGB\_534 &  -0.29 $\pm$ 0.08 & -0.03 $\pm$ 0.05 & -0.11 $\pm$ 0.05 &   0.00   \\
RGB\_546 &  -0.22 $\pm$ 0.13 & -0.15 $\pm$ 0.05 & -0.30 $\pm$ 0.06 &  -0.10   \\
RGB\_548 &  -0.30 $\pm$ 0.10 & -0.06 $\pm$ 0.06 & -0.27 $\pm$ 0.08 &  -0.19   \\
RGB\_565 &  -0.15 $\pm$ 0.10 &  0.20 $\pm$ 0.05 & -0.19 $\pm$ 0.04 &  -0.10   \\
RGB\_576 &  -0.43 $\pm$ 0.10 & -0.14 $\pm$ 0.05 & -0.14 $\pm$ 0.03 &   0.05   \\
RGB\_593 &  -0.24 $\pm$ 0.15 &  0.00 $\pm$ 0.05 & -0.24 $\pm$ 0.06 &  -0.05   \\
RGB\_599 &  -0.25 $\pm$ 0.12 & -0.28 $\pm$ 0.06 & -0.24 $\pm$ 0.09 &  -0.10   \\
RGB\_601 &   0.05 $\pm$ 0.10 & -0.03 $\pm$ 0.04 & -0.21 $\pm$ 0.04 &  -0.18   \\
RGB\_606 &   0.35 $\pm$ 0.12 &    -  $\pm$ 0.36 & -0.46 $\pm$ 0.07 &  $<$0.10  \\
RGB\_611 &  -0.18 $\pm$ 0.06 & -0.03 $\pm$ 0.08 & -0.28 $\pm$ 0.02 &  -0.36   \\
RGB\_614 &  -0.17 $\pm$ 0.07 &  0.24 $\pm$ 0.07 & -0.41 $\pm$ 0.10 &    -     \\
RGB\_620 &  -0.07 $\pm$ 0.18 &  0.11 $\pm$ 0.08 & -0.16 $\pm$ 0.11 &  $<$0.00  \\
RGB\_625 &  -0.11 $\pm$ 0.12 &  0.18 $\pm$ 0.08 & -0.18 $\pm$ 0.12 &   0.10   \\
RGB\_629 &  -0.45 $\pm$ 0.12 & -0.34 $\pm$ 0.04 & -0.24 $\pm$ 0.11 & $<$-0.20  \\
RGB\_631 &  -0.12 $\pm$ 0.12 & -0.02 $\pm$ 0.07 & -0.26 $\pm$ 0.06 &  -0.14   \\
RGB\_633 &  -0.20 $\pm$ 0.10 & -0.06 $\pm$ 0.02 & -0.23 $\pm$ 0.06 &  -0.04   \\
RGB\_640 &  -0.14 $\pm$ 0.14 & -0.02 $\pm$ 0.06 & -0.08 $\pm$ 0.05 & $<$-0.20   \\
RGB\_646 &  -0.38 $\pm$ 0.06 & -0.25 $\pm$ 0.06 & -0.06 $\pm$ 0.03 &  -0.10   \\
RGB\_651 &  -0.28 $\pm$ 0.12 & -0.16 $\pm$ 0.06 & -0.26 $\pm$ 0.07 &  -0.15   \\
RGB\_655 &  -0.09 $\pm$ 0.12 &  0.12 $\pm$ 0.06 & -0.13 $\pm$ 0.08 &  -0.20   \\
RGB\_656 &  -0.21 $\pm$ 0.12 &  0.02 $\pm$ 0.05 & -0.25 $\pm$ 0.10 &  -0.09   \\
RGB\_658 &  -0.56 $\pm$ 0.08 &  0.16 $\pm$ 0.07 & -0.04 $\pm$ 0.09 &    -     \\
RGB\_664 &  -0.29 $\pm$ 0.09 & -0.24 $\pm$ 0.07 & -0.36 $\pm$ 0.11 &  -0.27   \\
RGB\_666 &  -0.43 $\pm$ 0.09 & -0.38 $\pm$ 0.05 & -0.27 $\pm$ 0.06 &    -     \\
RGB\_671 &  -0.28 $\pm$ 0.12 & -0.08 $\pm$ 0.06 & -0.23 $\pm$ 0.07 &  -0.09   \\
RGB\_672 &  -0.21 $\pm$ 0.12 & -0.17 $\pm$ 0.05 & -0.21 $\pm$ 0.04 &  -0.22   \\
RGB\_679 &  -0.25 $\pm$ 0.14 & -0.02 $\pm$ 0.05 & -0.28 $\pm$ 0.06 &  -0.14   \\
RGB\_690 &  -0.12 $\pm$ 0.12 &  0.01 $\pm$ 0.09 & -0.19 $\pm$ 0.05 &  -0.22   \\
RGB\_699 &  -0.08 $\pm$ 0.06 &   -   $\pm$ 0.06 & -0.50 $\pm$ 0.17 &  -0.17   \\
RGB\_700 &  -0.48 $\pm$ 0.12 & -0.20 $\pm$ 0.08 & -0.34 $\pm$ 0.09 &  -0.20   \\
RGB\_701 &   0.03 $\pm$ 0.07 &  0.23 $\pm$ 0.08 & -0.12 $\pm$ 0.14 &  -0.12   \\
RGB\_705 &  -0.13 $\pm$ 0.09 & -0.26 $\pm$ 0.04 & -0.25 $\pm$ 0.09 &  -0.25   \\
RGB\_710 &  -0.24 $\pm$ 0.05 & -0.43 $\pm$ 0.04 & -0.34 $\pm$ 0.06 &  -0.20   \\
RGB\_720 &  -0.11 $\pm$ 0.06 &  0.05 $\pm$ 0.06 & -0.16 $\pm$ 0.12 &  -0.29    \\
RGB\_728 &  -0.17 $\pm$ 0.12 & -0.28 $\pm$ 0.08 & -0.20 $\pm$ 0.12 &  -0.08   \\
RGB\_731 &  -0.17 $\pm$ 0.10 & -0.29 $\pm$ 0.04 & -0.28 $\pm$ 0.02 &  -0.20   \\
RGB\_748 &   0.25 $\pm$ 0.15 & -0.18 $\pm$ 0.13 & -0.37 $\pm$ 0.13 &  -0.10   \\
RGB\_752 &  -0.18 $\pm$ 0.07 & -0.15 $\pm$ 0.07 & -0.31 $\pm$ 0.09 &  -0.13   \\
RGB\_756 &  -0.02 $\pm$ 0.09 &  0.13 $\pm$ 0.05 & -0.19 $\pm$ 0.09 &  -0.10   \\
RGB\_758 &  -0.27 $\pm$ 0.09 & -0.08 $\pm$ 0.09 & -0.34 $\pm$ 0.11 &    -     \\
RGB\_766 &  -0.05 $\pm$ 0.10 &  0.25 $\pm$ 0.07 & -0.08 $\pm$ 0.03 &  -0.09   \\
RGB\_773 &  -0.10 $\pm$ 0.07 &    -  $\pm$ 0.06 & -0.33 $\pm$ 0.04 &  -0.01   \\
RGB\_775 &  -0.18 $\pm$ 0.06 &    -  $\pm$ 0.03 & -0.37 $\pm$ 0.03 &  -0.22   \\
RGB\_776 &  -0.21 $\pm$ 0.14 & -0.26 $\pm$ 0.05 & -0.27 $\pm$ 0.06 &  -0.22   \\
RGB\_782 &  -0.13 $\pm$ 0.10 & -0.19 $\pm$ 0.07 & -0.22 $\pm$ 0.02 &    -     \\
RGB\_789 &  -0.19 $\pm$ 0.07 &  0.28 $\pm$ 0.05 & -0.21 $\pm$ 0.06 &  -0.20   \\
RGB\_793 &  -0.19 $\pm$ 0.13 & -0.14 $\pm$ 0.05 & -0.29 $\pm$ 0.06 &  -0.20   \\
RGB\_834 &  -0.13 $\pm$ 0.11 &  0.05 $\pm$ 0.05 & -0.26 $\pm$ 0.07 &     -    \\
RGB\_854 &  -0.15 $\pm$ 0.12 & -0.01 $\pm$ 0.09 & -0.09 $\pm$ 0.05 &   0.12   \\
RGB\_855 &  -0.07 $\pm$ 0.14 & -0.04 $\pm$ 0.10 & -0.36 $\pm$ 0.12 &  -0.04   \\
RGB\_859 &  -0.37 $\pm$ 0.11 & -0.05 $\pm$ 0.05 & -0.15 $\pm$ 0.03 &  -0.08   \\
RGB\_900 &  -0.14 $\pm$ 0.10 & -0.03 $\pm$ 0.06 & -0.18 $\pm$ 0.05 &  -0.12
\end{longtable}

\begin{longtable}{lcccclcccc}\label{AbHeavy} \\
\caption{Abundance ratios of the elements (continuation). Heavy and light $s$-process elements.} \\
\hline \hline
Star & [La/Fe]& [Ba/Fe] & [Y/Fe] & [Zr/Fe] & Star & [La/Fe]& [Ba/Fe] & [Y/Fe] & [Zr/Fe]   \\
\hline
\endhead
RGB\_1055 & $<$0.05 &  0.40  &$<$-0.40 &   -    &   RGB\_666  & $<$0.10 &  0.65  &$<$-0.20 &$<$-0.30 \\
RGB\_1105 &  0.15   &  0.40  &$<$-0.48 & -0.50  &   RGB\_671  &  0.15   &  0.70  &$<$-0.60 & -0.65  \\
RGB\_1118 &  1.12   &  1.00  &   0.14  &  0.00  &   RGB\_672  &  0.20   &  0.60  &  -0.30  & -0.62  \\
RGB\_499  &  0.48   &  0.35  &   0.27  &  0.00  &   RGB\_679  &  0.40   &  0.65  &  -0.24  & -0.60  \\
RGB\_512  &  0.50   &  0.50  &  -0.04  & -0.17  &   RGB\_690  &  0.66   &  0.95  &  -0.20  & -0.32  \\
RGB\_522  &  0.40   &  0.60  &   0.00  & -0.35  &   RGB\_699  &  0.12   &  0.60  &$<$-0.50 & -0.80  \\
RGB\_533  &  0.40   &  0.15  &  -0.20  & -0.38  &   RGB\_700  & $<$0.30 & -0.10  &  -0.45  & -0.90  \\
RGB\_534  & $<$0.20 &  0.20  & $<$0.10 &$<$-0.35&   RGB\_701 &  0.24   &  0.30  &   0.00  & -0.24  \\
RGB\_546  &  0.62   &  0.65  &  -0.20  & -0.20  &   RGB\_705  &  0.32   &  0.40  &  -0.37  & -0.80  \\
RGB\_548  &  0.32   &  0.70  &  -0.30  & -0.61  &   RGB\_710  &  0.40   &  0.60  &  -0.46  & -0.78  \\
RGB\_565  &  0.20   &  0.20  &  -0.18  & -0.30  &   RGB\_720  &   -     &  0.35  &    -    & -0.60  \\
RGB\_576  &  0.32   &  0.45  &$<$-0.20 & -0.28  &   RGB\_728  & $<$0.10 &  0.50  &  -0.20  &$<$-0.50\\
RGB\_593  & $<$0.20 &  0.00  &  -0.20  &$<$-0.45&   RGB\_731  &  0.05   &  0.25  &  -0.60  & -0.70  \\
RGB\_599  &  0.25   &  0.50  &$<$-0.40 & -0.40  &   RGB\_748  &$<$-0.30 &   0.4  &   -0.3  & $<$-0.8 \\
RGB\_601  &   -     &  0.80  &   0.00  & -0.43  &   RGB\_752  &  0.37   &  0.85  &  -0.45  & -0.75  \\
RGB\_606  &  0.30   &  0.80  &     -   &   -    &   RGB\_756  &  0.20   &  0.45  &  -0.33  & -0.40  \\
RGB\_611  &  0.30   &  0.60  &  -0.52  & -0.70  &   RGB\_758  &   -     &    -   &    -    &   -    \\
RGB\_614  &   -     &  0.40  &   0.00  &   -    &   RGB\_766  &  0.60   &  0.50  &$<$-0.25 &$<$-1.00\\
RGB\_620  & $<$0.50 &  0.50  &  -0.10  &$<$-0.40&   RGB\_773  &  0.25   &  0.00  &  -0.10  & -0.42  \\
RGB\_625  &  0.18   &  0.20  &  -0.30  & -0.24  &   RGB\_775  &  0.30   &  0.60  &$<$-0.80 & -0.40  \\
RGB\_629  & $<$0.00 &  0.20  &  -0.50  &$<$-0.60&   RGB\_776  &  0.40   &  0.55  &$<$-0.80 & -0.84  \\
RGB\_631  &  0.20   &  0.50  &$<$-0.50 &$<$-1.00&   RGB\_782  & $<$0.10 &  0.45  &  -0.52  & -0.87  \\
RGB\_633  &  0.23   &  0.35  &  -0.27  & -0.55  &   RGB\_789  &  0.00   &  0.55  &  -0.20  & -0.60  \\
RGB\_640  & $<$0.10 &  0.10  &$<$-0.50 &$<$-0.20&   RGB\_793  &  0.24   &  0.30  &  -0.30  & -0.49  \\
RGB\_646  &  0.47   &  0.50  &$<$-0.50 &$<$-0.40&   RGB\_834  &  0.17   &  0.25  &  -0.28  & -0.40  \\
RGB\_651  &  0.32   &  0.63  &  -0.50  & -0.77  &   RGB\_854  & $<$0.35 &  0.40  &  -0.25  & -0.50  \\
RGB\_655  &  0.40   &  0.70  &  -0.32  & -0.38  &   RGB\_855  &    -    &  0.55  &  -0.20  & -0.50  \\
RGB\_656  &  0.32   &  0.55  &  -0.40  & -0.60  &   RGB\_859  &  0.40   &  0.80  &$<$-0.50 & -0.56  \\
RGB\_658  &    -    &  0.80  &  -0.30  &  -     &   RGB\_900  &  0.40   &  0.50  &  -0.25  & -0.55  \\
RGB\_664  &  0.26   &  0.25  &  -0.68  & -0.80  &  & & & & \\
\hline
\end{longtable}

\begin{table}[htb]
\caption{Errors due to stellar parameters uncertainties.}
\label{ErrorPar}
\begin{tabular}{lccccc}
\hline
Element &   $\Delta$T$_{\rm eff}$=+100 K & $\Delta$logg=-0.4 &
$\Delta$V$_{t}$=+0.2 km/s &  $\Delta$[Fe/H]=-0.15 & $\delta_{tot}$  \\
\hline
\hline
$\rm[Fe\,I/H]$   &-0.01  &-0.10  &-0.12  &-0.03   & 0.16\\
$\rm[O\,I/FE]$   & 0.04  &-0.06  & 0.10  &-0.02   & 0.12\\
$\rm[V\,I/FE]$   & 0.16  & 0.06  & 0.00  & 0.04   & 0.17\\
$\rm[Y\,I/FE]$   & 0.19  & 0.08  & 0.07  & 0.04   & 0.22\\
$\rm[Ca\,I/FE]$  & 0.11  & 0.11  &-0.01  & 0.04   & 0.16\\
$\rm[Cr\,I/FE]$  & 0.12  & 0.07  & 0.00  & 0.04   & 0.15\\
$\rm[Fe\,II/Fe]$ &-0.18  &-0.16  & 0.08  &-0.07   & 0.26\\
$\rm[Mg\,I/Fe]$  & 0.01  & 0.06  & 0.01  & 0.01   & 0.06\\
$\rm[Na\,I/Fe]$  & 0.11  & 0.10  & 0.06  & 0.05   & 0.17\\
$\rm[Ni\,I/Fe]$  & 0.01  &-0.03  & 0.05  &-0.01   & 0.06\\
$\rm[Si\,I/Fe]$  &-0.07  &-0.01  & 0.09  &-0.01   & 0.12\\
$\rm[Ti\,I/Fe]$  & 0.15  & 0.08  & 0.01  & 0.04   & 0.17\\
$\rm[Ti\,II/Fe]$ &-0.05  &-0.10  & 0.04  &-0.04   & 0.12\\
$\rm[Zr\,I/Fe]$  & 0.19  & 0.06  & 0.03  & 0.03   & 0.20\\
$\rm[Ba\,II/Fe]$ & 0.04  &-0.04  &-0.04  & 0.00   & 0.07\\
$\rm[Co\,I/Fe]$  & 0.03  &-0.01  & 0.05  &-0.01   & 0.06\\
$\rm[Cu\,I/Fe]$  & 0.07  &-0.04  & 0.06  & 0.07   & 0.12\\
$\rm[La\,II/Fe]$ & 0.04  &-0.06  & 0.05  &-0.04   & 0.10\\
$\rm[Sc\,II/Fe]$ &-0.01  &-0.04  & 0.01  &-0.01   & 0.04\\
\hline
\end{tabular}\\[2pt]
\end{table}

\appendix{

\section{Comparison between equivalent width from DAOSPEC and from Splot-Iraf}

In order to evaluate the quality of the DAOSPEC estimates, we have derived by eye inspection
the EW of six stars, using the Splot Iraf task. We have chosen stars in a range of S/N ratio
typical of our total sample in order to better evaluate the errors: S/N = 54 for RGB\_522, 59 for RGB\_546, 47 for RGB\_664, 42 for RGB\_666, 26 for RGB\_720, and 47 for RGB\_1055. The detailed values are given in Table A.1. We have found two problems relative to the DAOSPEC results from GIRAFFE spectra: a) not all blends have been identified, nevertheless all lines with weak blends (which are most
of the lines) have been correctly analysed by the iterative process of the program; b) cosmic rays also have not been identified, and those lines too near CR features must be discarded. Therefore the use of  DAOSPEC requires spectra and line lists as clear as possible from blends and spectra as clean from cosmic hits as possible.
As can be seen in Table A.1, we have found a very good agreement between the program results and
those from the Splot manual measurements. The average differece EW(Dao) - EW(Splot) is 0.46 m{\AA} for the six stars, with no strong systematic trend in one direction or the other (the mean difference for each star ranges from $-$3.7 to +5~m{\AA}). However, the dispersion of the measurements around the mean are higher, between 9.4m{\AA} for RGB\_1055 and 22.2m{\AA} for RGB\_666. These dispersion seem to anticorrelate with S/N as expected (the two
stars with the highest dispersion are the two lowest quality
spectra), and may also correlate with temperature, although our
sample of 6 stars is not quite high enough to investigate these
dependancies any further.

We have checked for systematic trends on the EW with wavelength and with the EW strength by plotting the differences EW(Dao) - EW(Splot) vs. wavelength and EW(Dao) - EW(Splot) vs. EW(Dao) for four stars, as shown in Fig. A.1. No trends have been found and confort us in the validity of using DAOSPEC EW mesurement for atmospheric parameter determinations (effective temperature and microturbulence velocities).

In Table A.2 we give the differences for the abundances derived from DAOSPEC and Splot,
Ab(Dao) - Ab(Splot). As can be seen in this Table, the agreement between the two methods
is good for most of the elements, always better than the typical
errorbars given for our measurement of the corresponding elements, and
usually below 0.10~dex. The differences are higher for those elements with fewer
lines (e.g. Na I and Ti I), which increases the weight of the scatter
among lines. Let us note in particular that elements such as Ca~I
or Ti~I do not seem to be affected by the method
used for equivalent width measurement in a systematic direction, so
that the strong underabundances found for these elements (with
respect to the galactic trends) are robust against EW systematics.

\begin{figure}
\centerline{%
\begin{tabular}{c@{\hspace{2pc}}l}
\includegraphics[width=3 in]{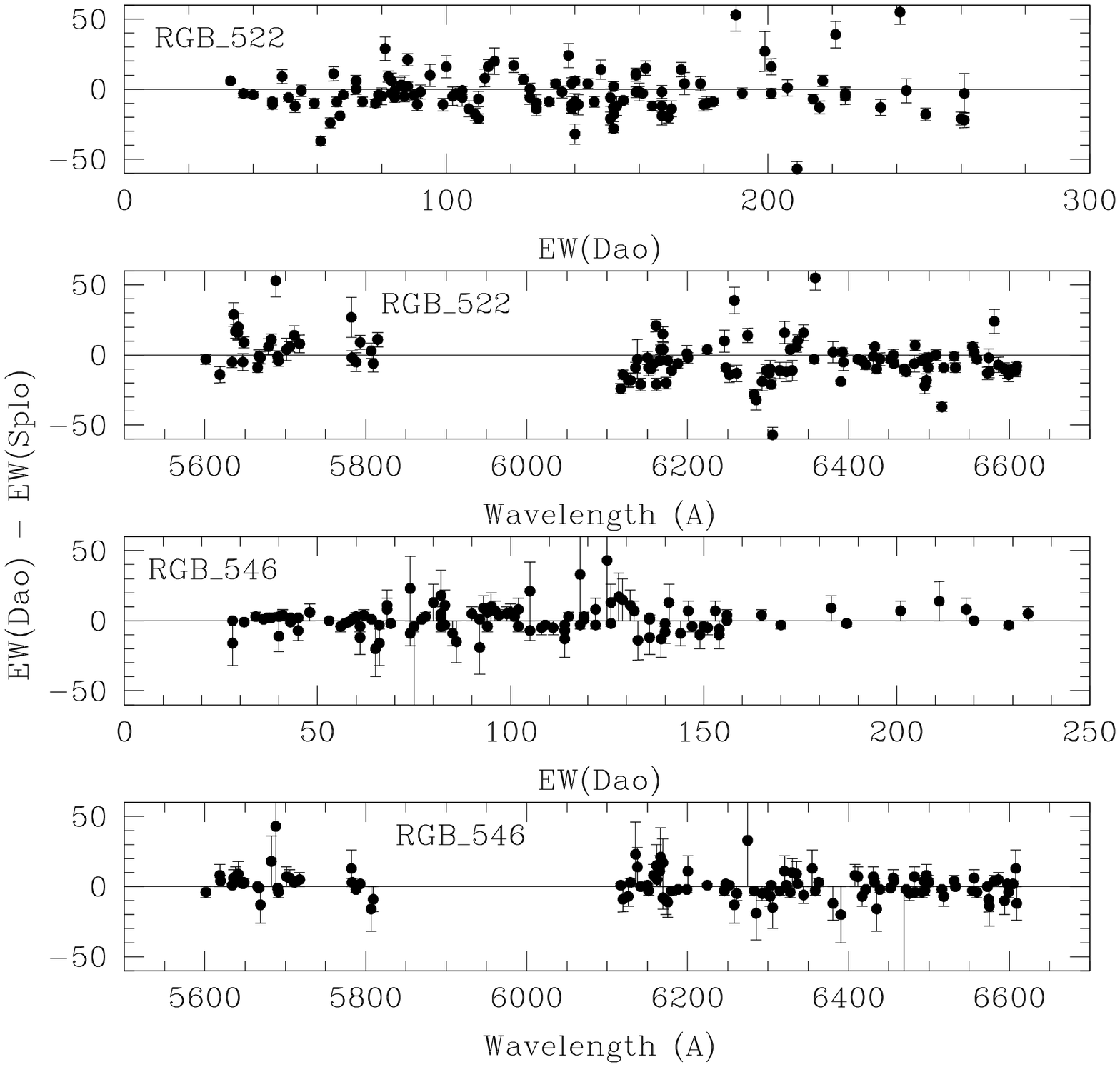} &
\includegraphics[width=3 in]{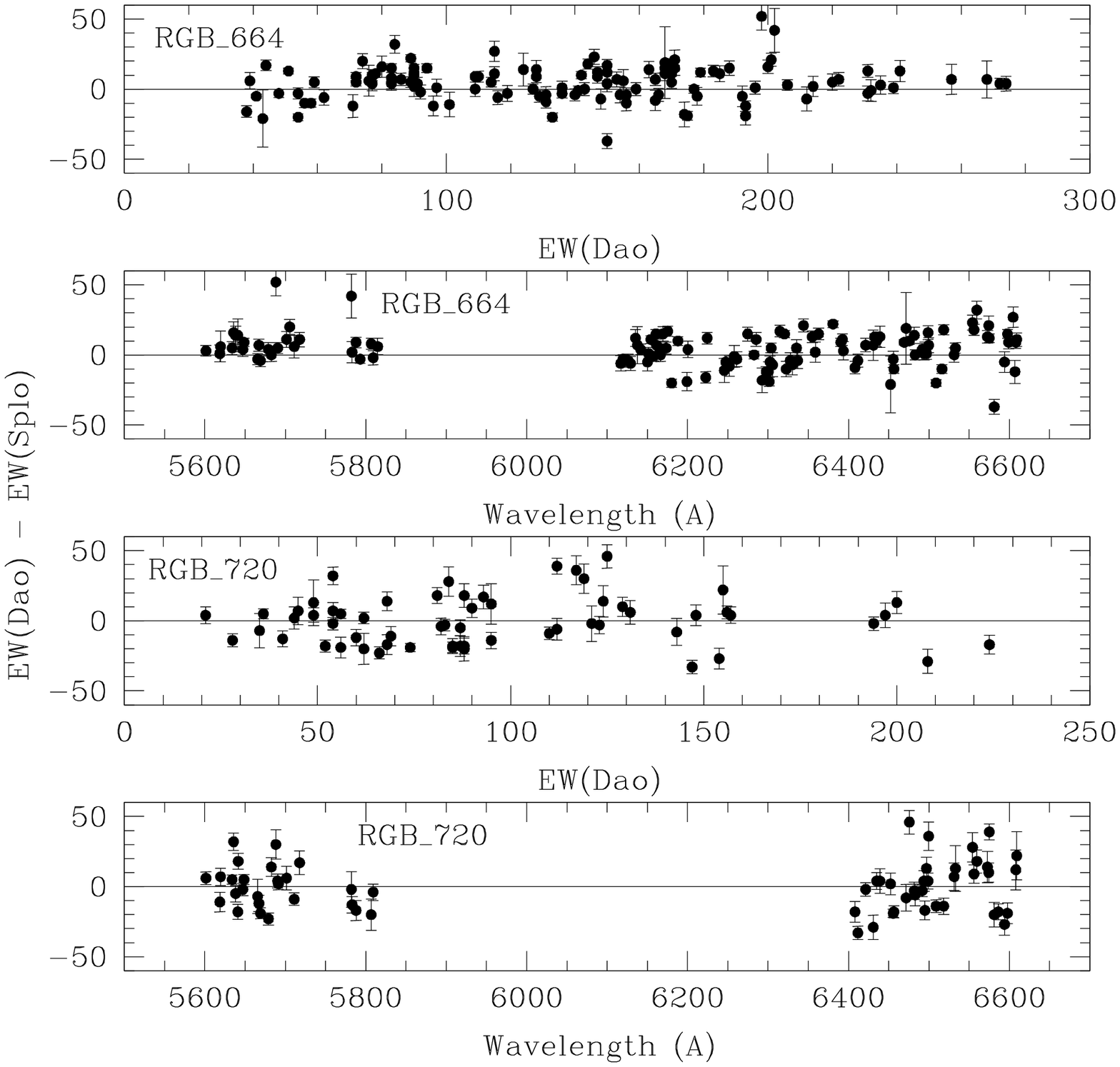} \\
\end{tabular}}
\caption{Difference between the EW derived using the DAOSPEC program and the
Iraf task Splot: trends with respect to the EW(Dao) and to the wavelenght of
the lines for RGB\_522 and RGB\_546 (left plot) and for RGB\_720 and RGB\_664
(right plot).}
\end{figure}
\hspace{\fill}

}

\scriptsize
\begin{longtable}{cccccccccccccc}
\caption{Comparison of the equivalent widths derived from DAOSPEC and Iraf-Splot task for
six of our sample stars.}\\
\label{Tab_EW} \\
\hline
\hline
     &    &   RGB\_522 &   &  RGB\_546 & & RGB\_664 &  & RGB\_666 &  & RGB\_720 & &  RGB\_1055 & \\
\hline
Line &  Element & Dao & Splot & Dao & Splot & Dao & Splot & Dao & Splot & Dao & Splot & Dao & Splot \\
\hline
\endhead
6300.31  &  O1 &   - &  -   &    -  &  -   &  96  & 108 &   - &    -  &   - &   -   &   -  &  -  \\
6274.66  &  V1 & 173 & 159  &  118  &  85  & 171  & 156 &  -  &    -  &   - &   -   &  75  &  54 \\
6285.17  &  V1 & 140 & 172  &   92  & 111  & 170  & 159 &  71 &   61  &   - &   -   &  61  &   - \\
6199.19  &  V1 & 206 & 205  &  126  & 128  & 193  & 212 &  90 &  186  &   - &   -   &  79  &  72 \\
6292.82  &  V1 & 167 & 186  &  111  & 116  & 174  & 192 &  70 &   90  &   - &   -   &   -  &  -  \\
6224.51  &  V1 & 134 & 130  &   77  &  76  & 147  & 135 &  44 &   54  &   - &   -   &  32  &  24 \\
6251.82  &  V1 & 170 & 184  &  119  & 118  & 165  & 173 &  69 &   24  &   - &   -   &   -  &  -  \\
6150.15  &  V1 & 167 & 169  &   92  &  91  & 178  & 183 &  44 &   44  &   - &   -   &  54  &  53 \\
6135.37  &  V1 & 146 & 155  &   74  &  51  & 150  & 138 &  47 &   51  &   - &   -   &  40  &   - \\
6119.53  &  V1 & 139 & 153  &   85  &  94  & 136  & 139 &  63 &   96  &   - &   -   &  65  &  89 \\
6452.32  &  V1 & 103 & 106  &   43  &  44  &  43  &  64 &  21 &   32  &  44 &  42   &   -  &  -  \\
6531.41  &  V1 & 105 & 106  &   41  &  37  & 109  & 109 &   - &   -   &  45 &  38   &   -  &  -  \\
6357.29  &  V1 &  37 &  40  &    -  &  -   &   -  &  -  &   - &   -   &   - &   -   &   -  &  -  \\
6222.58  &  Y1 &   - &  -   &    -  &  -   &  38  &  54 &   - &   -   &   - &   -   &   -  &  -  \\
6435.01  &  Y1 &  78 &  88  &   28  &  44  &  77  &  67 &  22 &   47  &  21 &  17   &   -  &  -  \\
6613.73  &  Y2 &   - &  -   &    -  &  -   &   -  &  -  &   - &   -   &  94 &  -    &   -  &  -  \\
6496.90  & BA2 & 249 & 267  &  218  & 210  & 239  & 238 & 209 &  206  & 200 &   187 &   -  &  -  \\
6141.73  & BA2 & 260 & 281  &  220  & 220  & 272  & 268 & 201 &  207  &  -  &   -   &   -  &  -  \\
6572.80  & CA1 & 216 & 229  &  156  & 156  & 231  & 218 & 125 &  145  & 124 &   110 & 107  & 120 \\
6162.19  & CA1 & 308 & 334  &  234  & 229  & 312  & 319 & 211 &  215  &  -  &   -   & 198  & 194 \\
6169.56  & CA1 & 179 & 175  &  140  & 148  & 188  & 173 & 143 &  167  &  -  &   -   & 134  & 142 \\
6169.04  & CA1 & 162 & 147  &  128  & 111  & 163  & 138 & 119 &  137  &  -  &   -   & 106  &  99 \\
5601.29  & CA1 & 201 & 204  &  150  & 154  & 206  & 203 & 125 &  129  & 156 &   150 & 117  & 129 \\
6493.79  & CA1 & 192 & 195  &  165  & 161  & 196  & 195 & 117 &  107  & 148 &   144 & 130  & 125 \\
6166.44  & CA1 & 144 & 140  &  105  &  84  & 143  & 143 &  98 &   99  &  -  &   -   & 104  & 111 \\
6499.65  & CA1 & 159 & 161  &  115  & 112  & 153  & 146 & 114 &  116  & 117 &   81  & 101  & 106 \\
6161.30  & CA1 & 151 & 172  &  129  & 114  & 165  & 158 &  75 &   55  &  -  &   -   &  81  &  88 \\
6455.61  & CA1 & 126 & 126  &   94  &  88  & 119  & 122 &  85 &   74  & 74  &  93   &  76  &  76 \\
6439.08  & CA1 & 224 & 227  &  187  & 189  & 241  & 228 & 182 &  163  & 197 &   193 & 163  & 163 \\
6471.67  & CA1 & 164 & 176  &  140  & 142  & 168  & 149 & 117 &  116  & 143 &   151 & 122  & 134 \\
6508.84  & CA1 &  72 &  72  &    -  &  -   &  54  &  74 &   - &   -   & 28  &  42   &   -  &  -  \\
6282.60  & CO1 & 152 & 180  &  109  & 112  & 177  & 177 &  -  &    -  &  -  &   -   &  55  &  22 \\
6117.00  & CO1 &   - &  -   &    -  &  -   &   -  &  -  &   - &   -   &  -  &   -   &  29  &  18 \\
5647.24  & CO1 &  80 &  85  &   45  &  43  &  77  &  73 &  36 &   39  & 54  &  56   &  31  &  29 \\
6330.10  & CR1 & 141 & 152  &   95  &  85  & 148  & 155 &  75 &   44  &  -  &   -   &  69  &  74 \\
6362.88  & CR1 &   - &  -   &  101  &  98  & 168  & 153 &  64 &   29  &  -  &   -   &  66  &  55 \\
5787.93  & CR1 & 102 & 107  &   69  &  71  & 110  & 101 &  39 &    -  & 68  &  85   &  47  &  51 \\
5783.07  & CR1 &  92 &  94  &   60  &  57  &  90  &  88 &  25 &    -  & 41  &  54   &  32  &   - \\
5782.13  & CU1 & 199 & 172  &  126  & 113  & 202  & 160 & 100 &   98  & 121 &   123 &   -  &  -  \\
6358.69  & FE1 & 241 & 186  &  170  & 173  & 214  & 212 & 119 &    -  &  -  &   -   & 164  &   - \\
6498.95  & FE1 & 183 & 192  &  156  & 152  & 200  & 184 & 137 &  134  & 157 &   153 & 126  & 131 \\
6574.25  & FE1 & 160 & 162  &  144  & 153  & 171  & 150 & 112 &  108  & 129 &   119 & 107  & 115 \\
6581.22  & FE1 & 138 & 114  &   97  &  93  & 150  & 187 &  92 &  110  & 88  &  108  &  75  &  72 \\
6430.86  & FE1 & 243 & 244  &  201  & 194  & 257  & 250 & 195 &  203  & 208 &   237 & 164  & 155 \\
6151.62  & FE1 & 140 & 149  &  118  & 121  & 136  & 135 & 100 &   13  &  -  &   -   &  92  &  98 \\
6335.34  & FE1 & 217 & 211  &  183  & 174  & 220  & 215 & 170 &  158  &  -  &   -   & 148  & 144 \\
6297.80  & FE1 & 180 & 191  &  147  & 151  & 193  & 205 & 137 &  160  &  -  &   -   & 123  & 117 \\
6173.34  & FE1 & 169 & 189  &  149  & 159  & 170  & 165 & 135 &  146  &  -  &   -   & 124  & 126 \\
6421.35  & FE1 & 214 & 221  &  187  & 189  & 222  & 215 & 200 &  197  & 194 &   196 & 157  & 159 \\
6481.88  & FE1 & 151 & 157  &  132  & 125  & 163  & 149 & 126 &  128  & 123 &   126 & 108  & 102 \\
6392.54  & FE1 &  88 &  86  &    -  &  -   &  90  &  79 &  32 &   26  &  -  &   -   &   -  &  -  \\
6392.54  & FE1 &  88 &  86  &    -  &  -   &  90  &  79 &  32 &   26  &  -  &   -   &   -  &  -  \\
6608.04  & FE1 &  99 & 110  &   80  &  67  & 128  & 119 &  32 &   32  & 95  &  83   &   -  &  -  \\
6494.99  & FE1 & 261 & 283  &  229  & 232  & 274  & 270 & 222 &  220  & 224 &   241 & 174  & 183 \\
6393.61  & FE1 & 224 & 229  &    -  &  -   & 235  & 232 & 188 &    -  &  -  &   -   & 167  & 161 \\
6344.16  & FE1 & 201 & 185  &  154  & 160  & 201  & 180 & 115 &  127  &  -  &   -   &   -  &  -  \\
6593.87  & FE1 & 181 & 191  &  154  & 164  & 192  & 197 & 141 &  177  & 154 &   181 & 120  & 117 \\
5701.56  & FE1 & 174 & 170  &  146  & 139  & 185  & 174 & 140 &  142  & 131 &   125 & 130  & 122 \\
6609.12  & FE1 & 155 & 163  &  136  & 148  & 168  & 157 &  87 &   94  & 155 &   133 & 122  & 108 \\
6475.63  & FE1 & 134 & 132  &  108  & 113  & 142  & 132 &  97 &   94  & 125 &   79  &  89  &  94 \\
6137.70  & FE1 & 261 & 264  &  211  & 197  & 268  & 261 & 200 &  218  &  -  &   -   & 174  & 168 \\
6322.69  & FE1 & 153 & 165  &  136  & 135  & 156  & 166 & 135 &  174  &  -  &   -   & 116  & 108 \\
6575.04  & FE1 & 167 & 179  &  133  & 147  & 179  & 167 & 137 &  161  & 112 &   73  & 102  & 107 \\
6200.32  & FE1 & 160 & 162  &  131  & 120  & 150  & 146 & 121 &  129  &  -  &   -   & 128  & 120 \\
6180.21  & FE1 & 141 & 152  &  114  & 117  & 133  & 153 &  94 &  133  &  -  &   -   &  94  & 106 \\
6518.37  & FE1 & 132 & 141  &  114  & 121  & 144  & 126 & 112 &  125  & 95  &  109  &  93  &  84 \\
6355.04  & FE1 &   - &  -   &  141  & 128  & 183  & 170 & 130 &  102  &  -  &   -   & 127  & 117 \\
6411.66  & FE1 & 161 & 164  &  153  & 146  & 156  & 160 & 137 &  136  & 147 &   180 & 118  & 118 \\
6301.51  & FE1 & 152 & 165  &    -  &  -   & 175  & 194 & 150 &  144  &  -  &   -   & 109  & 106 \\
6302.50  & FE1 & 128 & 138  &  105  & 112  & 129  & 134 &  87 &  110  &  -  &   -   &  89  &  87 \\
6336.83  & FE1 & 159 & 149  &  136  & 134  & 154  & 158 & 138 &  139  &  -  &   -   & 125  &  85 \\
6408.03  & FE1 &   - &  -   &  122  & 114  & 131  & 140 & 118 &  117  & 87  &  105  & 111  & 106 \\
5809.22  & FE1 & 105 & 111  &   74  &  83  &  92  &  94 &  60 &   63  & 82  &  86   &  62  &  51 \\
6188.02  & FE1 &  84 &  90  &   69  &  71  &  90  &  80 &  57 &   58  &  -  &   -   &  51  &  42 \\
6157.73  & FE1 & 126 & 132  &  102  &  94  & 141  & 142 &  93 &   94  &  -  &   -   &  86  &  88 \\
6165.36  & FE1 &  79 &  83  &   68  &  57  &  72  &  67 &  77 &  104  &  -  &   -   &  42  &  58 \\
6380.75  & FE1 &  87 &  85  &   61  &  73  &  89  &  67 &  66 &   12  &  -  &   -   &   -  &  -  \\
6380.75  & FE1 &  87 &  85  &   61  &  73  &  89  &  67 &  66 &   12  &  -  &   -   &   -  &  -  \\
5618.63  & FE1 & 107 & 121  &   68  &  60  &  97  &  96 &  61 &   65  & 69  &  80   &  60  &  56 \\
5638.27  & FE1 & 121 & 104  &   96  &  89  & 128  & 114 &  96 &   90  & 87  &  92   &  82  &  70 \\
5635.82  & FE1 &  81 &  52  &   48  &  42  &  80  &  64 &  45 &   47  & 54  &  22   &  40  &   - \\
5641.45  & FE1 & 115 &  95  &   93  &  84  &   -  &  -  &  77 &   75  & 81  &  63   &  69  &  63 \\
5814.81  & FE1 &  65 &  54  &    -  &  -   &  39  &  33 &   - &   -   &  -  &   -   &  25  &   - \\
5717.83  & FE1 & 112 & 104  &   90  &  85  & 115  & 104 &  77 &   70  & 93  &  76   &  67  &  66 \\
5705.47  & FE1 &  72 &  66  &   48  &  42  &  74  &  54 &  33 &   31  &  -  &   -   &  38  &  38 \\
5691.50  & FE1 &  90 &  94  &   56  &  60  &  79  &  73 &  36 &   35  & 62  &  60   &  43  &  41 \\
5619.61  & FE1 &   - &  -   &   41  &  37  &  76  &  70 &  44 &   34  & 54  &  47   &  29  &  23 \\
5806.73  & FE1 &  86 &  83  &   66  &  82  &  83  &  75 &  50 &    -  & 62  &  82   &  45  &  46 \\
5679.02  & FE1 &  83 &  77  &    -  &  -   &  83  &  79 &  59 &   54  & 66  &  89   &  52  &  46 \\
6597.56  & FE1 &  53 &  65  &   37  &  35  &  83  &  68 &  40 &   32  & 56  &  75   &  37  &  27 \\
6469.19  & FE1 & 128 & 138  &   75  & 213  & 109  & 100 &  62 &   70  & 92  &  200  &  55  &  58 \\
5633.95  & FE1 &  87 &  92  &   64  &  63  &  83  &  78 &  57 &   58  & 56  &  51   &  52  &  46 \\
6516.08  & FE2 &  61 &  98  &   57  &  59  &  58  &  68 &  -  &    -  &  -  &   -   &  60  &  64 \\
6432.68  & FE2 &  33 &  27  &   40  &  37  &  51  &  38 &  47 &   51  & 53  &  56   &   -  &  -  \\
6149.25  & FE2 &   - &  -   &    -  &  -   &   -  &  -  &  31 &   54  &  -  &   -   &   -  &  -  \\
6247.56  & FE2 &  46 &  55  &   43  &  41  &  41  &  46 &  57 &   50  &  -  &   -   &  38  &  53 \\
6456.39  & FE2 &  51 &  57  &   62  &  58  &  56  &  66 &  71 &   55  & 52  &  70   &  64  &  63 \\
6320.43  & LA2 & 100 &  84  &   83  &  72  &  90  &  75 &  40 &    -  &  -  &   -   &   -  &  -  \\
6390.48  & LA2 &  67 &  86  &   65  &  85  &  72  &  63 &  43 &   72  &  -  &   -   &   -  &  -  \\
6390.48  & LA2 &  67 &  86  &   65  &  85  &  72  &  63 &  43 &   72  &  -  &   -   &   -  &  -  \\
5711.09  & MG1 & 148 & 134  &  119  & 116  & 155  & 149 & 116 &  116  & 110 &  119  & 104  & 103 \\
5688.22  & NA1 & 190 & 137  &  125  &  82  & 198  & 146 & 100 &   76  & 119 &   89  &  37  &   - \\
5682.65  & NA1 & 159 & 148  &   82  &  64  & 159  & 159 &  57 &   68  & 68  &   54  &  45  &  45 \\
6160.75  & NA1 &  88 &  67  &    -  &  -   &  94  &  79 &  32 &   42  &  -  &   -   &   -  &  -  \\
6154.23  & NA1 &  59 &  69  &    -  &  -   &  78  &  67 &  24 &   20  &  -  &   -   &   -  &  -  \\
6327.60  & NI1 & 139 & 135  &  102  & 106  & 131  & 135 &  84 &  110  &  -  &   -   &  73  &  71 \\
6128.98  & NI1 & 109 & 127  &   78  &  75  & 116  & 122 &  68 &   70  &  -  &   -   &  60  &  56 \\
6314.67  & NI1 & 139 & 150  &  122  & 125  & 150  & 133 & 116 &  165  &  -  &   -   & 100  &  91 \\
6482.81  & NI1 & 124 & 117  &   94  &  98  & 127  & 127 &  89 &   88  & 112 &  118  &  77  &  77 \\
6532.89  & NI1 &  74 &  83  &   53  &  53  &  89  &  84 &  33 &   36  & 49  &  36   &   -  &  -  \\
6586.32  & NI1 & 110 & 117  &   99  &  94  &   -  &  -  & 105 &  125  & 88  &  106  &  82  &  96 \\
6175.37  & NI1 &  68 &  72  &   40  &  51  &  44  &  27 &  36 &   57  &  -  &   -   &   -  &  -  \\
6305.67  & SC1 & 209 & 266  &   86  & 101  & 212  & 219 &  58 &   60  &  -  &   -   &  29  &  28 \\
6604.60  & SC2 &  91 & 102  &   82  &  80  & 115  &  88 &  31 &   48  & 125 &   -   &  60  &  53 \\
5640.99  & SC2 & 113 &  97  &   82  &  77  & 124  & 110 &  85 &   79  & 85  &  103  &  72  &  68 \\
5669.04  & SC2 & 136 & 138  &  114  & 127  & 140  & 144 &  93 &   96  & 85  &  104  &  82  &  87 \\
5667.15  & SC2 &  84 &  85  &   58  &  59  &  86  &  79 &  69 &   53  & 60  &  72   &  54  &  51 \\
6245.62  & SC2 &  95 &  85  &   66  &  69  & 101  & 112 &  76 &  101  &  -  &   -   &  57  &  64 \\
5665.56  & SI1 &  66 &  75  &   28  &  28  &  54  &  57 &  32 &   41  & 35  &  42   &  31  &  16 \\
5690.43  & SI1 &  55 &  56  &   31  &  32  &  59  &  54 &  29 &    -  & 49  &  45   &  36  &  39 \\
5793.07  & SI1 &  49 &  40  &   38  &  36  &  48  &  51 &  39 &   39  &  -  &   -   &  25  &  30 \\
6599.11  & TI1 & 128 & 142  &   82  &  86  & 147  & 138 &   - &   -   &  -  &   -   &  34  &  44 \\
6126.22  & TI1 & 152 & 170  &  114  & 121  & 166  & 170 &  81 &  130  &  -  &   -   &  72  &  92 \\
6261.11  & TI1 & 235 & 248  &  151  & 156  & 231  & 234 &  75 &  120  &  -  &   -   & 103  & 104 \\
6554.24  & TI1 & 140 & 134  &   83  &  86  & 146  & 123 &  68 &   66  & 84  &  56   &  61  &  64 \\
6303.77  & TI1 & 110 & 131  &   59  &  58  & 114  & 109 &  50 &    -  &  -  &   -   &   -  &  -  \\
6258.10  & TI1 & 221 & 182  &  139  & 152  & 232  & 233 & 114 &    -  &  -  &   -   & 108  & 110 \\
6556.08  & TI1 & 152 & 150  &  100  &  94  & 169  & 151 &  83 &   71  & 90  &  81   &  64  &  58 \\
5648.58  & TI1 &  82 &  73  &   34  &  31  &  77  &  68 &   - &    -  & 36  &  31   &  23  &  18 \\
6559.59  & TI2 &  83 &  86  &   61  &  65  &  84  &  52 &  60 &   49  & 88  &  70   &  57  &  48 \\
6491.56  & TI2 & 103 & 107  &   75  &  79  &  91  &  87 &  67 &   81  & 83  &  86   &  59  &  63 \\
6606.95  & TI2 &  46 &  57  &    -  &  -   &  71  &  83 &   - &    -  &  -  &  -    &   -  &  -  \\
\hline
\hline
\end{longtable}

\begin{table}[htb]
\caption{Absolute differences Ab(Dao) - Ab(Splot) and the average value (see text).}
\begin{tabular}{cccccccc}
\hline
Element & RGB\_522 & RGB\_546 & RGB\_664 & RGB\_666 & RGB\_720 & RGB\_1055 & Average Difference\\								
\hline
CA1  & -0.03 & 0.13& -0.04 & 0.03& -0.06& -0.09 &-0.01 $\pm$ 0.08\\
CR1  & -0.08 & 0.03&  0.10 &-0.05& -0.05&  0.11 & 0.01 $\pm$ 0.08\\
FE1  & -0.11 & 0.01&  0.10 &-0.01& -0.06&  0.00 &-0.02 $\pm$ 0.07\\
FE2  & -0.13 &-0.06& -0.02 & 0.04& -0.05& -0.10 &-0.05 $\pm$ 0.06\\
NA1  &  0.08 & 0.09&  0.18 &-0.06&  0.06& -0.06 & 0.05 $\pm$ 0.09\\
NI1  & -0.03 &-0.06&  0.09 & 0.04& -0.12& -0.09 &-0.03 $\pm$ 0.08\\
SI1  & -0.05 & 0.01& -0.01 & 0.07&  0.20& -0.03 & 0.03 $\pm$ 0.09\\
TI1  & -0.06 & 0.00&  0.14 & 0.02& -0.16& -0.10 &-0.03 $\pm$ 0.10\\
TI2  & -0.18 &-0.07&  0.13 & 0.0 &  0.13&  0.06 & 0.01 $\pm$ 0.12\\
 V1  & -0.05 & 0.02&  0.10 &-0.22& -0.10& -0.09 &-0.06 $\pm$ 0.11\\
\hline
\end{tabular}\\[2pt]
\end{table}

\end{document}